\documentclass[11pt]{article}
\pdfoutput=1
\usepackage{aas_macros,amsmath,amssymb,comment,cite,esint,graphicx,mathtools}
\usepackage[margin=.8in,letterpaper]{geometry}
\usepackage[colorlinks=true]{hyperref}
\usepackage[affil-it]{authblk}
\usepackage{subcaption}
\usepackage[utf8]{inputenc}
\usepackage{mathrsfs}
\usepackage{appendix}
\usepackage{amssymb}
\usepackage{float}
\usepackage{color}
\usepackage{cite}
\usepackage{hyperref}
\hypersetup{pageanchor=false}
\usepackage{indentfirst}
\usepackage{url}
\usepackage{float}
\usepackage{caption}
\usepackage[numbers,square,comma,sort&compress,merge]{natbib}
\usepackage{esint}
\usepackage{overpic}
\usepackage{graphicx}
\usepackage{epsf,amsmath,bbold,amsfonts,stmaryrd}
\usepackage{textcomp}
\usepackage{ulem}
\usepackage{tikz}
\usepackage{amsfonts}
\usepackage{array}
\usepackage{longtable}
\usepackage[utf8]{inputenc}
\usepackage{multirow}

\numberwithin{equation}{section}
\let\originalleft\left
\let\originalright\right
\renewcommand{\left}{\mathopen{}\mathclose\bgroup\originalleft}
\renewcommand{\right}{\aftergroup\egroup\originalright}
\mathcode`\*="8000
{\catcode`\*=\active\gdef*{\mathclose{}\,\mathopen{}}}

\newcommand{\be}{\begin{equation}}
\newcommand{\ee}{\end{equation}}
\newcommand{\bea}{\setlength\arraycolsep{2pt} \begin{eqnarray}}
\newcommand{\eea}{\end{eqnarray}}
\newcommand{\nn}{\nonumber}

\newcommand{\mA}{{\mathcal A}}
\newcommand{\mI}{{\mathcal I}}
\newcommand{\mJ}{{\mathcal J}}
\newcommand{\mK}{{\mathcal K}}
\newcommand{\cir}{{\text{cir}}}
\newcommand{\IS}{{\text{ISCO}}}
\newcommand{\OS}{{\text{OSCO}}}
\newcommand{\ma}{{\text{max}}}
\newcommand{\mi}{{\text{min}}}

\def\d{\delta}
\def\D{\Delta}
\def\f{\frac}

\def\lm{\lambda} %
\def\l{\left}
\def\L{\Lambda}

\def\n{\nu} %
\def\nn{\nonumber}
\def\om{\omega}

\def\p{\phi} %
\def\r{\right}

\def\S{\Sigma}
\def\t{\theta}

\def\be{\begin{equation}}
\def\ee{\end{equation}}
\def\bag{\begin{aligned}}
\def\eag{\end{aligned}}
\def\bea{\begin{eqnarray}}
\def\eea{\end{eqnarray}}
\def\ba{\begin{array}}
\def\ea{\end{array}}

\def\bc{\begin{center}}
\def\ec{\end{center}}

\begin{document}
\title{Image of Kerr-Melvin black hole with thin accretion disk}

\author{
Yehui Hou$^{1}$, Zhenyu Zhang$^{1}$, Haopeng Yan$^{2}$, Minyong Guo$^{3\ast}$,
Bin Chen$^{1,2,4}$}
\date{}

\maketitle

\vspace{-10mm}

\begin{center}
{\it
$^1$Department of Physics, Peking University, No.5 Yiheyuan Rd, Beijing
100871, P.R. China\\\vspace{4mm}

$^2$Center for High Energy Physics, Peking University,
No.5 Yiheyuan Rd, Beijing 100871, P. R. China\\\vspace{4mm}

$^3$ Department of Physics, Beijing Normal University,
Beijing 100875, P. R. China\\\vspace{4mm}

$^4$ Collaborative Innovation Center of Quantum Matter,
No.5 Yiheyuan Rd, Beijing 100871, P. R. China\\\vspace{2mm}
}
\end{center}

\vspace{8mm}

\begin{abstract}
In this present work, we study the observational appearance of Kerr-Melvin black hole (KMBH) illuminated by an accretion disk. The accretion disk is assumed to be located on the equatorial plane and be thin both geometrically and optically. Considering the fact that outside the innermost stable circular orbit (ISCO) the accretion flow moves in prograde or retrograde circular orbit and falls towards the horizon along plunging orbit inside the ISCO, we develop the numerical backward ray-tracing method and obtain the images of KMBH accompanying with the accretion disk for various black hole spins, strengths of magnetic fields and  inclination angles of observers. We present the intensity distribution horizontally and longitudinally and show the profiles of the red-shift for the direct and lensed images. Our study suggests that the inner shadow and critical curves can be used to estimate the magnetic field around a black hole without degeneration.
\end{abstract}

\vfill{\footnotesize $\ast$ Corresponding author: minyongguo@bnu.edu.cn}

\maketitle

\newpage
\baselineskip 18pt
\section{Introduction}\label{sec1}
The existence of black holes in the universe has been supported by lots of evidences, including the gravitational waves detected by the LIGO/Virgo \cite{LIGOScientific:2016aoc} and the black hole images photographed by the Event Horizon Telescope (EHT) \cite{EventHorizonTelescope:2019dse, EventHorizonTelescope:2022xnr}. In particular,  the polarized image of the supermassive black hole in the center of the M$87$ galaxy reveals a strong magnetic field around the black hole \cite{EventHorizonTelescope:2021bee, EventHorizonTelescope:2021srq}. In fact, there could exist a fairly strong magnetic field in the environment containing an astrophysically realistic black hole \cite{Eatough:2013nva}. For example, the strongest magnetic field around a black hole can attain $B=1.6\times10^{14}$ Gauss which is provided by a magnetar companion close to Sagittarius A* (Sgr A*), known as SGR J1745-29 \cite{Kennea:2013dfa, Olausen:2013bpa}. Moreover, in addition to synchrotron radiations around a black hole, there are also significant phenomena closely related to the magnetic fields, such as the jets and energy extraction \cite{1977Electromagnetic}, magnetic accretion disks \cite{Thorne:1974ve, Abramowicz:2011xu} and so on. 

Thus, it is interesting to model a black hole surrounded by a magnetic field and explore the effects of the magnetic field. Along this line, ignoring the backreaction to spacetime, Wald found a solution to the source-free Maxwell equation which can describe a weakly uniform magnetic field in Schwarzschild and Kerr black hole spacetimes \cite{Wald:1974np}. When the backreaction of the magnetic field is considered, Ernst found analytical solutions to the Einstein-Maxwell equations, which represent Schwarzschild and Kerr black holes embedded in the Melvin universe \cite{Ernst:1976mzr, Ernst1976Kerr}, also dubbed as Schwarzschild-Melvin black hole (SMBH) and Kerr-Melvin black hole (KMBH). These black holes admit a vertical magnetic field which involves in the metric of the corresponding spacetime, called the Melvin magnetic field in some literatures. The SMBH and KMBH have been extensively studied in various aspects. The geodesics in the SMBH and KMBH spacetimes were investigated in \cite{Stuchlik:1999mro, 1983Null, 2007Geodesics}. The motions of charged particles were discussed in \cite{1979Trajectories, Li:2018wtz, Shaymatov:2022enf}. In \cite{Junior:2021dyw, Wang:2021ara}, the authors studied the light rings and critical curves in SMBH and KMBH spacetimes. In \cite{Zhu:2022amy}, the synchrotron radiations and polarized images of SMBH was studied. 

On the other hand, motivated by the black hole images produced by the EHT, the black hole shadow and photon ring have been considered as promising tools to estimate the parameters of black holes, such as the mass, spin, as well as the magnetic field and the accetion disk in the environment \cite{Virbhadra:2008ws, Takahashi:2004xh, Gralla:2019xty, Gralla:2020srx, Li:2020drn, Gan:2021xdl, Meng:2022kjs}. In addition, the black hole shadow has been found some interesting connections with other properties of black hole spacetimes \cite{Cardoso:2008bp, Yang:2012he, Li:2021zct, Stefanov:2010xz, Cunha:2017eoe, Zhang:2019glo}. Besides, the analysis of shadow can impose limits on the modified gravities \cite{Grenzebach:2014fha, Amir:2016cen, Abdujabbarov:2016hnw, Wang:2017hjl, Wang:2018prk, Guo:2020zmf, Amarilla:2011fx, Vagnozzi:2019apd, Hou:2021okc, Pal:2021nul, Zeng:2021mok, Kazempour:2022asl, Mizuno:2018lxz}. Recently, there was an interesting work suggesting that the ``inner shadow" and photon ring in the Kerr black hole image illuminated by an accretion disk can be used to estimate the black hole mass and spin, respectively \cite{Chael:2021rjo}. Thus, it is natural to ask if the inner shadow and photon ring can be applied to estimate the magnetic field outside a black hole. With this question, we would like to calculate and discuss the images of the KMBH illuminated by a geometrically and optically thin accretion disk in this work.

The remaining parts of this paper are organized as follows. In Sec. \ref{sec2}, we review the Kerr-Melvin spacetime and discuss its innermost stable circular orbits (ISCOs). In Sec. \ref{sec3}, we set up our problems and construct our accretion disk model. In Sec. \ref{sec4} we discuss the appearance of the KMBH illuminated by the disk model. In Sec. \ref{sec5}, we summarize and conclude this work. 

\section{KMBH spacetime and timelike geodesics}\label{sec2}
In this section, we would like to give a brief review on the KMBH spacetime and exhibit some important timelike geodesic orbits around the KMBH. 

\subsection{The spacetime}
The spacetime of a rotating black hole immersed in a strong uniform magnetic field is described by the KMBH metric, which is a stationary and axisymmetric solution of the Einstein-Maxwell equations \cite{Ernst1976Kerr},
\bea
ds^2= |\L|^2\Sigma \l[ -\f{\D}{\mA}dt^2 + \f{dr^2}{\D} + d\t^2 \r] + \f{\mA}{\S|\L|^2}\sin^2{\t}(d\phi-\om dt)^2 , \label{KM}
\eea
where
\bea
&&\S= r^2 + a^2\cos^2{\t}, \quad \mA = (r^2+a^2)^2 - \D a^2\sin^2{\t}, \nn \\
&&\D= r^2 -2Mr + a^2, \quad
\L = 1+ \f{1}{4}B^2\f{\mA}{\S}\sin^2{\t} - \f{i}{2}B^2Ma\cos{\t}(3-\cos^2{\t}+\f{a^2}{\S}\sin^4{\t}), \nn \\
\l. \r.
\eea
and 
\begin{align}
	\om &= \f{a}{r^2+a^2}\Bigg\{ (1-B^4M^2a^2)-\D \bigg[ \f{\S}{\mA} + \f{B^4}{16} \Big( -8Mr\cos^2{\t}(3-\cos^2{\t})^2 -6Mr\sin^4{\t}  \nn \\
	&+ \f{2Ma^2\sin^6{\t}}{\mA}[r(r^2+a^2)+2Ma^2] + \f{4M^2a^2\cos^2{\t}}{\mA}[ (r^2+a^2)(3-\cos^2{\t})^2-4a^2\sin^2{\t} ] \Big) \bigg] \Bigg\}\,.
\end{align}
The parameters $M, a$ denote the mass and spin parameter of the black hole, $B$ is the strength of the magnetic field aligned along the symmetry axis of black hole, and $i$ is the imaginary unit, that is, $i^2=-1$. From $\Delta=0$, we can see that the coordinate singularity is at $r_{\pm}=M\pm\sqrt{M^2-a^2}$, which is the same with the Kerr black hole spacetime. Though the concept of the event horizon for such non-asymptotically flat spacetime is not completely clear, we can still regard $r_H=r_{+}$ as the apparent horizon of KMBH, which remains as a marginally outer trapped horizon for the KMBH solution generated from the event horizon in a seed Kerr-Newman solution via the Harrison transform \cite{Booth:2015nwa}.

From the metric Eq.~\eqref{KM} , we can see that the KMBH spacetime reduces to the Kerr BH spacetime when the magnetic field is vanishing. On one hand, near the horizon, the strong gravity is mainly contributed by the mass of the KMBH when $BM \ll 1$. On the other hand, when moving out from the black hole, the curvature of the spacetime is gradually dominated by the magnetic field, which is described by the metric of the Melvin universe  \cite{Melvin:1965zza}. In order to have an intuitive understanding of the strength of the Melvin magnetic field in practice, we transfer the field strength to the Gaussian unit as follows
\bea
B_{\text{Gauss}} = \f{c^4}{G^{3/2}M}B \approx 2.36\times10^{19} \frac{M_{\odot}}{M}(BM)  \quad \text{Gauss}.
\eea

In the case of supermassive black holes $M \sim10^9 M_{\odot}$, a dimensionless $BM  = 0.01$ corresponds to $B_{\text{Gauss}} \sim10^8 \text{Gauss}$ in the Gaussian unit, which is already a very strong magnetic field in astronomical observations. Therefore, we mostly choose $BM =0.01$ to display our results in the following sections \footnote{In \cite{Junior:2021dyw, Wang:2021ara}, it is very interesting that there appear chaos in the images of spherical source when $BM$ is large. For comparison, we show in Appendix. \ref{appB} that there is chaos in the image of accretion disk as well, if the magnetic field is strong enough.}. In addition, for simplicity and without loss of generality, we set $M = 1$.

\subsection{Timelike geodesics}

The geodesics in the KMBH spacetime have been investigated in some interesting works \cite{Stuchlik:1999mro, 1983Null, 2007Geodesics}. Considering that the metric of the KMBH spacetime is very complicated and the Hamilton-Jacobi equation cannot be separated, one is not allowed to study the geodesics in the KMBH spacetime analytically in general. Nevertheless, if the magnetic field is not very strong such that we have $1\ll r\ll 1/B$,  a transition region exists between the near region with $Br\ll 1$ and the far region with $r\gg 1$. The near region is approximately described by a Kerr metric, while the far region is governed by the Melvin universe metric, and the two regions overlap in the transition region. We argue that one can calculate the particle motions separately in the near and far regions, and paste the two geodesics via the matched asymptotic expansion technique\footnote{The matched asymptotic expansion technique has been successfully applied to analytically study the photon emissions in near-horizon extremal rotating black hole spacetimes and Myers-Perry black hole spacetimes in the large dimension limit. The interested readers are suggested to see more details in \cite{Porfyriadis:2016gwb, Guo:2019pte}}. However, our present work is not limited to the case $1\ll r\ll 1/B$, and thus we have to appeal to the numerical method to solve the geodesic problems. In the present work, we are interested in the geometrically thin disk which is located at the equatorial plane with $\theta=\pi/2$,  it is enough for us to focus on the equatorial timelike geodesics.

The KMBH spacetime has two killing vectors originating from the time translational and rotational invariance, thus a massive neutral particle with four-velocity $u^a$ has two conserved quantities along a geodesic, namely
\bea
E = -u_t , \quad  L = u_\p. 
\eea
where $E$ and $L$ are the energy per unit mass and angular momentum per unit mass, respectively. Combining with the normalization condition $u^au_a=-1$, we can obtain the radial equation of the motion
\bea
u^r = - \sqrt{-\f{V(r,   E,  L )}{g_{rr}}}\,,\label{eq motion}
\eea
where
\be
V(r,   E,  L ) = \l(1+g^{tt}E^2+g^{\phi\phi}L^2-2g^{t\phi}EL\r)\bigg|_{\t=\f{\pi}{2}}
\ee
is defined as the effective potential function. 

\begin{figure}[t]
	\centering
	\includegraphics[scale=0.42]{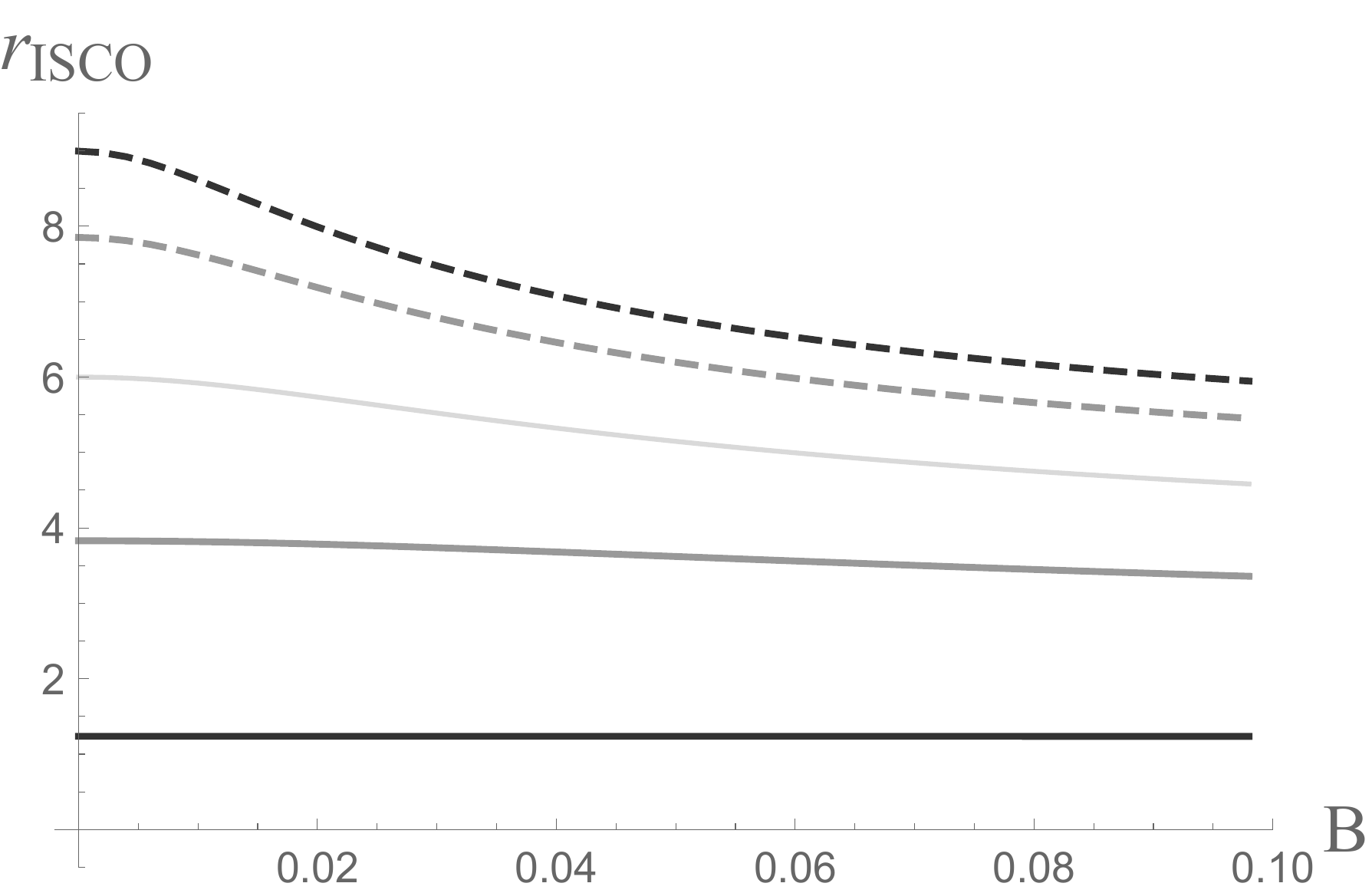} \
	\includegraphics[scale=0.42]{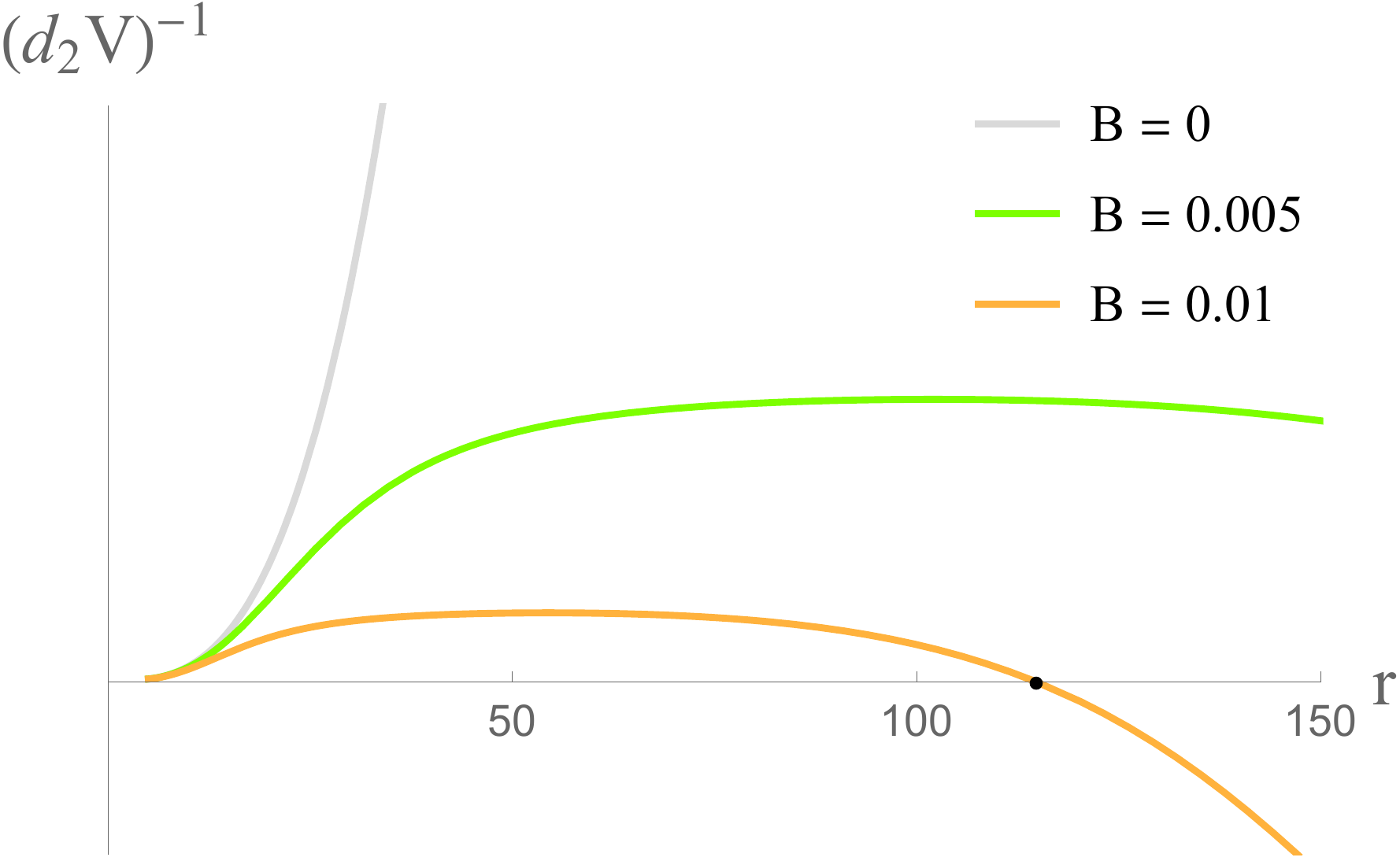}
	\caption{Left: ISCOs as the functions of $B$. From light to dark gray, $a$ is taken to be $0, 0.6, 0.998$. The solid and dashed lines represent prograde and retrograde orbits, respectively. Right: $(d_2V)^{-1}$ of prograde orbits. The spin is $a = 0.6$. The OSCO of $B = 0.01$ is located at $r_{\OS} \approx 114.8$. The retrograde orbits have similar behavior at a large $r$ and have very similar OSCOs.}
	\label{fig:isco}
\end{figure}

At first, we would like to consider the circular orbits which satisfy $V = \partial_r V = 0$. From these two equations, we can determine the conserved quantities $E = E_{\cir}(r)$ and $L = L_{\cir}(r)$ for the circular orbits. To discuss the stability of circular orbits, it is convenient to define 
\be
d_2V(r) =  \partial^2_r V \big|_{E=E_{\cir}(r) , L = L_{\cir}(r)}\,,
\ee
and a stable circular orbit requires $d_2V \geq 0$. For usual asymptotically flat spacetimes, there always exists the ISCO which satisfies $d_2V = 0$. Recall that the KMBH spacetime is no longer asymptotically flat due to the Melvin magnetic field, there are another bound  where $(d_2V)^{-1}=0$, which means that in addition to the ISCO near the horizon, there appears an outermost stable circular orbit (OSCO) \footnote{Here the OSCO is different from the ISCO, since outside the OSCO there are no circular orbits while inside the ISCO there are still unstable circular orbits. For example, in Eq. (\ref{ELM}) we can see that the condition $4-3B^2r^2\ge0$ should be satisfied and $r_{\text{OSCO}=\frac{2}{\sqrt{3}B}}$ corresponds to $(d_2V)^{-1}=0$.} far from the horizon of the KMBH. Thus, by definition the stable circular orbits only exist between the ISCO and OSCO. Furthermore, one can generally demonstrate that if there is an ISCO in the KMBH spacetime, then there has to be an OSCO as well. The reason is that when $r \gg 1$, the KMBH spacetime approaches the Melvin universe and in Appendix. \ref{app:Melvin}, we demonstrate that all the circular orbits are unstable in the Melvin universe. In Fig.~\ref{fig:isco}, we show some examples of the variations of ISCOs and $(d_2V)^{-1}$ with respect to $B$ and $r$, respectively. From the Fig. \ref{fig:isco}, we can see that the radius of the OSCO $r_{\OS}\gg r_H$ when $B$ is not big. Therefore, in this work we take the OSCO and the horizon as the outer and inner boundaries of the KMBH spacetime, respectively, and we place the observer $(t_o, r_o, \theta_o, \phi_o)$ far from the horizon and inside the OSCO, that is, $r_H\ll r_o< r_\OS$.

Moreover, for the purpose in the next section, we also take into consideration the plunging orbits which start from the ISCO and fall into the horizon on the equatorial plane. Hence, we are allowed to let these critical plunging orbits carry the conserved quantities $E = E_{\IS}$, $L = L_{\IS}$, then from Eq.~\eqref{eq motion}, the radial equation of motion is given by
\bea\label{cpo}
u^r_c = - \sqrt{-\f{V(r, \  E_{\IS}, \ L_{\IS} )}{g_{rr}}} \bigg|_{\t = \f{\pi}{2}}, 
\eea
where the minus sign in front of the square root means ingoing motions and $r$ is confined as $r_H<r<r_\IS$.

\section{Accretion disk and imaging method}\label{sec3}

In this section we turn to present our accretion disk model and the method to take photographs of the KMBH illuminated by the accretion disk. 

In our study, the accretion disk is taken to be geometrically thin on the equatorial plane, so that the compositions of the accretion disk can be regarded as free electrically neutral plasma which move along the equatorial timelike geodesics. Also, the accretion disk has a width: its external radius is larger than that of the ISCO, and it is able to extend to the horizon of the KMBH. Thus, the ISCO can be seen as the dividing line, outside of which the particles of the accretion disk move in stable circular orbits and inside of which they would travel in critical plunging orbits. More precisely, for the particles in the accretion disk, the radial motion is determined by the equations $V=\partial_r V=0$ when $r\ge r_\IS$, while for $r_H<r<r_\IS$, the radial motion is governed by the Eq.~\eqref{cpo}.

Next, we consider the photons emitted from the accretion disk and arriving at a distant observer. Due to the symmetries in $t$ and $\p$ directions, We consider a zero-angular-momentum observer (ZAMO) at $(t_o=0, r_o, \theta_o, \phi_o=0)$, whose tetrad reads
\bea\label{ted}
&e_{(0)}= \xi (1,0,0,-\gamma)\,, \quad e_{(1)} = (0, -\f{1}{\sqrt{g_{rr}}},0,0)\,, \nn \\
&e_{(2)}= (0, 0 , \f{1}{\sqrt{g_{\t\t}}},0)\,, \quad e_{(3)} = (0,0,0, -\f{1}{\sqrt{g_{\phi\phi}}})\,,
\eea
where
\be
\xi =\sqrt{\f{-g_{\phi\phi}}{g_{tt}g_{\phi\phi}-g^2_{t\phi}}} \,, \hspace{3ex}\gamma = \f{g_{t\phi}}{g_{\phi\phi}}\,,
\ee
and we have included a minus sign in each of $e_{(1)}$ and $e_{(2)}$ to facilitate the backward ray-tracing method such that we can take the light path of the photons emanating from the ZAMO to be reversible. In the ZAMO's frame, the four-momentum of photons reads
\bea
p_{(\mu)}=k_\nu e^\nu_{(\mu)}\,,
\eea
where $e^\nu_{(\mu)}$ is given in Eq.~\eqref{ted}. On the other hand, in the frame of the ZAMO we can  define the celestial coordinates $\Theta$ and $\Psi$ to label each light ray. We follow the convention in \cite{Hu:2020usx} where the relation between the celestial coordinates and the four-momentum of photons $p_{(\mu)}$ is given by
\bea
\cos\Theta=\frac{p^{(1)}}{p^{(0)}}\,,\quad \tan\Psi=\frac{p^{(3)}}{p^{(2)}}\,.
\eea
Then, on the screen of the ZAMO, considering a Cartesian coordinates system we have
\bea\label{xyd}
x=-2\tan\frac{\Theta}{2}\sin\Psi\,,\quad y=-2\tan \frac{\Theta}{2}\cos \Psi\,,
\eea
which in turn determine the initial values of the momentum of the photons at the ZAMO with the initial values of the position $(0, r_o, \theta_o, 0)$. Then combining with the Hamiltonian canonical equations of the null geodesics we can numerically obtain the complete geodesic trajectories. Note that when tracing back the light ray, it may pass through the equatorial plane many times, with the radii of the intersections $r_n(x, y), n = 1,2,...N_{\text{max}}(x, y)$, where $N_{\text{max}}(x, y)$ is the maximum number of intersections. Thus we can see that the profiles of $\{r_n(x, y)\}$ are discrete for different $n$ on the screen. In addition, $r_n(x, y)$ is  called the transfer function which actually give the shape of the $n$-th image of the disk. For example, $n=1$ is the ``direct'' image, and $n = 2$ is the ``lensed'' image. It is worth mentioning that the transfer function depends on the observational angle $\t_o$.

Then, we turn to discuss the intensity of the KMBH image illuminated by the accretion disk. Considering a complete light ray connecting the light source, i.e., the accretion disk around the KMBH, with the screen in the frame of the ZAMO, the intensity would change due to the emission and absorption when the light ray interacts with the accretion disk. For simplicity, we assume the disk medium have ignorable refraction effect, then the change of the intensity is determined by the following equation \cite{Lindquist:1966igj}
\be 
\f{d}{d\lambda}\l( \f{I_\nu}{\nu^3} \r) = \f{J_\nu-\kappa_\nu I_\nu}{\nu^2} ,\label{transfer}
\ee
where $\lambda$ is the affine parameter of null geodesics, $I_\nu, \ J_\nu, \ \kappa_\nu$ are the specific intensity, emissivity and absorption coefficient at the frequency $\nu$, respectively. When the light propagates in vacuum, both  $J_\nu$ and $\kappa_\nu$ are 0, thus $I_\nu/\nu^3$ is conserved along the geodesics.

We assume that the accretion disk is steady, axisymmetic, and have $Z_2$ symmetry about the equatorial plane. Recall that the disk is geometrically thin so that the emissivity and absorption coefficient keep unchanged when light rays pass through it. By integrating Eq.~\eqref{transfer} along traced-back trajectories we obtain the formula of intensity at each location on the observer's screen which takes the form (a detailed derivation can be found in Appendix.~\ref{app:ray-tracing})
\bea
I_{\nu_o} = \sum_{n=1}^{N_{max}} \ \l(\f{\nu_o}{\nu_n}\r)^3 \f{J_n}{\tau_{n-1}} \l[ \f{1-e^{-\kappa_n f_n}}{\kappa_n} \r], \label{Io}
\eea
where $\nu_o = \mathcal{E}_o=-p_{(0)}|_{r=r_o}$ is the observed frequence on the screen and $\nu_n=\mathcal{E}_n=-k_\mu u^\mu|_{r=r_n}$ is the frequence observed by the local rest frames comoving with the accretion disk. For simplicity, we call the class of the frames $\{F_n\}$, where $n=1...N_{max}$ is the number of times that the ray crosses the equatorial plane, and we use the subscript $n$ to denote the corresponding measurements in the local rest frames $F_n$. The quantity $\tau_m$ is the optical depth of photons emitted at $m$
\bea
\tau_m =
\begin{cases}
	\exp \l[\sum_{n=1}^{m}\kappa_n f_n \r]& \ \ \text{if}\,\,\, \ m\geq 1, \\
	1& \ \ \text{if}\,\,\, \ m=0
\end{cases}
\eea
with $f_n = \nu_n \Delta \lambda_n $ being the ``fudge factor'' which needs to be further specified for specific models of accretion disk. In the fudge factor, $\Delta \lm_n$ is the change in the affine parameter when the ray passing through the disk medium at $F_n$. When the absorption can be neglected, that is, the accretion disk is assume to be optically thin, the Eq.~\eqref{Io} would reduce to 
\be
I_{\nu_o} = \sum_{n=1}^{N_{max}} \ f_n g_n^3 J_n\,, \label{Io1}
\ee
where we have introduce the redshift factor
\be
g_n = \f{\nu_o}{\nu_n}.\label{redshift}
\ee
Note that Eq.~\eqref{Io1}  had been used to study the intensity of the image of an optically thin disk in previous works \cite{Hadar:2020fda, Chael:2021rjo}. Moreover, the specific parameters of the emission of the accretion disk,  $J_n$ and $f_n$ appeared in Eq.~\eqref{Io1}, are still undetermined. Considering the fact that the images of M87* and Sgr A* are photographed at the observing wavelength of $1.3$ mm ($230$ GHz), we choose the emissivity to be a second-order polynomial  in log-space
\be
J = \exp{\bigg(-\f{1}{2}z^2-2z\bigg)},  \ \ \  z = \log{\f{r}{r_H}}
\ee
which has been also used in \cite{Chael:2021rjo}, to fit the 230 GHz images. Notice that the emission profile is isotropic and axisymmetric, and decreases rapidly with increasing radius, e.g., $J$ at $5r_H$ reduces to one percent of that at $r_H$. However, different with the choices in \cite{Chael:2021rjo},  we normalize all the fudge factors $f_n$ to $1$. This is because our interest is focused on the effects of the magnetic fields on the images and  in practice the values of $f_n$ mainly changes the strength of the narrow photon ring which has limited effect on overall image. Therefore, we can back-trace all the light ray and determine their positions on the $n$-th point and use Eq.~\eqref{Io1} to plot the image. 

In addition, we would like to give a more detailed expression for the redshift factor $g_n$ in Eq.~\eqref{redshift} in order to make it easier to understand. Recall that in our model, the accretion flow consists of electrically neutral plasma, which moves along timelike geodesics with conserved quantities $E$ and $L$. Outside the ISCO, the flow move along circular orbits with angular velocity $\Omega_n(r)=(u^\phi/u^t)|_{r=r_n}$. Then the redshift factor can be rewritten as
\be\label{gnd}
g_n= \f{e}{\zeta(1-\Omega_n b)}, \ \ \  r_n\ge r_\IS\,,
\ee
where we have introduced 
\bea
b=\frac{\mathcal{L}}{\mathcal{E}}=\frac{k_\phi}{-k_t}\,,\quad e=\frac{\mathcal{E}_o}{\mathcal{E}}=\frac{p_{(0)}}{k_t}=\xi(1+b\gamma)\,,\quad\zeta =\left.\sqrt{ \f{-1}{g_{tt}+2g_{t\phi}\Omega_n + g_{\phi\phi}\Omega_n^2} }\right|_{r=r_n}
\eea
to make Eq.~\eqref{gnd} more compact. Thereinto, $b$ is the impact parameter of photons with $\mathcal{E}=-k_t$ and $\mathcal{L}=k_\phi$ being the conserved energy and angular momentum along null geodesics, and $e$ is the ratio of the observed energy on the screen to the conserved energy along a null geodesic. Note that $e=1$ for asymptotically flat spacetimes when we set $r_o\to\infty$. However, the KMBH spacetime is not asymptotically flat so that there always is  $e<1$. For example, when the observer is far from the black hole and we set $r_o \geq 1,Br_o = 1$, then $e \sim 1/\L_0 \sim 4/(4+\sin^2{\t_o})$, which decreases with the increasing observational angle. For $\t_o = 17^{\circ} (163^{\circ})$, we have $e \approx 0.932$, and for $\t_o = 80^{\circ}$ we find $e \approx 0.802$.

On the other hand, the flows inside the ISCO should move along critical plunging orbits with radial velocity $u^r_c$ in Eq. (\ref{cpo}). In this case the redshift factor becomes
\be
g_n = - \f{e}{ u^r_c k_r/\mathcal{E} +E_{\IS}(g^{tt}-g^{t\phi}b) + L_{\IS} (g^{\phi\phi}b-g^{t\phi})},  \ \ \  r_n< r_\IS\,,
\ee
where $u^r_c$, $g^{tt}$, $g^{t\phi}$ and $g^{\phi\phi}$ are evaluated at $r=r_n$. The same treatment was used to study the images of Kerr black holes in \cite{Cunningham:1975zz, Gralla:2020srx}.

\section{Results } \label{sec4}
In this section, we would like to show the results by fixing the observational distance $r_o=100$. As for the radial range of the accretion disk, we set the outside radius of the accretion disk at $r_{\text{or}}=20$ and extend the disk to the event horizon, that is, the inside radius of the disk $r_{\text{ir}}=r_H$. Considering the emission profile of the accretion disk decreases quickly with its radius, the matters of the accretion disk far from the event horizon would give less contributions to  the images of the KMBH. 

\begin{figure}[h!]
	\centering
	\begin{tikzpicture}
	\node at (-6,2) {\includegraphics[scale=0.18]{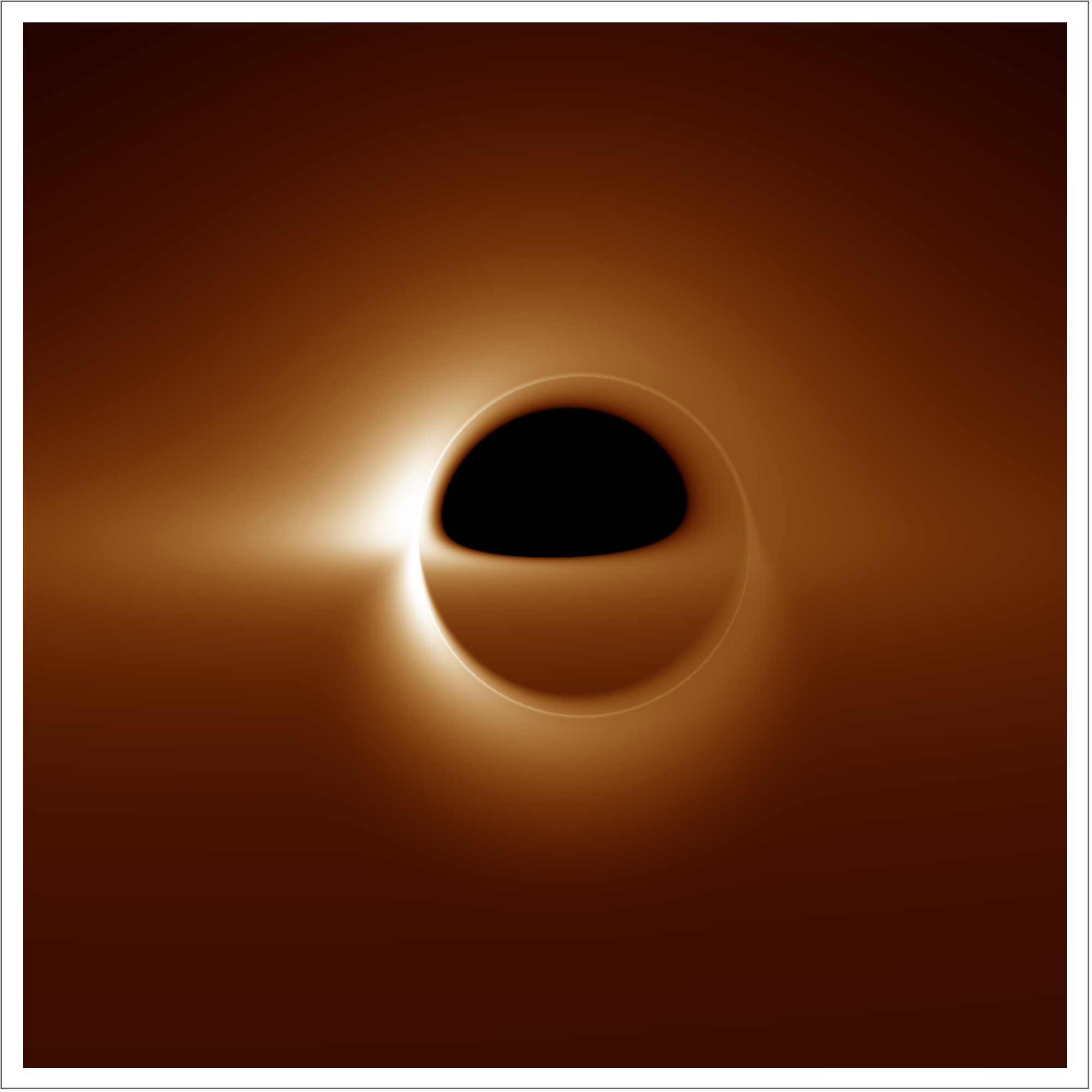}};
	\node[below left] at (-5.8,1) {\tiny\color{white} $a=0.6\ B=0$};
	\node[below left] at (-6.38,0.7) {\tiny\color{white} $\t_o=80^{\circ}$};
	
	\node at (-2,2) {\includegraphics[scale=0.18]{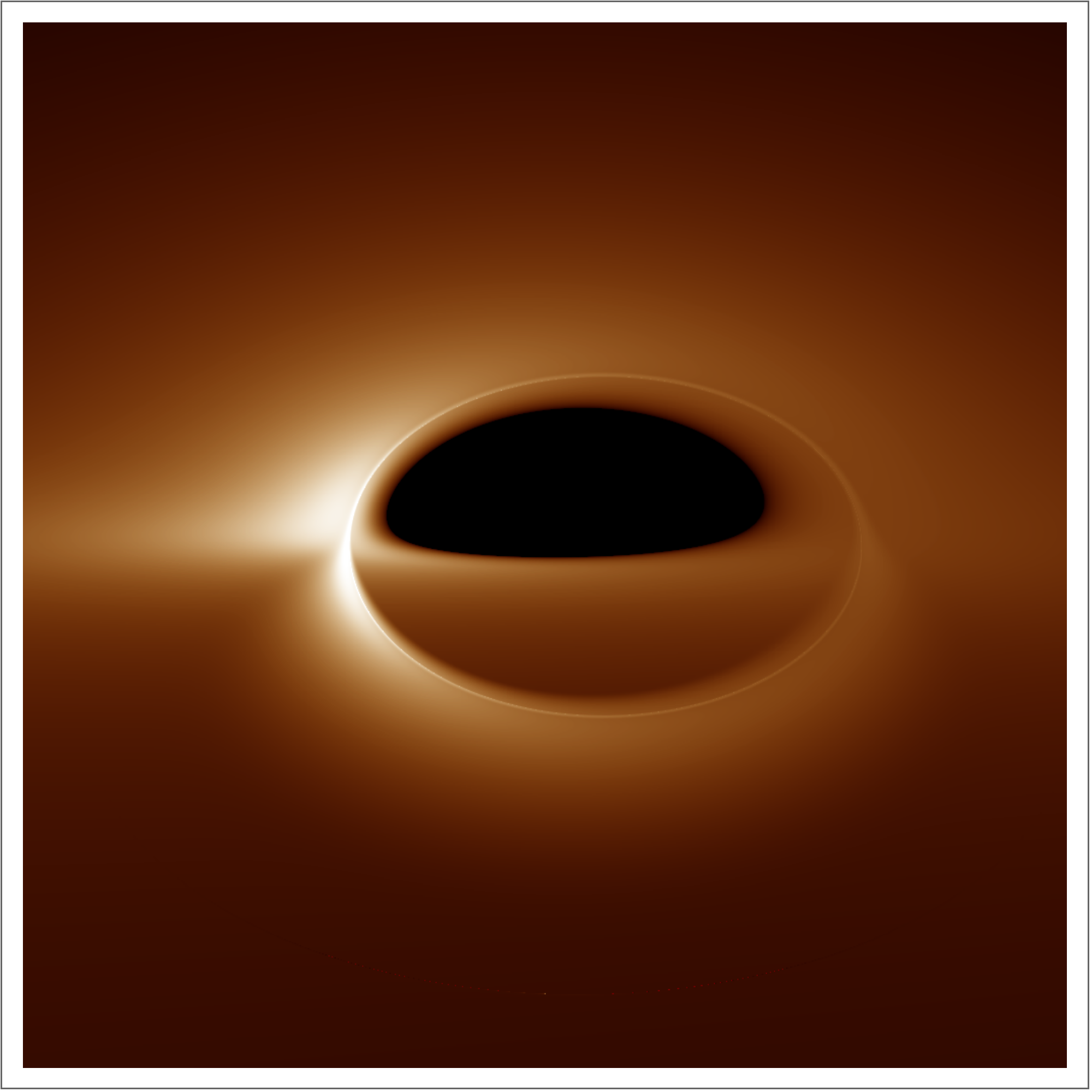}};
	\node[below left] at (-1.47,1) {\tiny\color{white} $a=0.6\ B=0.01$};
	\node[below left] at (-2.38,0.7) {\tiny\color{white}$\t_o=80^{\circ}$};
		
	\node at (2,2) {\includegraphics[scale=0.18]{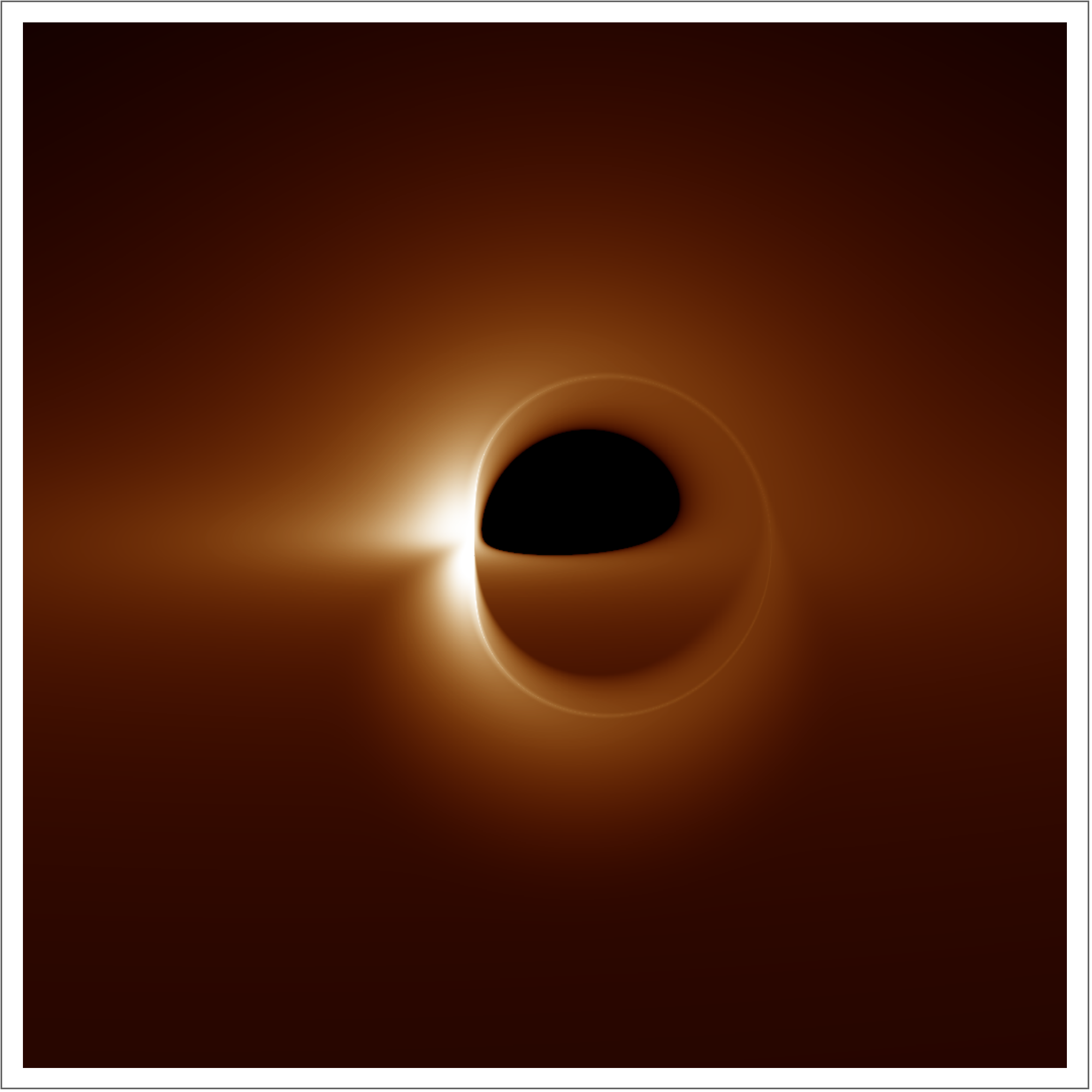}}; 
	\node[below left] at (2.46,1) {\tiny\color{white} $a=0.998\ B=0$};
	\node[below left] at (1.62,0.7) {\tiny\color{white}$\t_o=80^{\circ}$};
	
	\node at (6,2) {\includegraphics[scale=0.18]{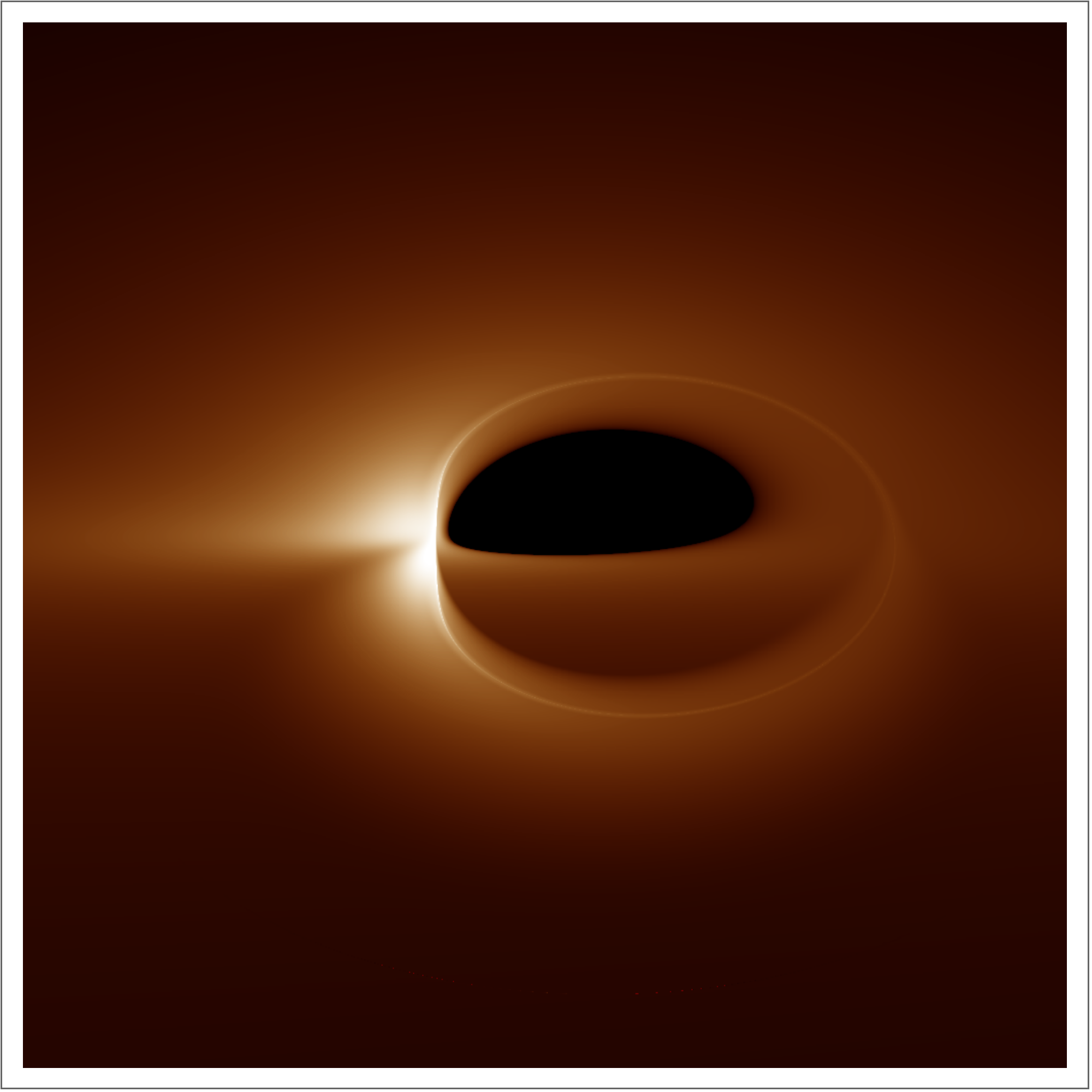}};
	\node[below left] at (6.8,1) {\tiny\color{white} $a=0.998\ B=0.01$};
	\node[below left] at (5.62,0.7) {\tiny\color{white}$\t_o=80^{\circ}$};
	
	\node at (-6,-2) {\includegraphics[scale=0.18]{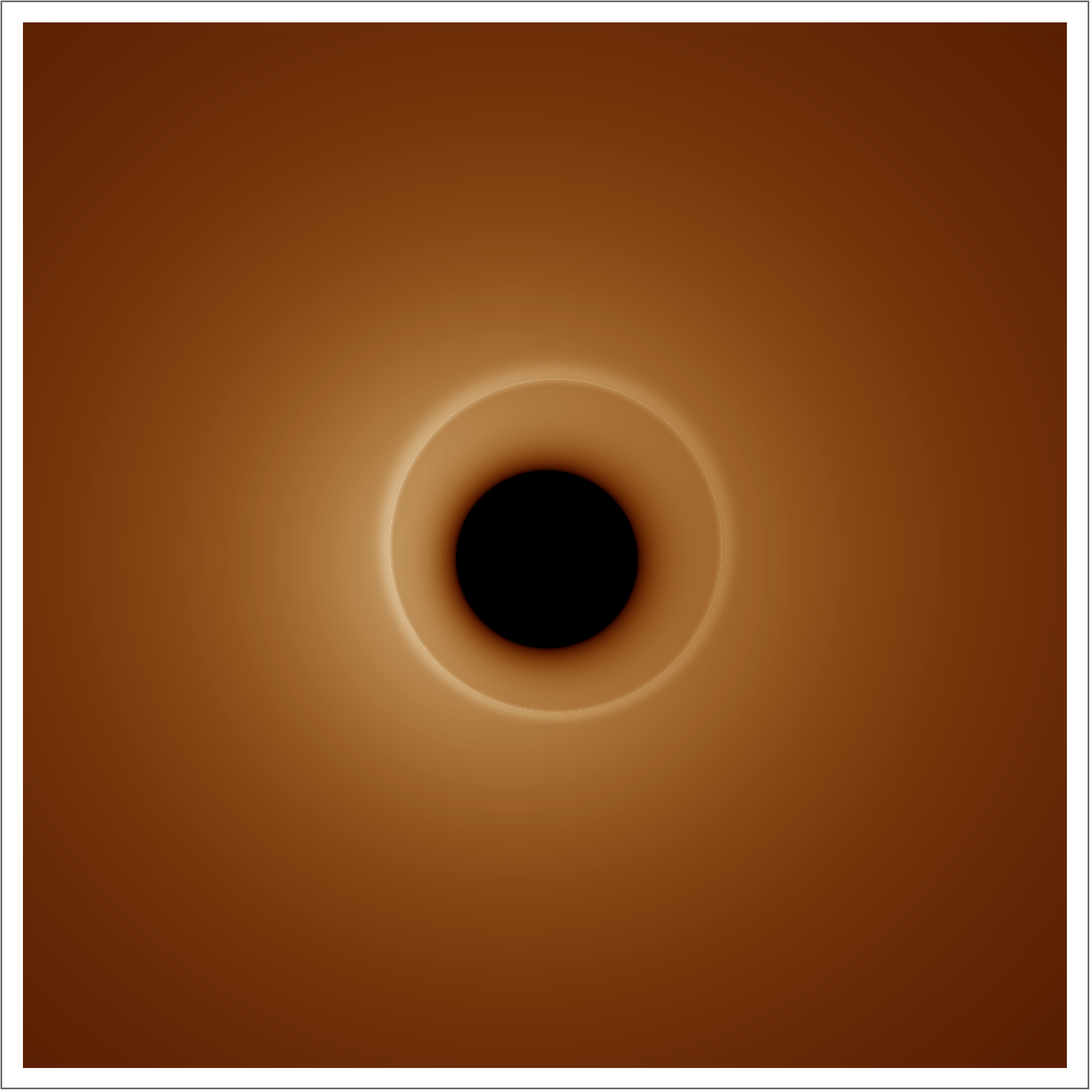}};
	\node[below left] at (-5.8,-3) {\tiny\color{white} $a=0.6\ B=0$};
	\node[below left] at (-6.25,-3.3) {\tiny\color{white}$\t_o=163^{\circ}$};
	
	\node at (-2,-2) {\includegraphics[scale=0.18]{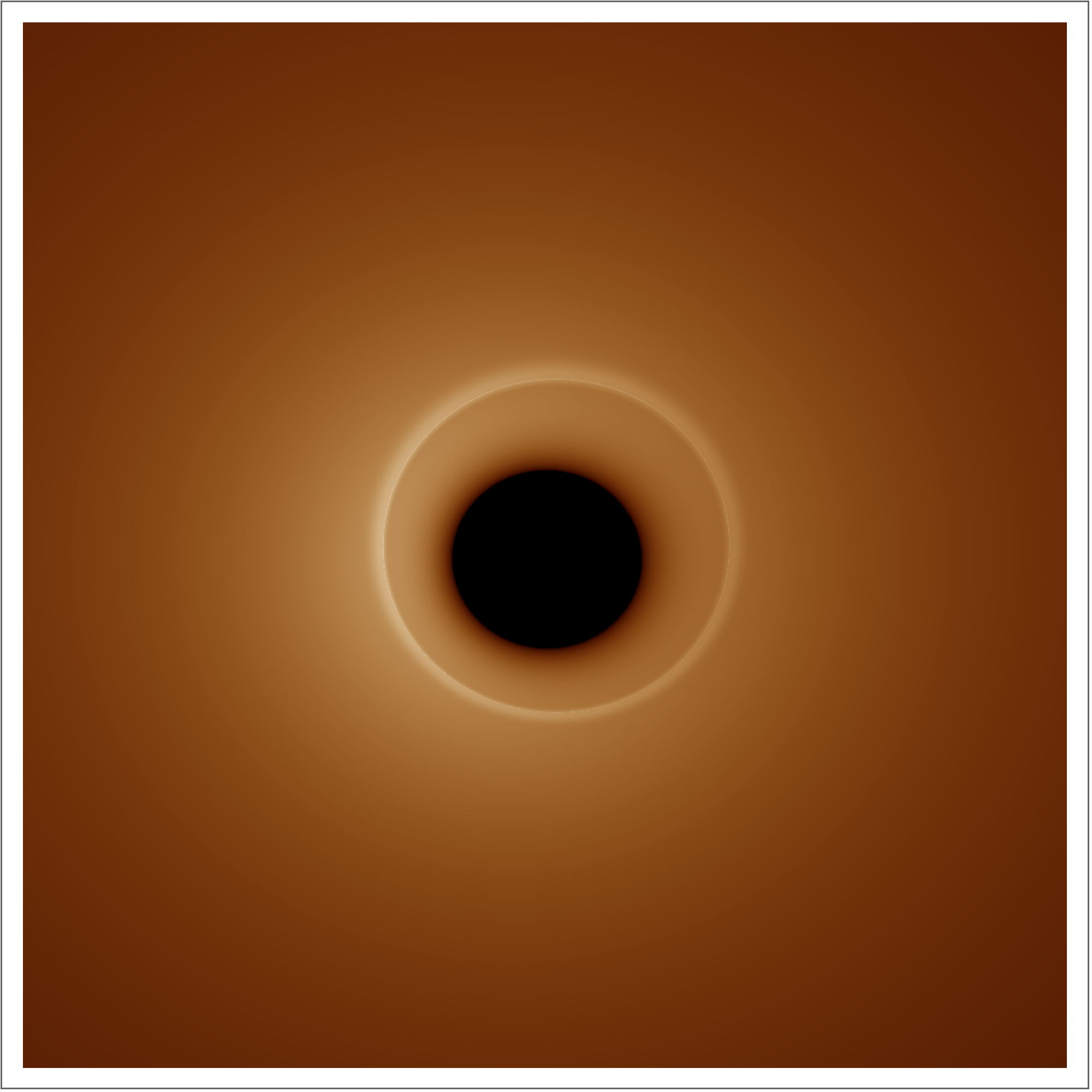}};
	\node[below left] at (-1.45,-3) {\tiny\color{white} $a=0.6\ B=0.01$};
	\node[below left] at (-2.25,-3.3) {\tiny\color{white}$\t_o=163^{\circ}$};
	
	\node at (2,-2) {\includegraphics[scale=0.18]{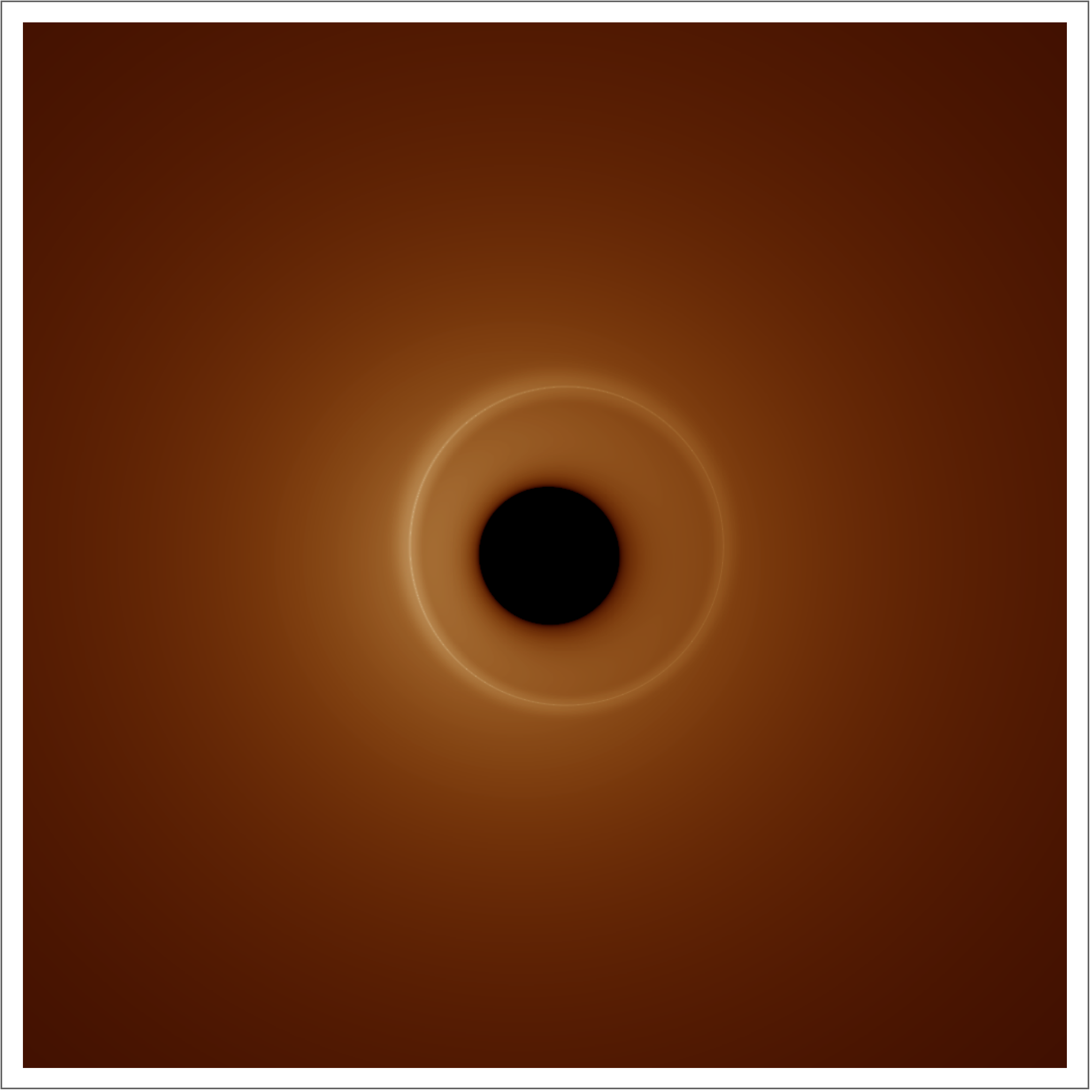}};
	\node[below left] at (2.46,-3) {\tiny\color{white} $a=0.998\ B=0$};
	\node[below left] at (1.75,-3.3) {\tiny\color{white}$\t_o=163^{\circ}$};
	
	\node at (6,-2) {\includegraphics[scale=0.18]{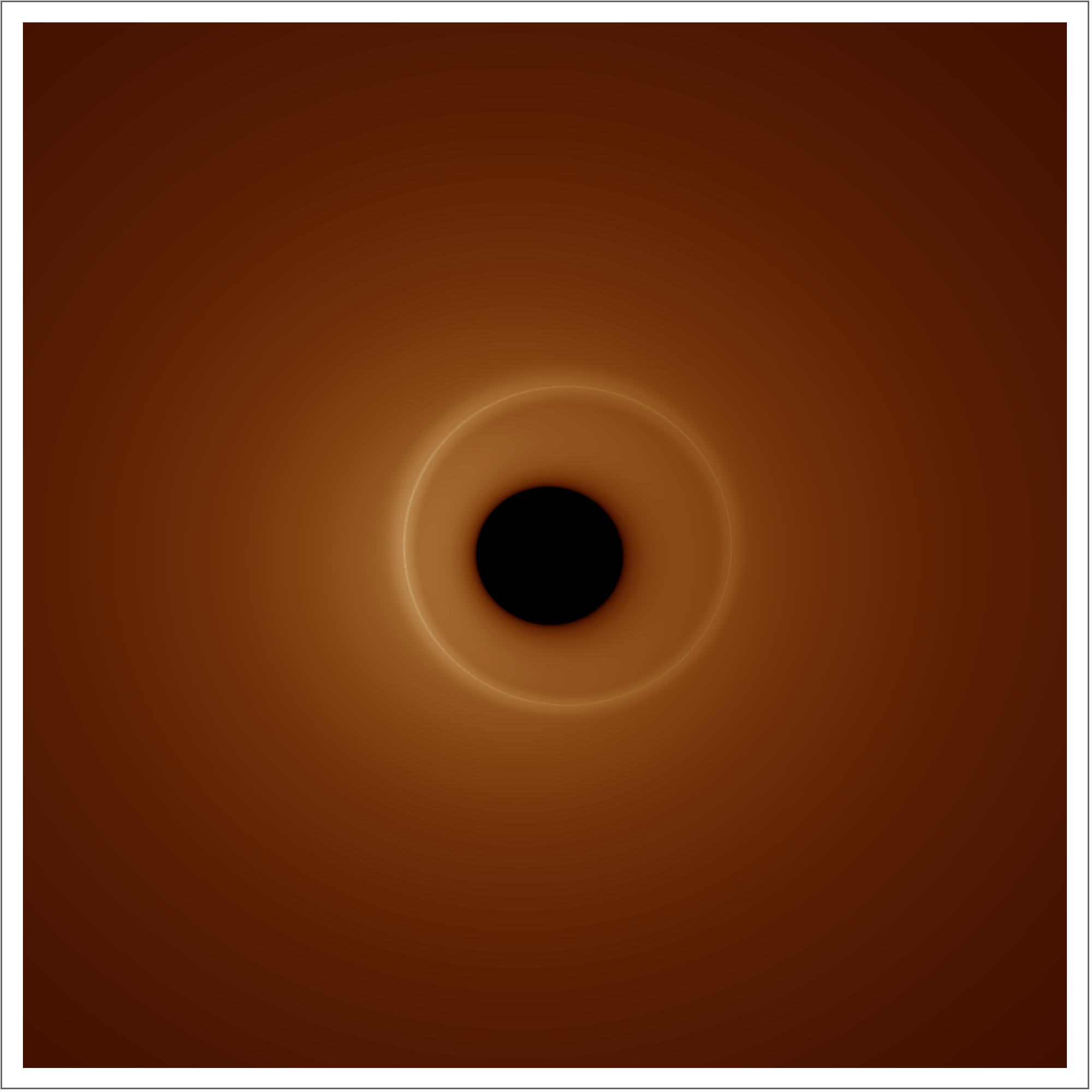}};
	\node[below left] at (6.8,-3) {\tiny\color{white} $a=0.998\ B=0.01$};
	\node[below left] at (5.75,-3.3) {\tiny\color{white}$\t_o=163^{\circ}$};
	\end{tikzpicture}
	\caption{Images of KMBH illuminated by prograde flows. The top row of the images is observed at $\t_o = 80^{\circ}$, and the bottom row is observed at $\t_o = 163^{\circ}$.}
	\label{fig:intensity1}
\end{figure}
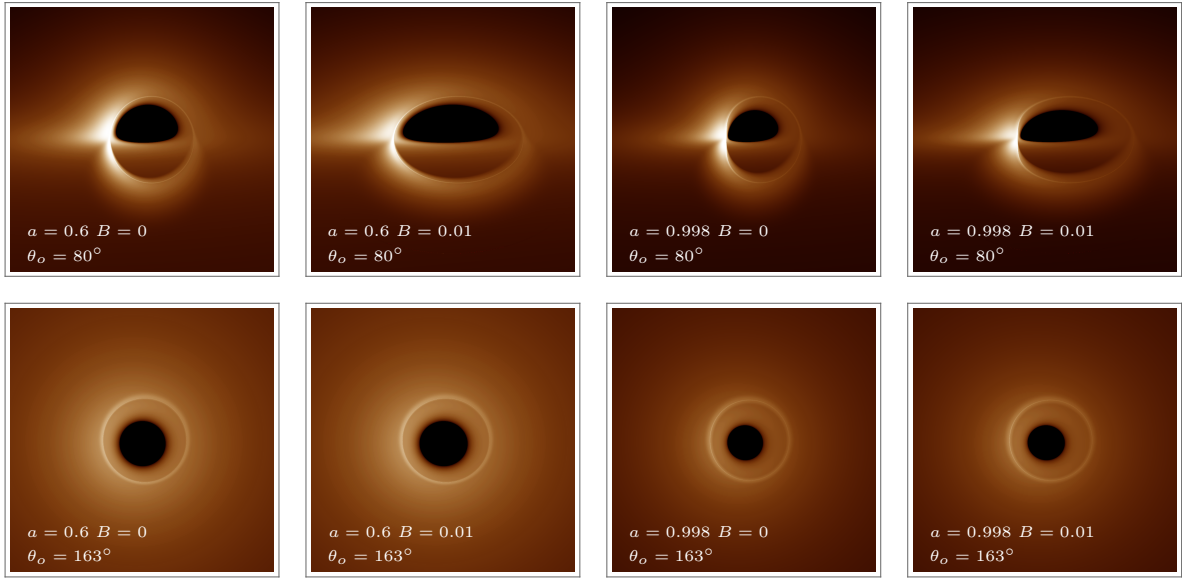

\begin{figure}[h!]
	\centering
	\begin{tikzpicture}
	\node at (-6,2) {\includegraphics[scale=0.18]{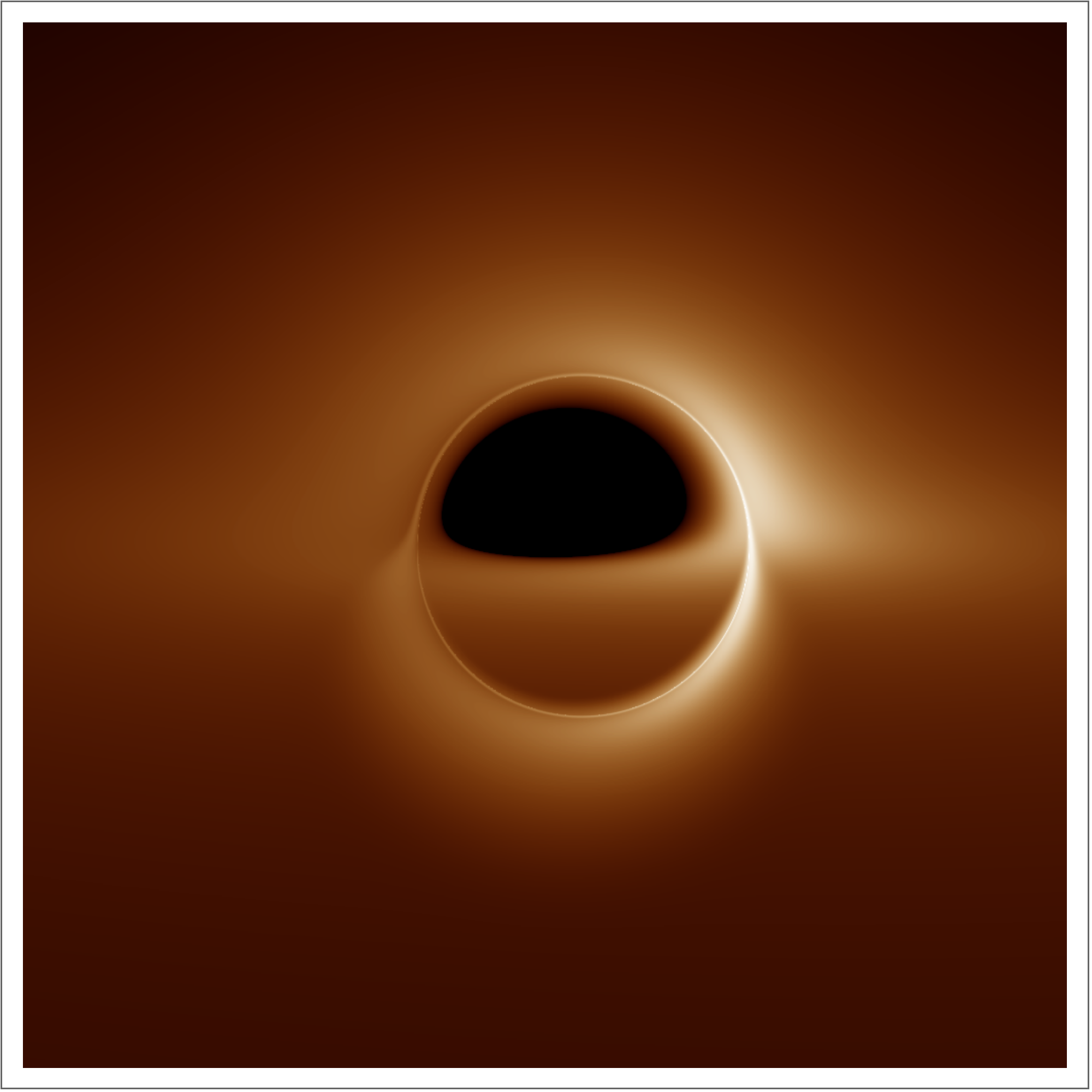}};
	\node[below left] at (-5.8,1) {\tiny\color{white} $a=0.6\ B=0$};
	\node[below left] at (-6.38,0.7) {\tiny\color{white} $\t_o=80^{\circ}$};
	
	\node at (-2,2) {\includegraphics[scale=0.18]{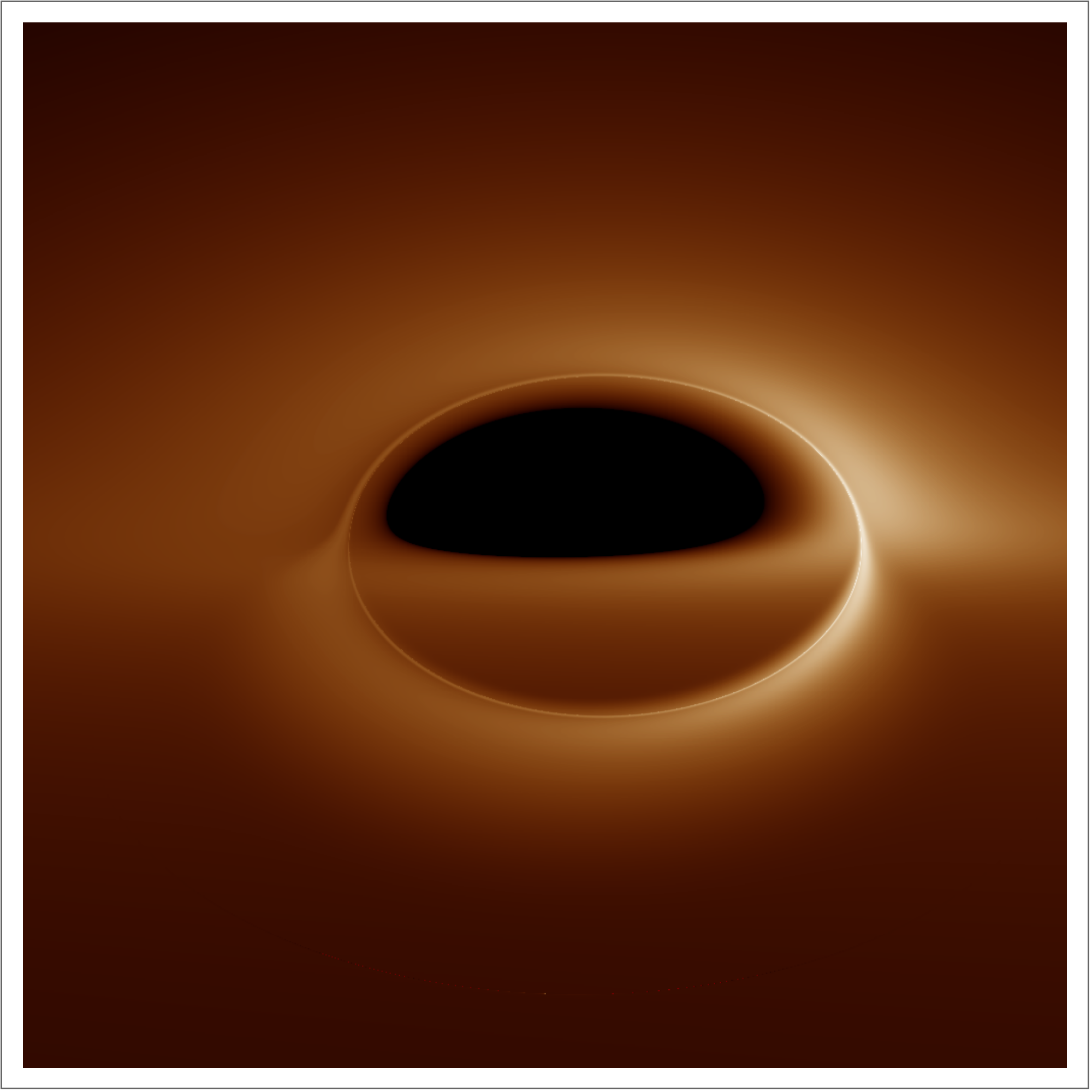}};
	\node[below left] at (-1.47,1) {\tiny\color{white} $a=0.6\ B=0.01$};
	\node[below left] at (-2.38,0.7) {\tiny\color{white}$\t_o=80^{\circ}$};
	
	\node at (2,2) {\includegraphics[scale=0.18]{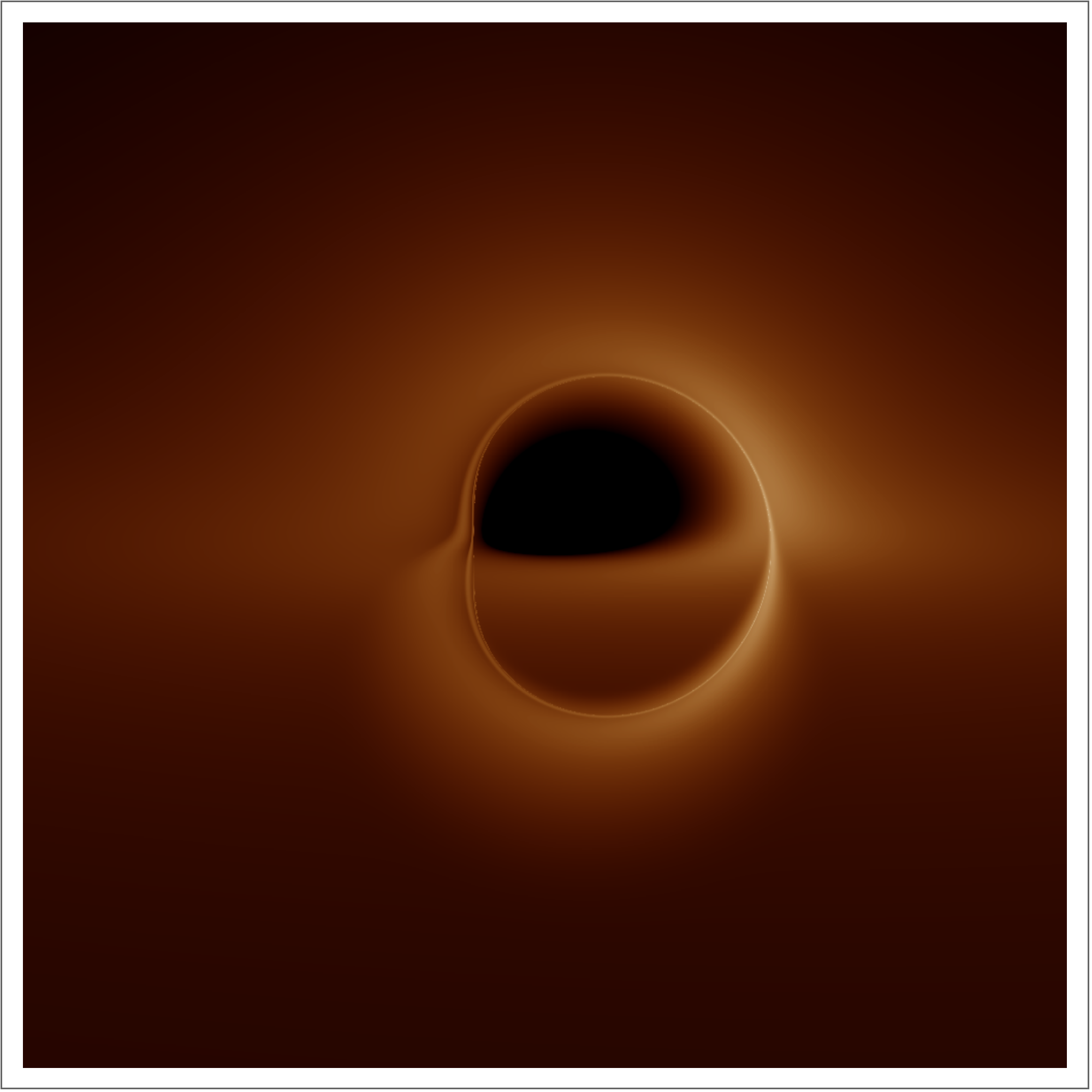}}; 
	\node[below left] at (2.46,1) {\tiny\color{white} $a=0.998\ B=0$};
	\node[below left] at (1.62,0.7) {\tiny\color{white}$\t_o=80^{\circ}$};
	
	\node at (6,2) {\includegraphics[scale=0.18]{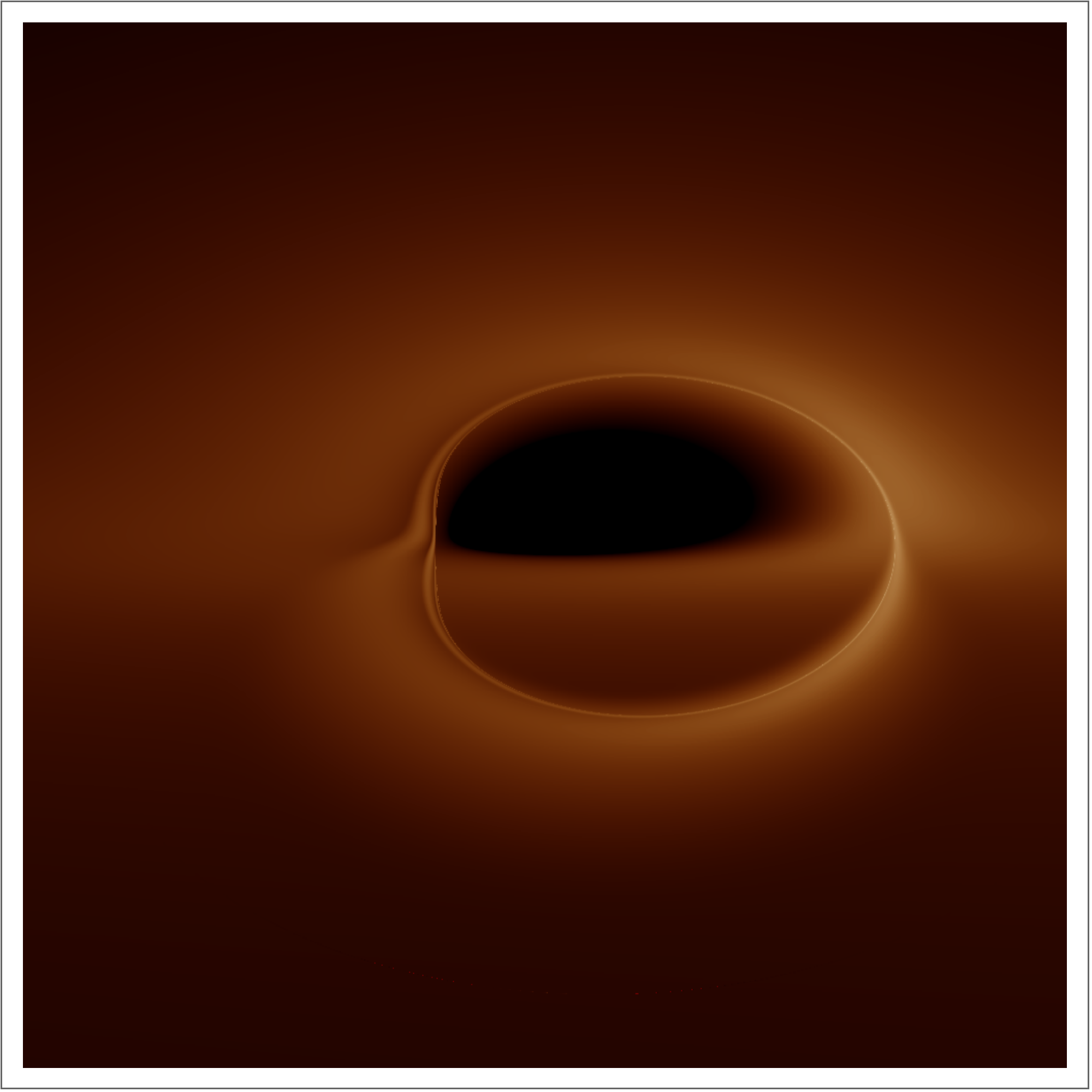}};
	\node[below left] at (6.8,1) {\tiny\color{white} $a=0.998\ B=0.01$};
	\node[below left] at (5.62,0.7) {\tiny\color{white}$\t_o=80^{\circ}$};
	
	\node at (-6,-2) {\includegraphics[scale=0.18]{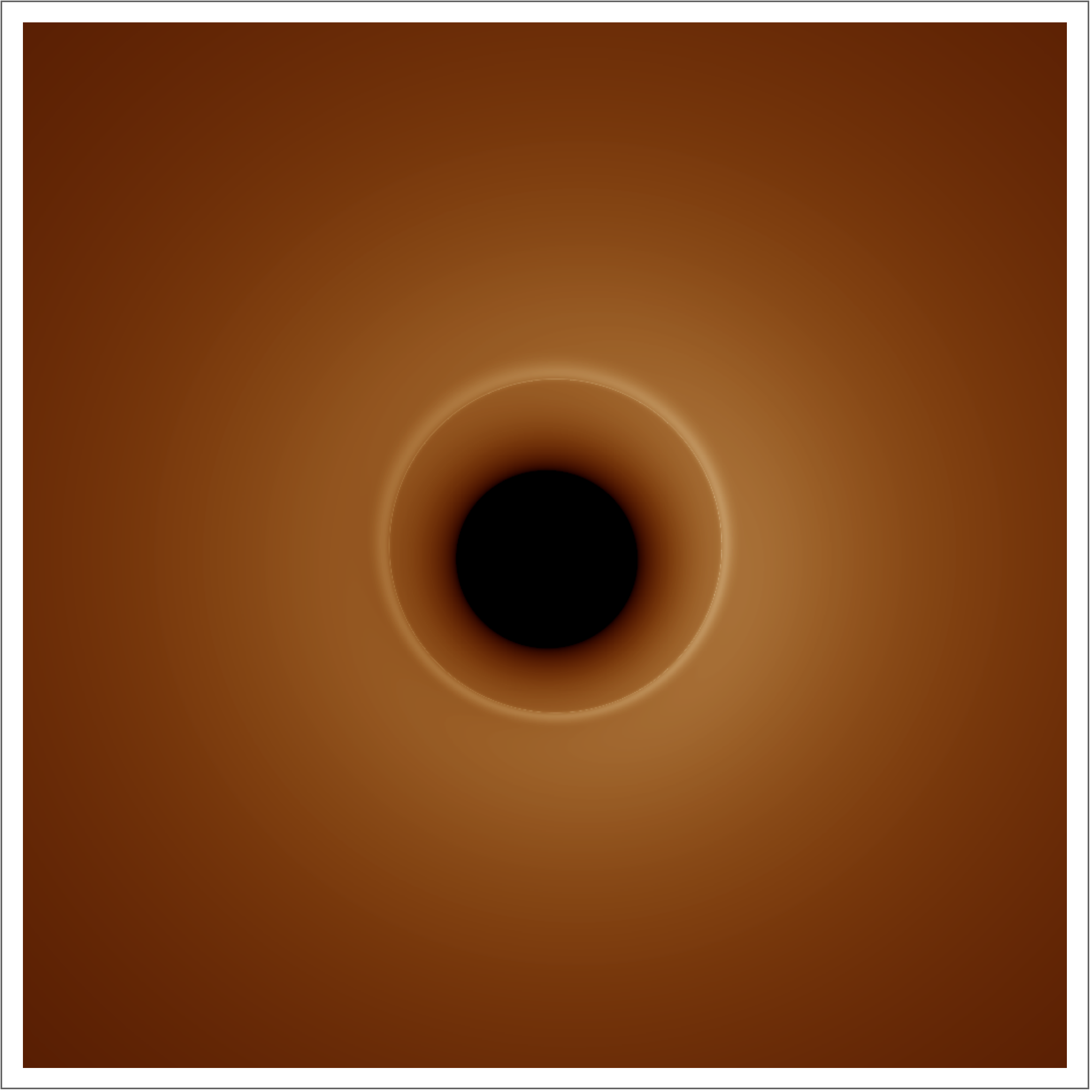}};
	\node[below left] at (-5.8,-3) {\tiny\color{white} $a=0.6\ B=0$};
	\node[below left] at (-6.25,-3.3) {\tiny\color{white}$\t_o=163^{\circ}$};
	
	\node at (-2,-2) {\includegraphics[scale=0.18]{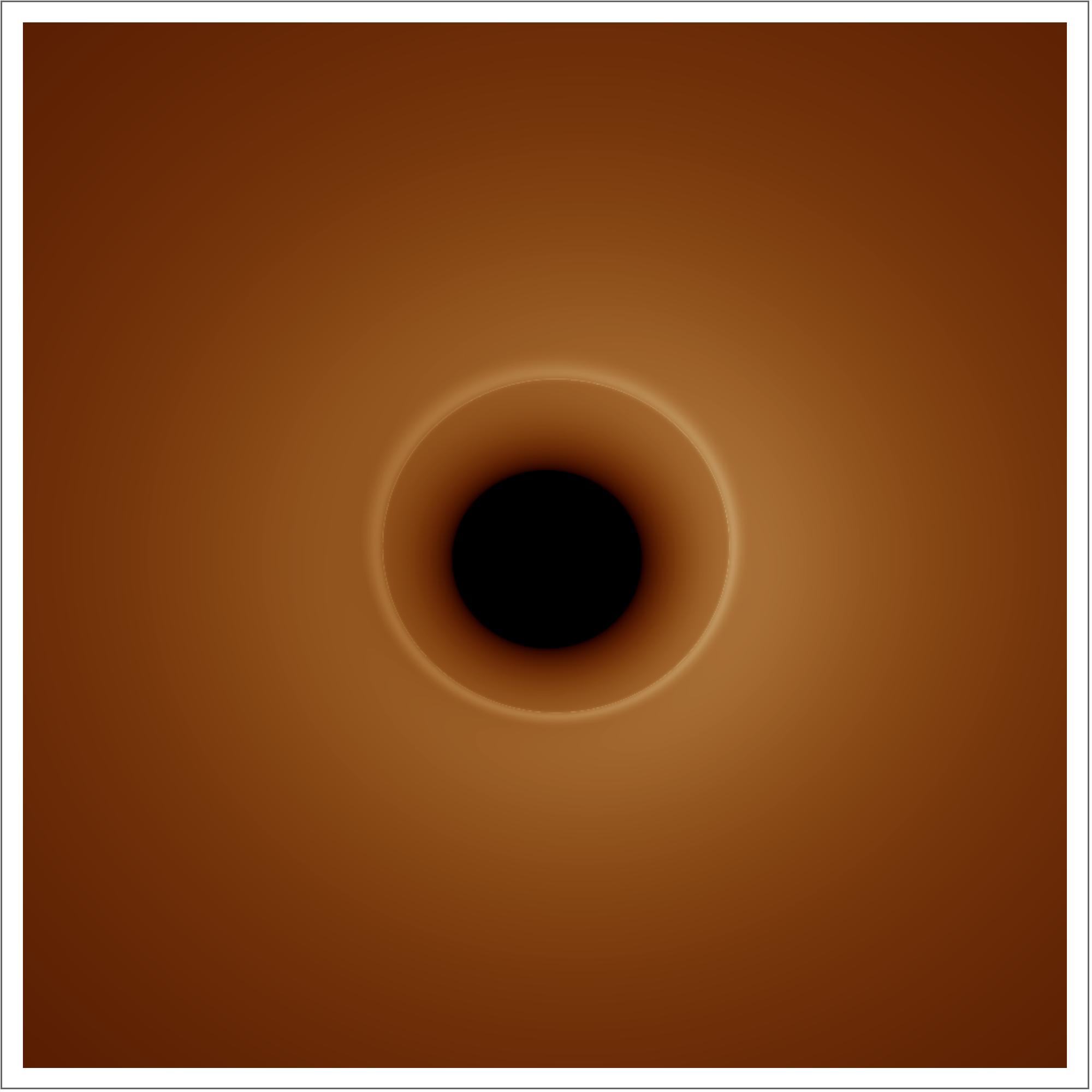}};
	\node[below left] at (-1.45,-3) {\tiny\color{white} $a=0.6\ B=0.01$};
	\node[below left] at (-2.25,-3.3) {\tiny\color{white}$\t_o=163^{\circ}$};
	
	\node at (2,-2) {\includegraphics[scale=0.18]{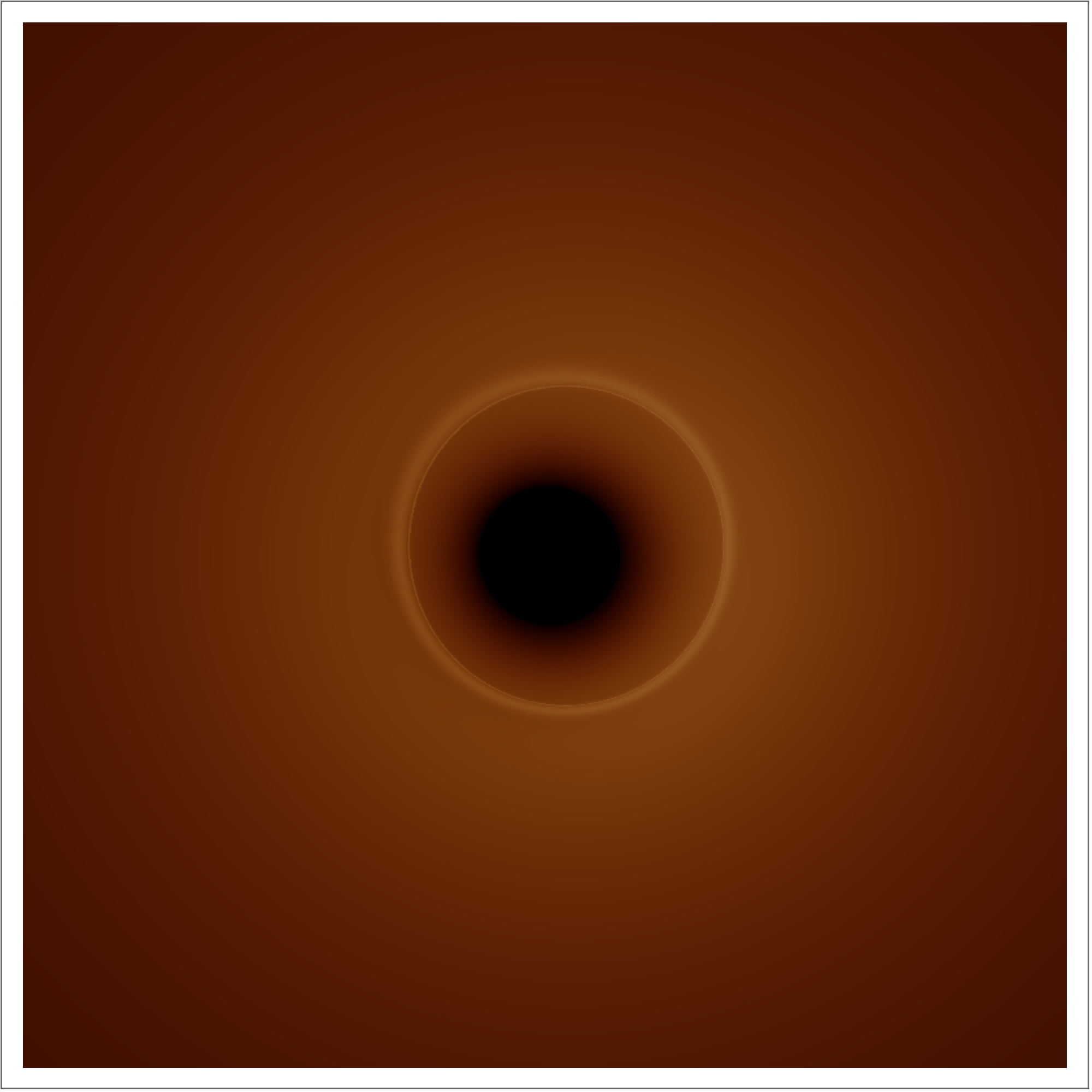}};
	\node[below left] at (2.46,-3) {\tiny\color{white} $a=0.998\ B=0$};
	\node[below left] at (1.75,-3.3) {\tiny\color{white}$\t_o=163^{\circ}$};
	
	\node at (6,-2) {\includegraphics[scale=0.18]{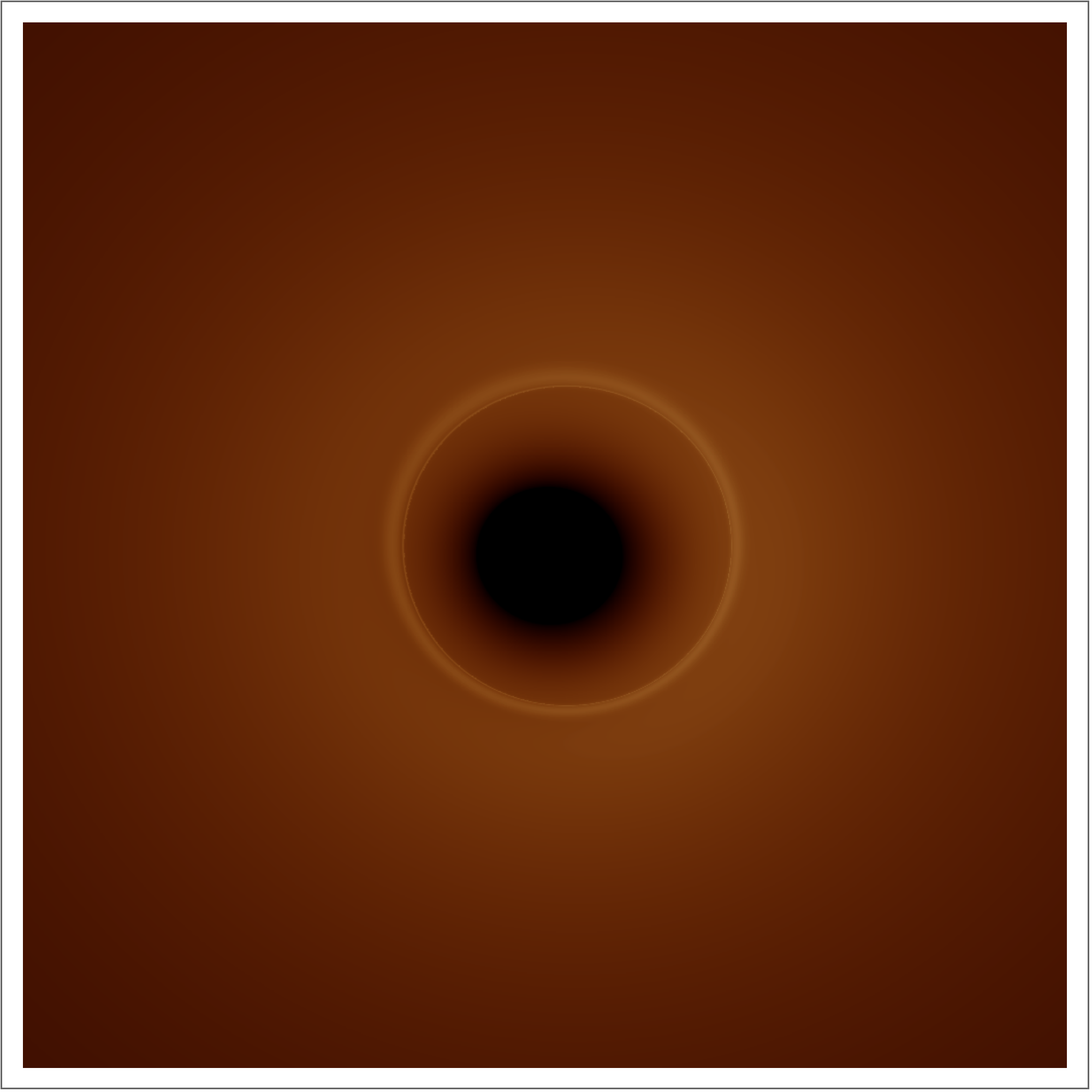}};
	\node[below left] at (6.8,-3) {\tiny\color{white} $a=0.998\ B=0.01$};
	\node[below left] at (5.75,-3.3) {\tiny\color{white}$\t_o=163^{\circ}$};
	\end{tikzpicture}
	\caption{Images of KMBH illuminated by retrograde flows. The top row of the images is observed at $\t_o = 80^{\circ}$, and the bottom row is observed at $\t_o = 163^{\circ}$.}
	\label{fig:intensitym1}
\end{figure}

\begin{figure}[h!]
	\centering
	\includegraphics[scale=0.4]{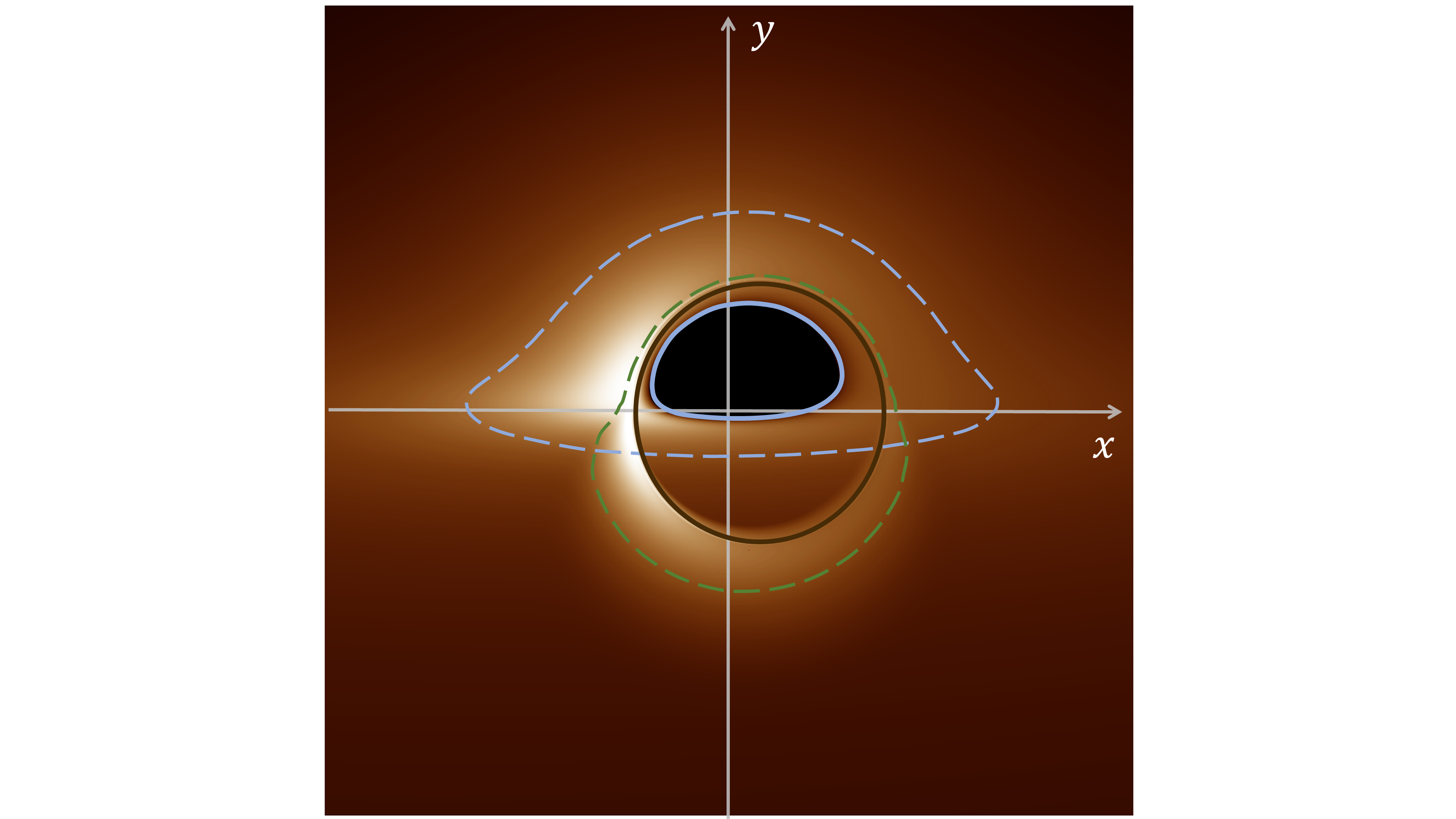} 
	\caption{An illustration of four characteristic curves in an image of the KMBH illuminated by an prograde accretion disk with $\theta_o=80^\circ$, $a=0.6$ and $B=0$. Blue dashed and solid lines are the direct images of the accretion disk at $r=9$ and $r=r_H$, respectively. The green dashed curve is the lensed image of the accretion disk at $r=9$ and the black solid curve is the critical curve of the KMBH.}
	\label{dia}
\end{figure}
In Fig. \ref{fig:intensity1} and Fig. \ref{fig:intensitym1}, we show the images of the KMBH illuminated by an accretion disk with prograde and retrograde flows, respectively. Top images of both figures are taken at $\theta_o=80^\circ$ and bottom ones are given at $\theta_o=163^\circ$. For each row in Fig. \ref{fig:intensity1} and Fig. \ref{fig:intensitym1}, we let the spin of the KMBH $a=0.6$ for the first two images and $a=0.998$ for the last two ones, while for the first and third we consider $B=0$ and for the others we take $B=0.01$. \footnote{When generating the images, the color-function is taken as RGBColor$(I^{1/8}, I^{1/4}, I^{1/2}, (I+1)^{-3})$ for better visual effect.}

We illustrate four characteristic curves in the image of the KMBH in Fig. \ref{dia}. The dashed curves are images of the accretion disk at $r=9$, of which the blue one is the direct image and the green one is the lensed. The blue solid curve is the image of the accretion disk at $r=r_H$ which is also called the inner shadow in \cite{Chael:2021rjo} and the black solid is the critical curve for the KMBH which is also called the photon ring in the literatures.

From the images in Fig. \ref{fig:intensity1} and Fig. \ref{fig:intensitym1}, we can see that the direct image and lensed image can be clearly distinguished when $\t_o = 80^{\circ}$ no matter the flow of the accretion disk is prograde or retrograde, while for $\t_o=163^\circ$ the intensity distribution is no longer so sharp that the direct and lensed images are hard to tell apart. Nevertheless, the inner shadow and the critical curve can be clearly observed for both $\theta_o=80^\circ$ and $\t_o=163^\circ$. Note that the inner shadow is from the fact that a light ray cannot gain any intensity if falling into the horizon without crossing the equatorial plane, i.e., $N_{max} = 0$. It is a main feature of the geometrically thin disks but not limited to them. The center intensity depression is observable as long as the disk is not too thick, i.e., $H \lesssim R$, where $H$ and $R$ are the thickness and width of the disk, respectively. In addition, with the increasing of the strength of the magnetic field $B$, the images are stretched significantly. Similar results have been found for the black holes immersed in Melvin magnetic fields \cite{Junior:2021dyw, Wang:2021ara} and Wald magnetic fields \cite{Zhong:2021mty} illuminated by the spherical extended source at infinity . Moreover,  there are significant Doppler effects on the left side of the screen due to the forward rotation of the prograde accretion disk. In contrast,  even though the Doppler effects on the right side of the screen for the retrograde accretion disk are obvious, they are weaker due to the dragging effect of the KMBH.

 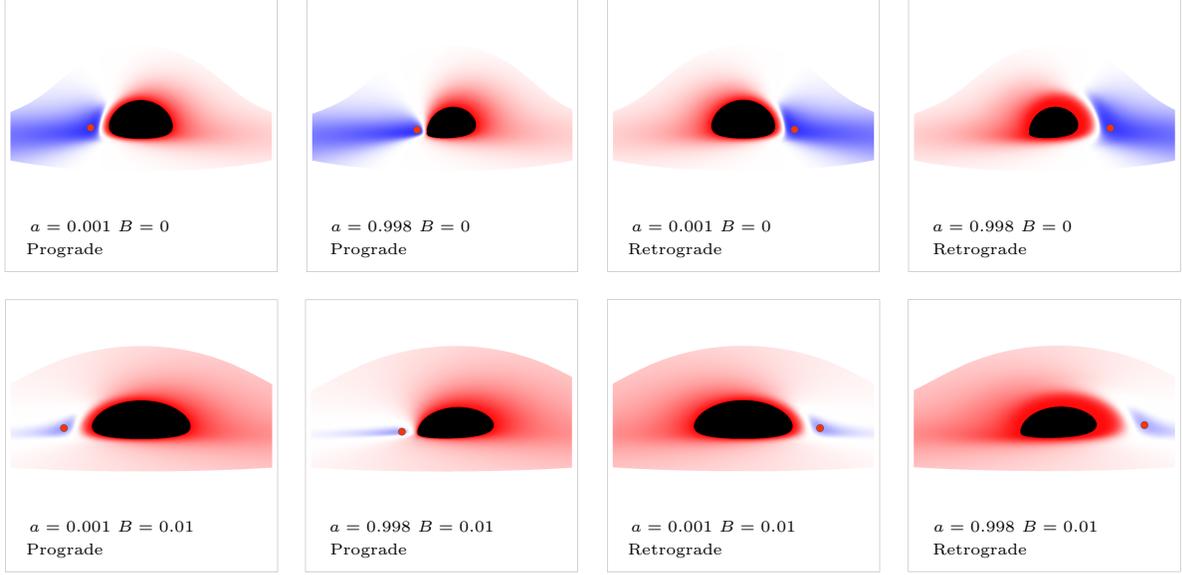
\begin{figure}[h!]
\centering
\begin{tikzpicture}
\node at (-6,2) {\includegraphics[scale=0.19]{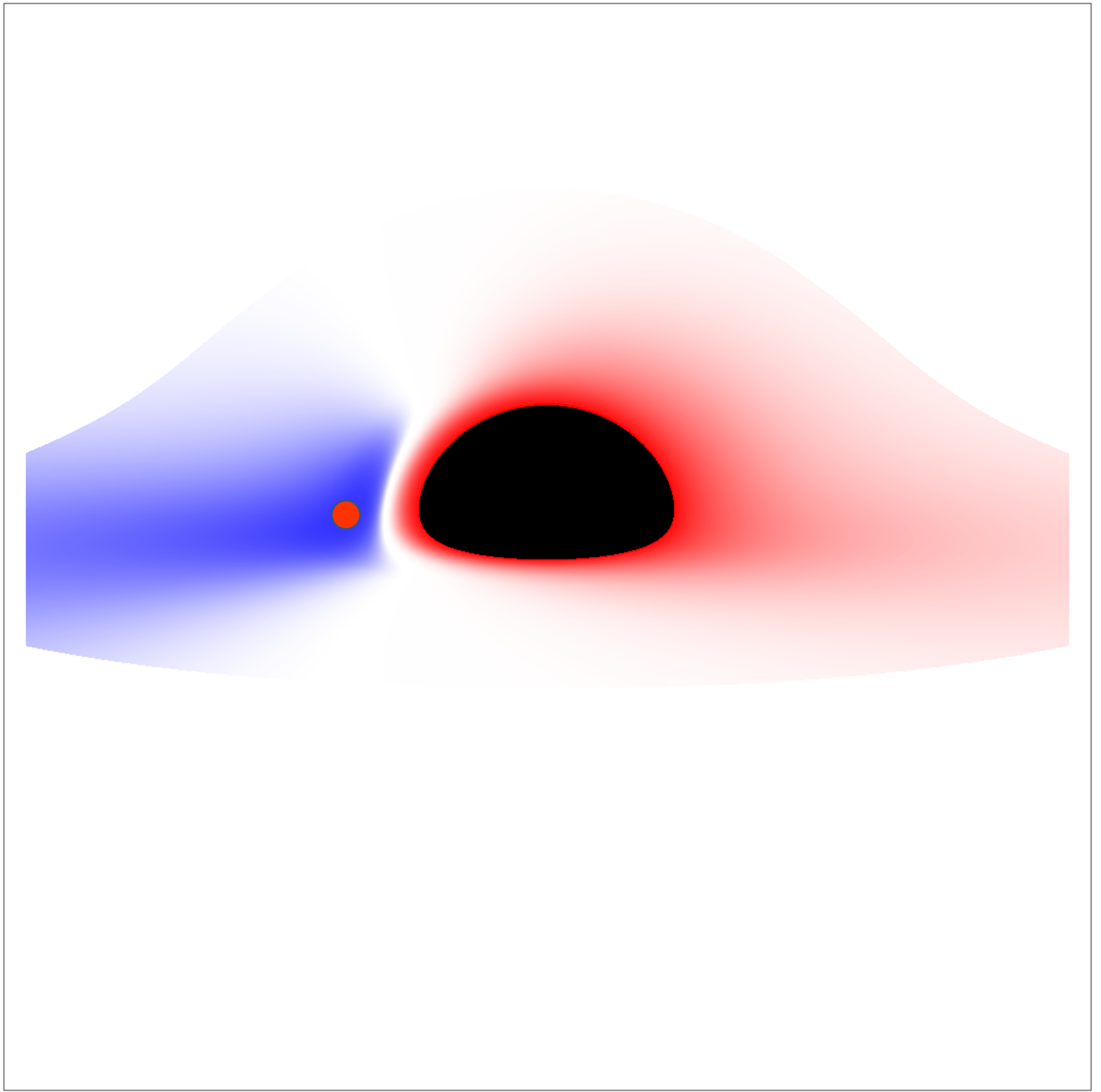}};
\node[below left] at (-5.5,1) {\tiny\color{black} $a=0.001\ B=0$};
\node[below left] at (-6.38,0.7) {\tiny\color{black} Prograde};

\node at (-2,2) {\includegraphics[scale=0.19]{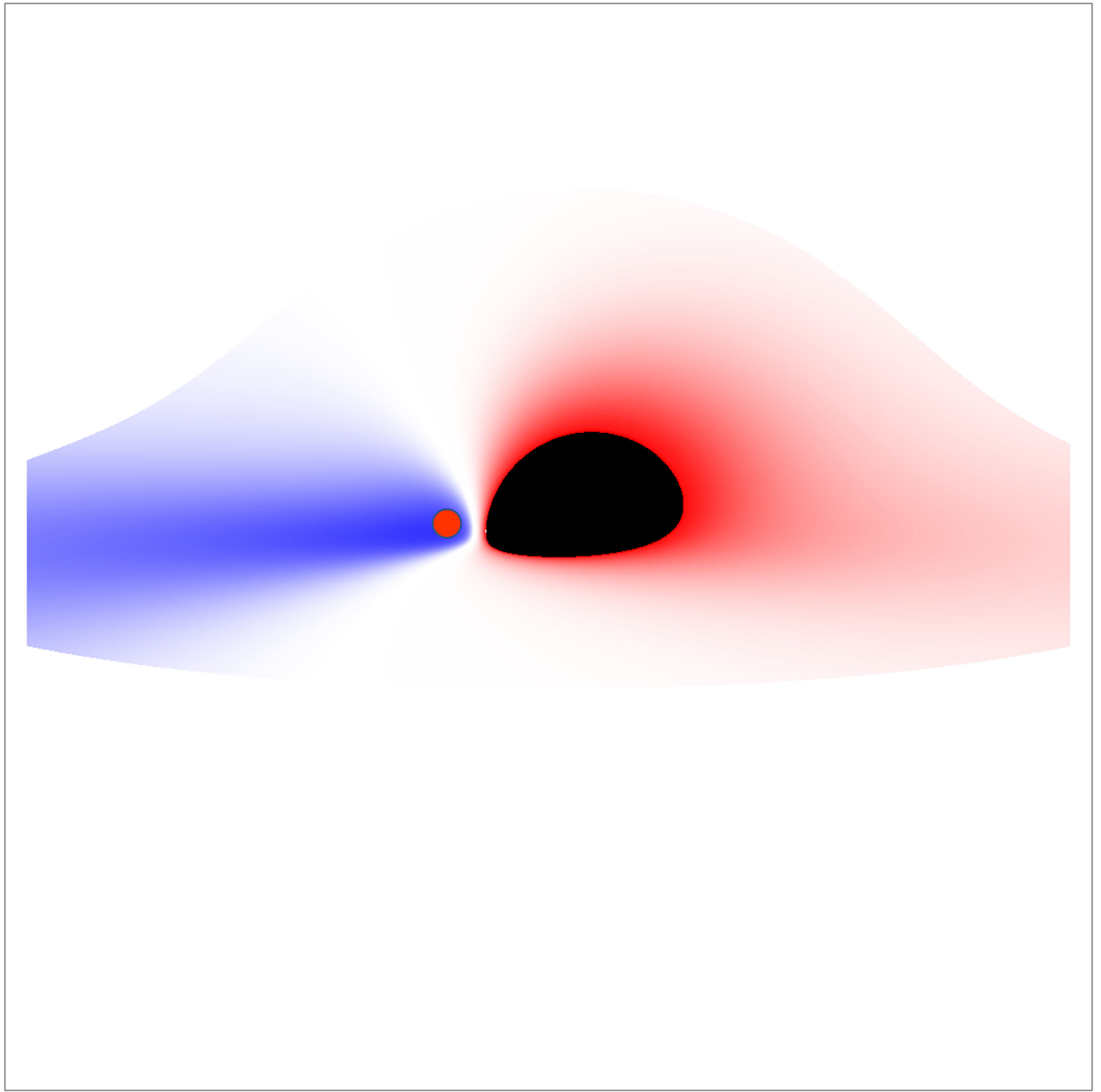}};
\node[below left] at (-1.5,1) {\tiny\color{black} $a=0.998\ B=0$};
\node[below left] at (-2.34,0.7) {\tiny\color{black} Prograde};

\node at (2,2) {\includegraphics[scale=0.19]{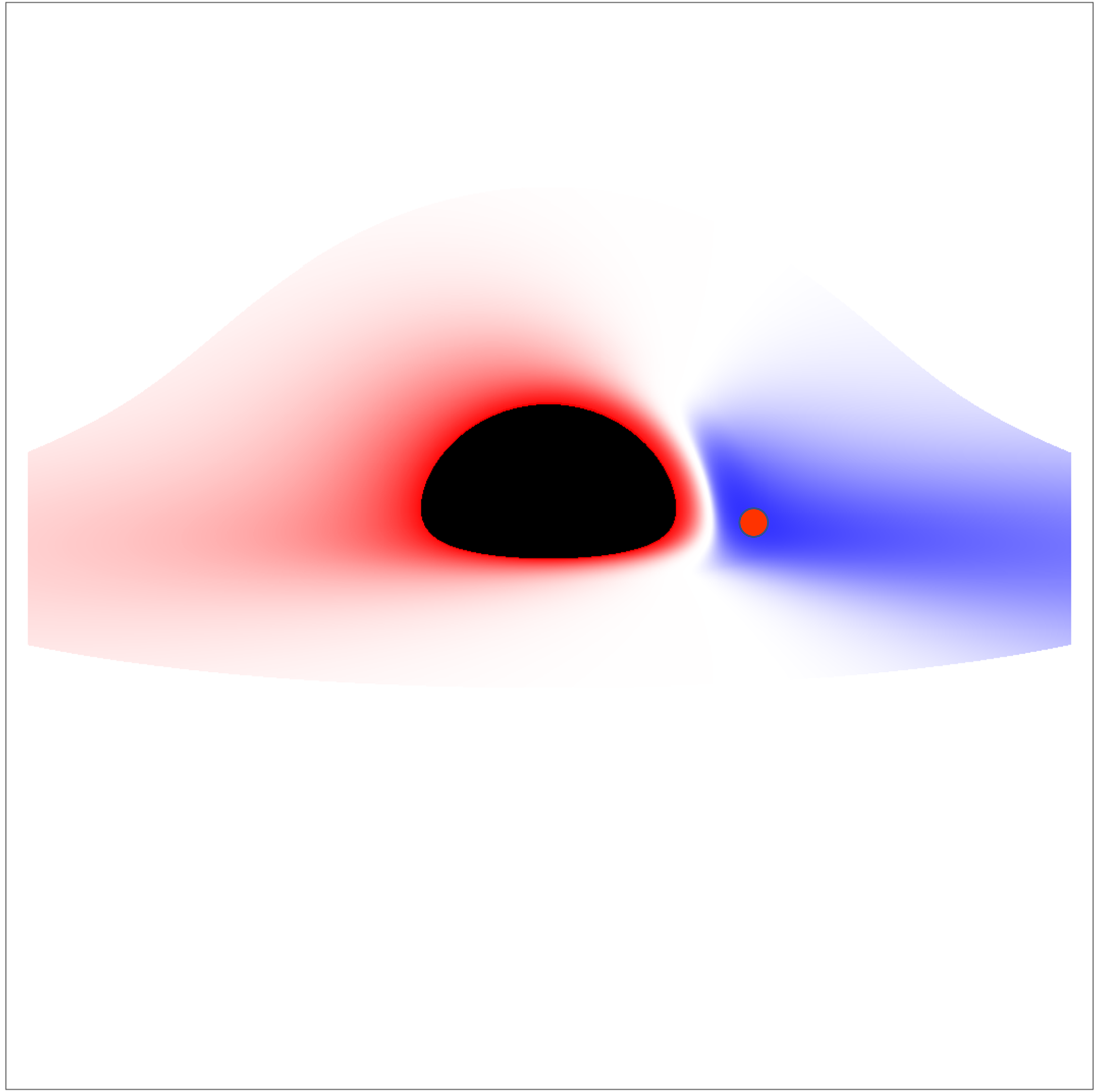}}; 
\node[below left] at (2.5,1) {\tiny\color{black} $a=0.001\ B=0$};
\node[below left] at (1.85,0.7)  {\tiny\color{black} Retrograde};

\node at (6,2) {\includegraphics[scale=0.19]{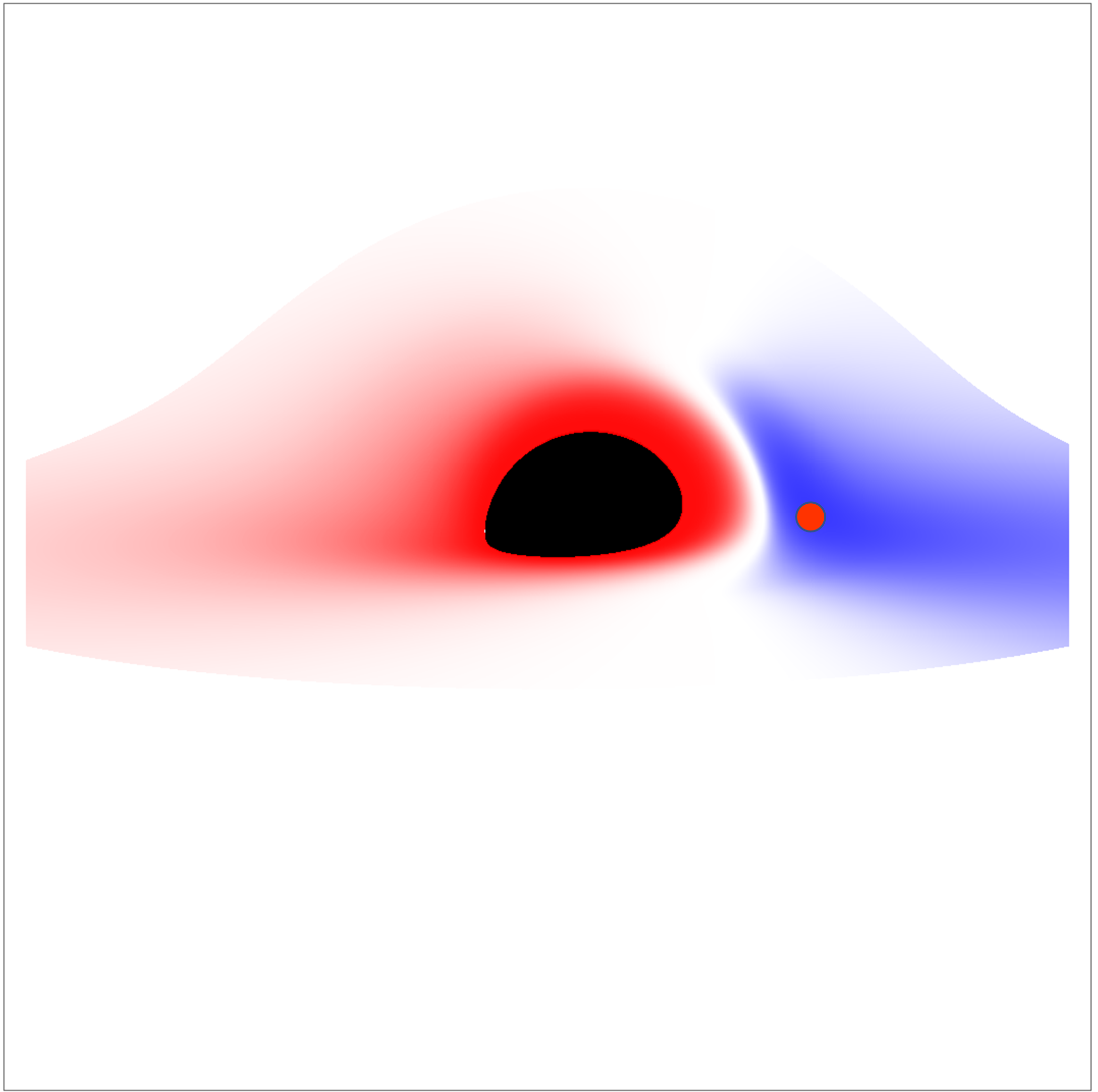}};
\node[below left] at (6.5,1) {\tiny\color{black} $a=0.998\ B=0$};
\node[below left] at (5.9,0.7) {\tiny\color{black} Retrograde};

\node at (-6,-2) {\includegraphics[scale=0.19]{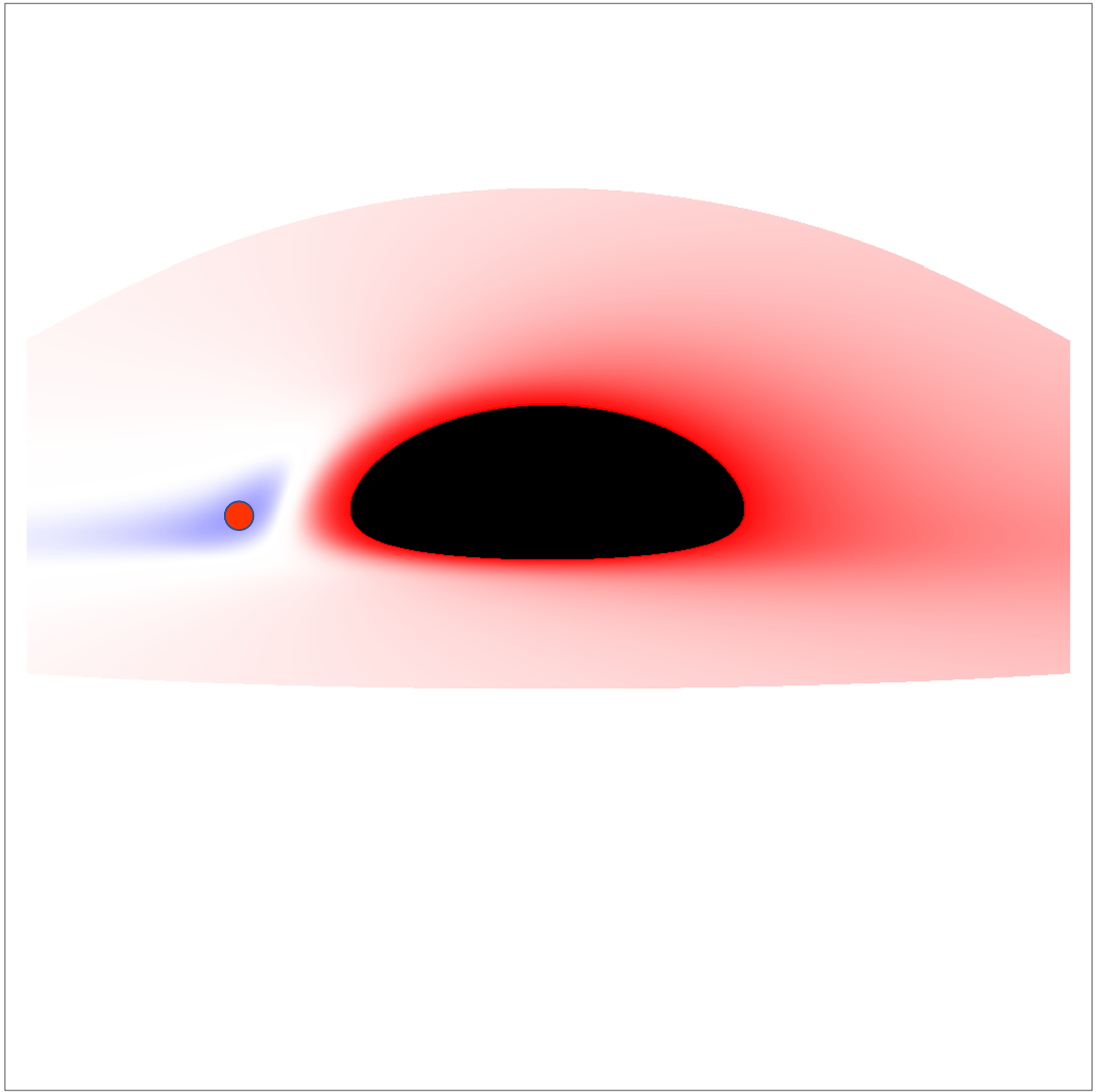}};
\node[below left] at (-5.17,-3) {\tiny\color{black} $a=0.001\ B=0.01$};
\node[below left] at (-6.38,-3.3) {\tiny\color{black} Prograde};

\node at (-2,-2) {\includegraphics[scale=0.19]{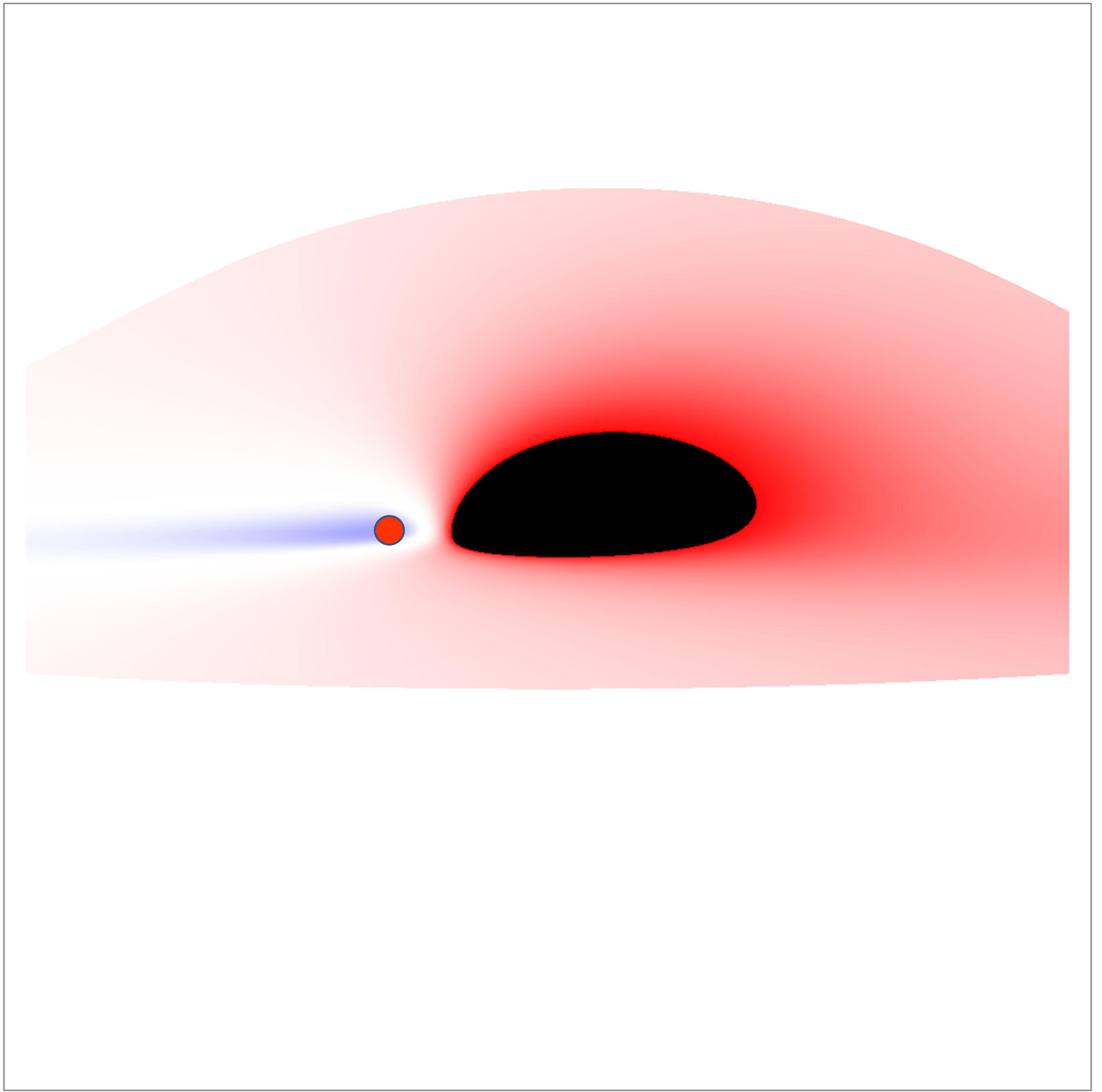}};
\node[below left] at (-1.18,-3) {\tiny\color{black} $a=0.998\ B=0.01$};
\node[below left] at (-2.34,-3.3) {\tiny\color{black} Prograde};

\node at (2,-2) {\includegraphics[scale=0.19]{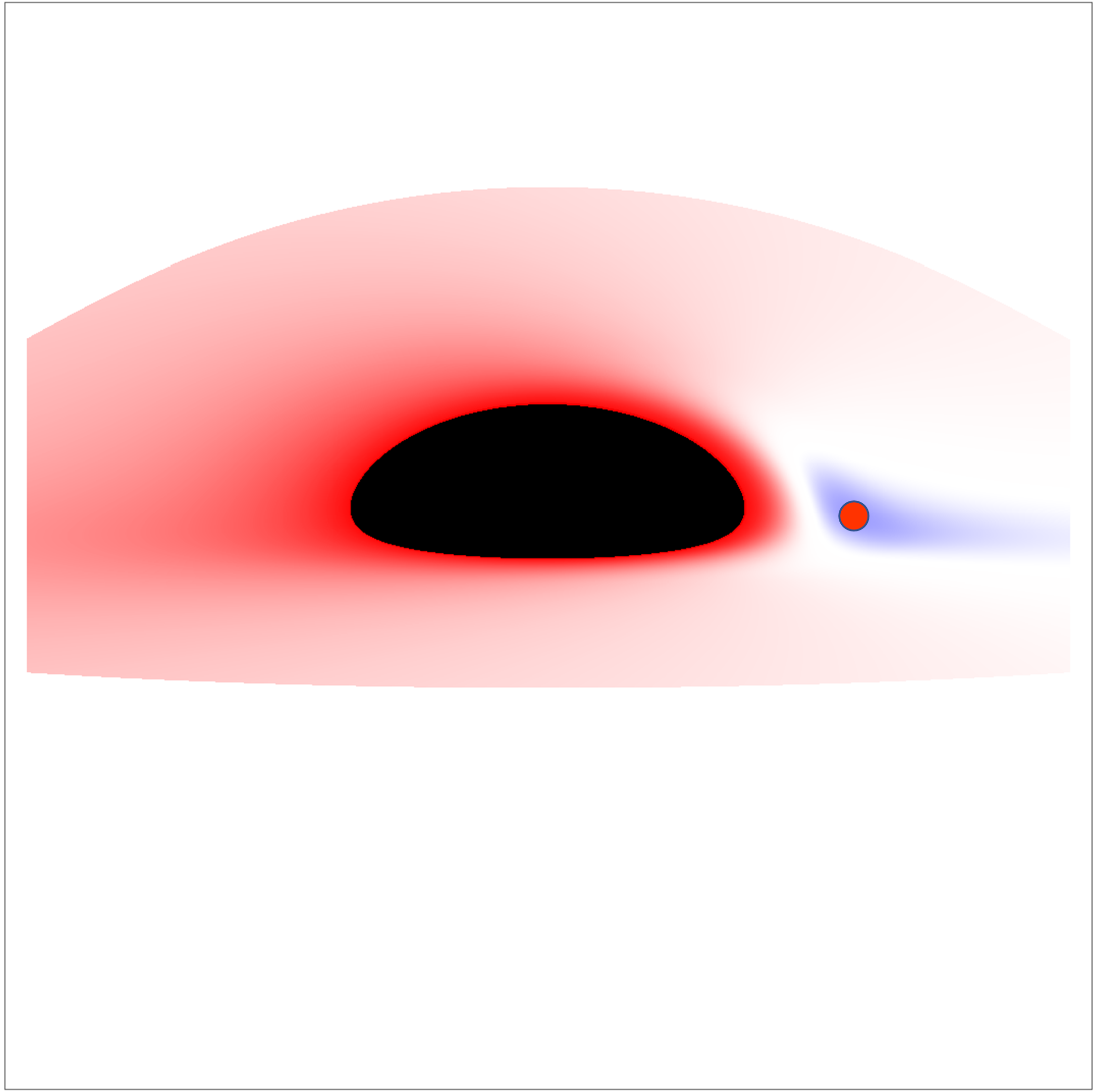}};
\node[below left] at (2.83,-3) {\tiny\color{black} $a=0.001\ B=0.01$};
\node[below left] at (1.85,-3.3) {\tiny\color{black} Retrograde};

\node at (6,-2) {\includegraphics[scale=0.19]{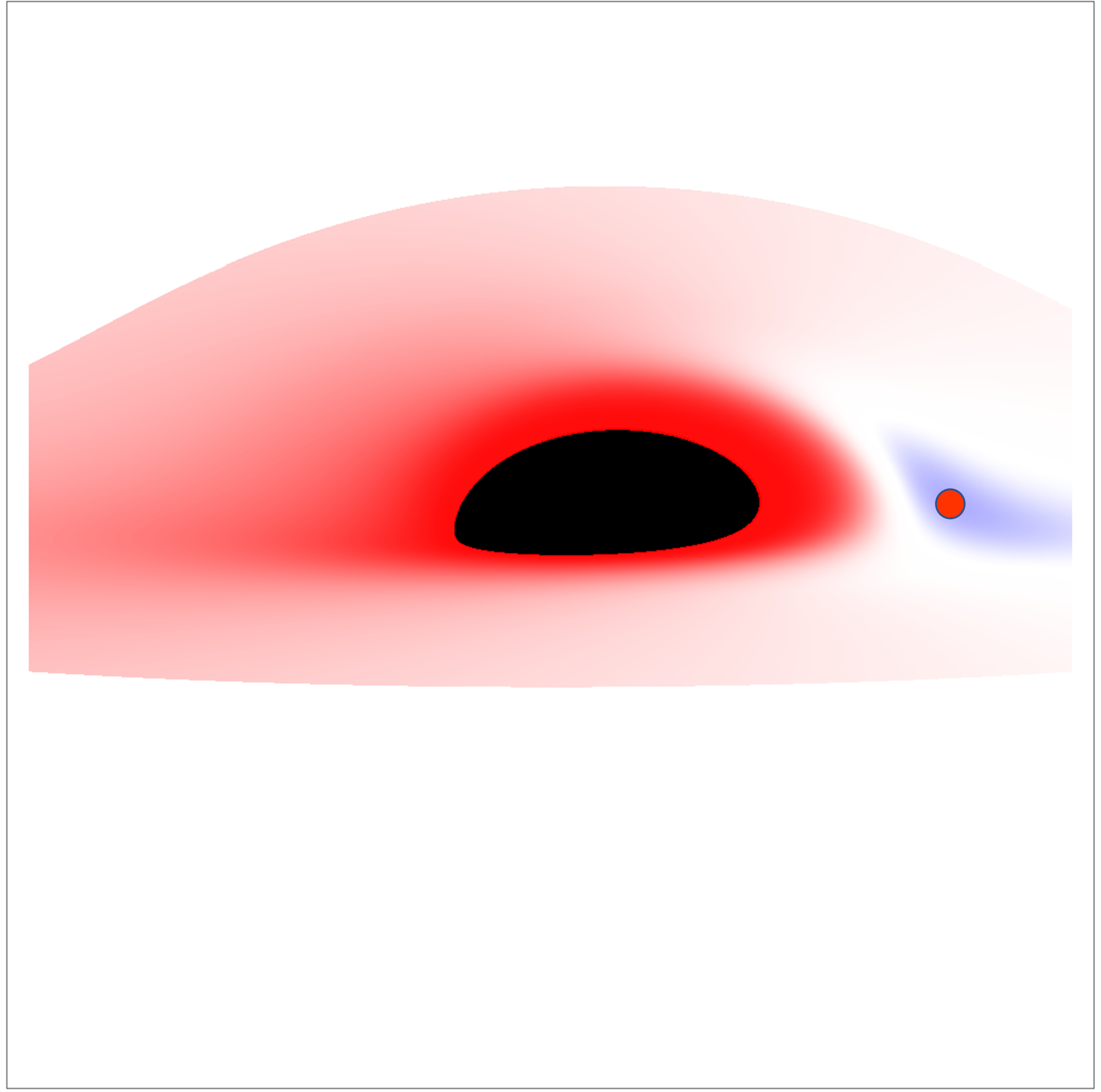}};
\node[below left] at (6.85,-3) {\tiny\color{black} $a=0.998\ B=0.01$};
\node[below left] at (5.9,-3.3) {\tiny\color{black} Retrograde};
\end{tikzpicture}
\caption{ The redshift factors of direct images of the accretion disk model. The radii range from $r_H$ to $20$. Red and blue represent redshift and blueshift, respectively, and the maximal blueshift points are indicated by red dots. The black regions are the inner shadows.}
\label{fig:rf1}
\end{figure}
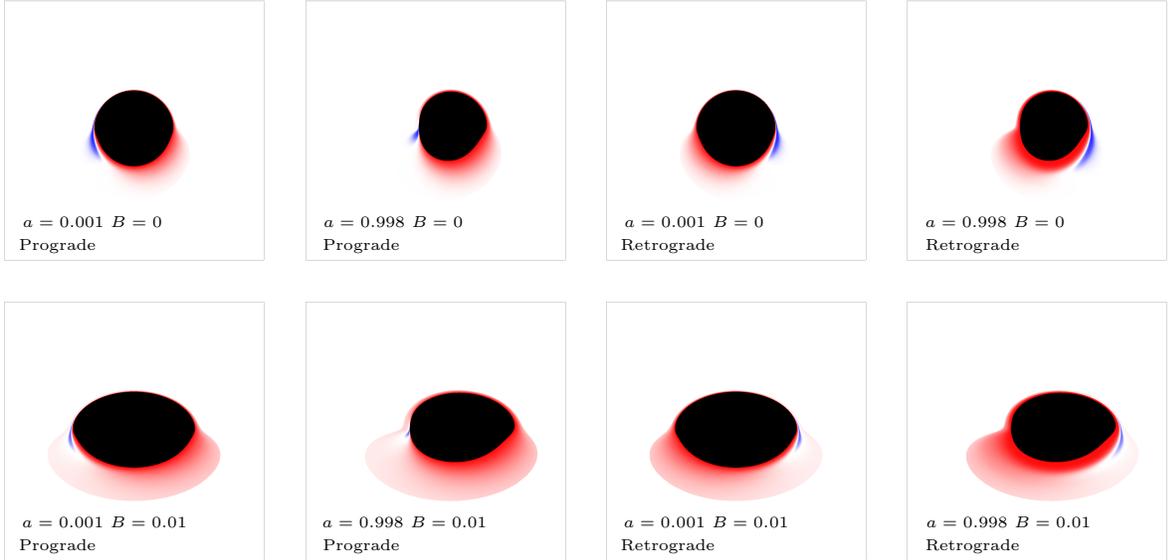
\begin{figure}[h!]
	\centering
	\begin{tikzpicture}
	\node at (-6,2) {\includegraphics[scale=0.17]{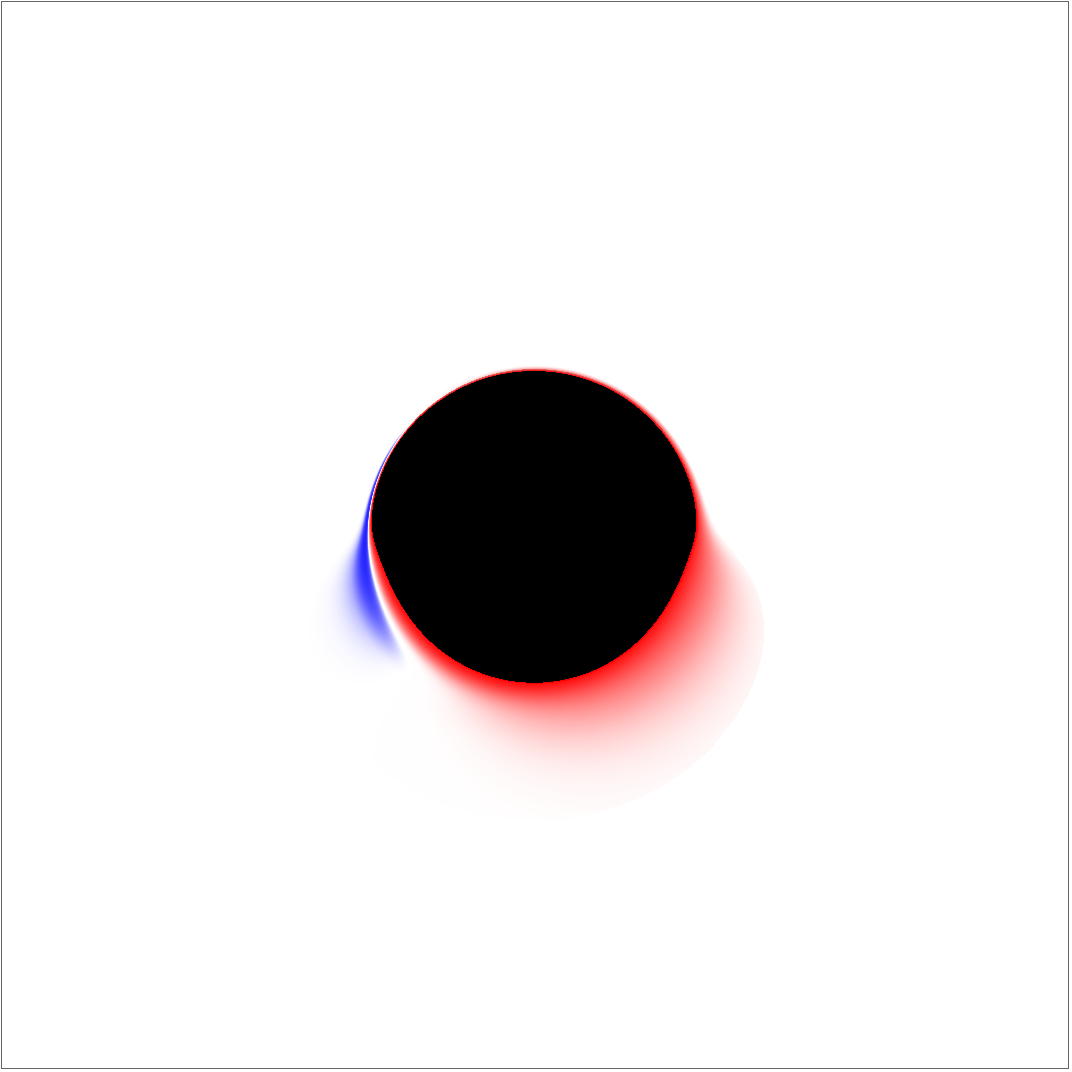}};
	\node[below left] at (-5.5,1) {\tiny\color{black} $a=0.001\ B=0$};
	\node[below left] at (-6.38,0.7) {\tiny\color{black} Prograde};
	
	\node at (-2,2) {\includegraphics[scale=0.17]{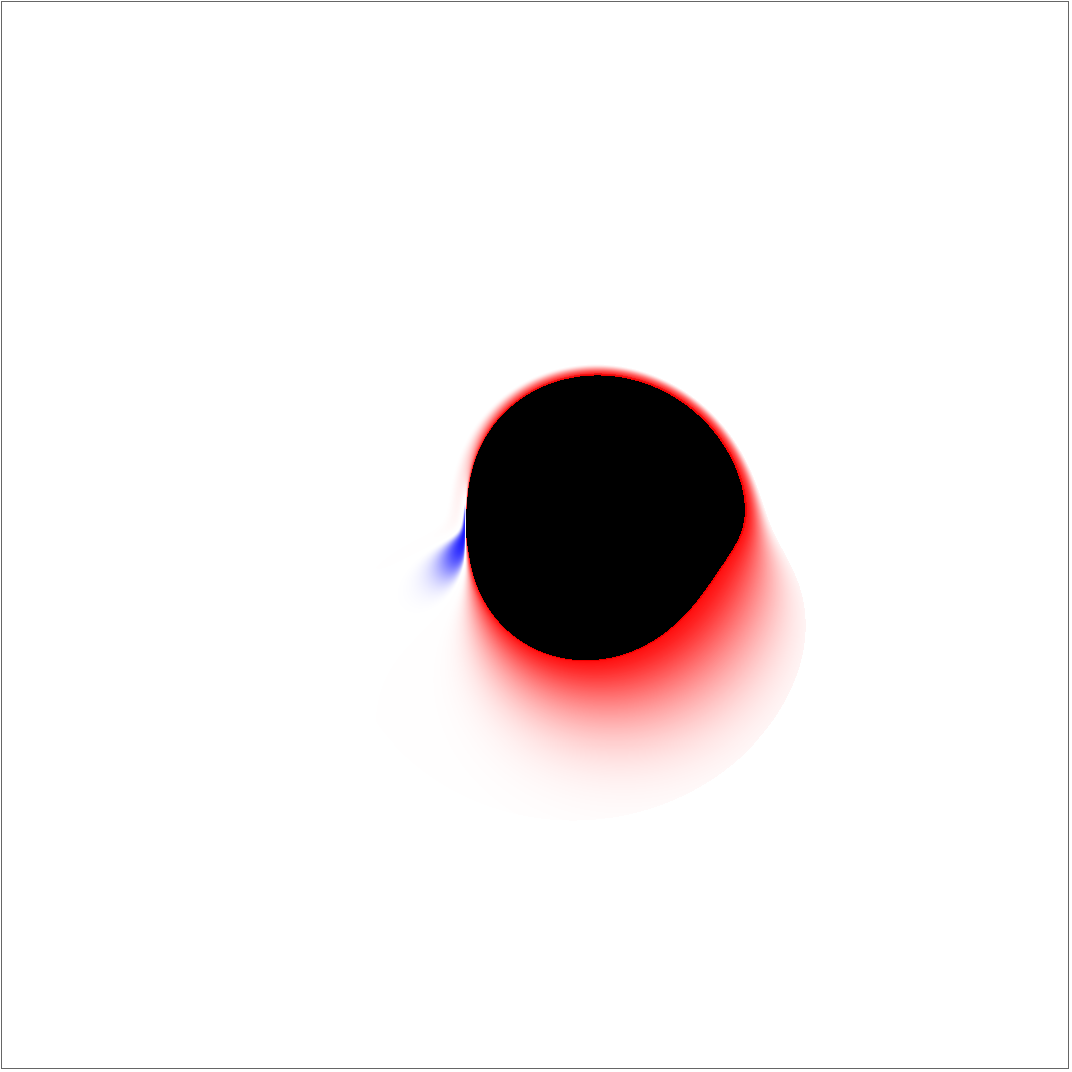}};
	\node[below left] at (-1.5,1) {\tiny\color{black} $a=0.998\ B=0$};
	\node[below left] at (-2.34,0.7) {\tiny\color{black} Prograde};
	
	\node at (2,2) {\includegraphics[scale=0.17]{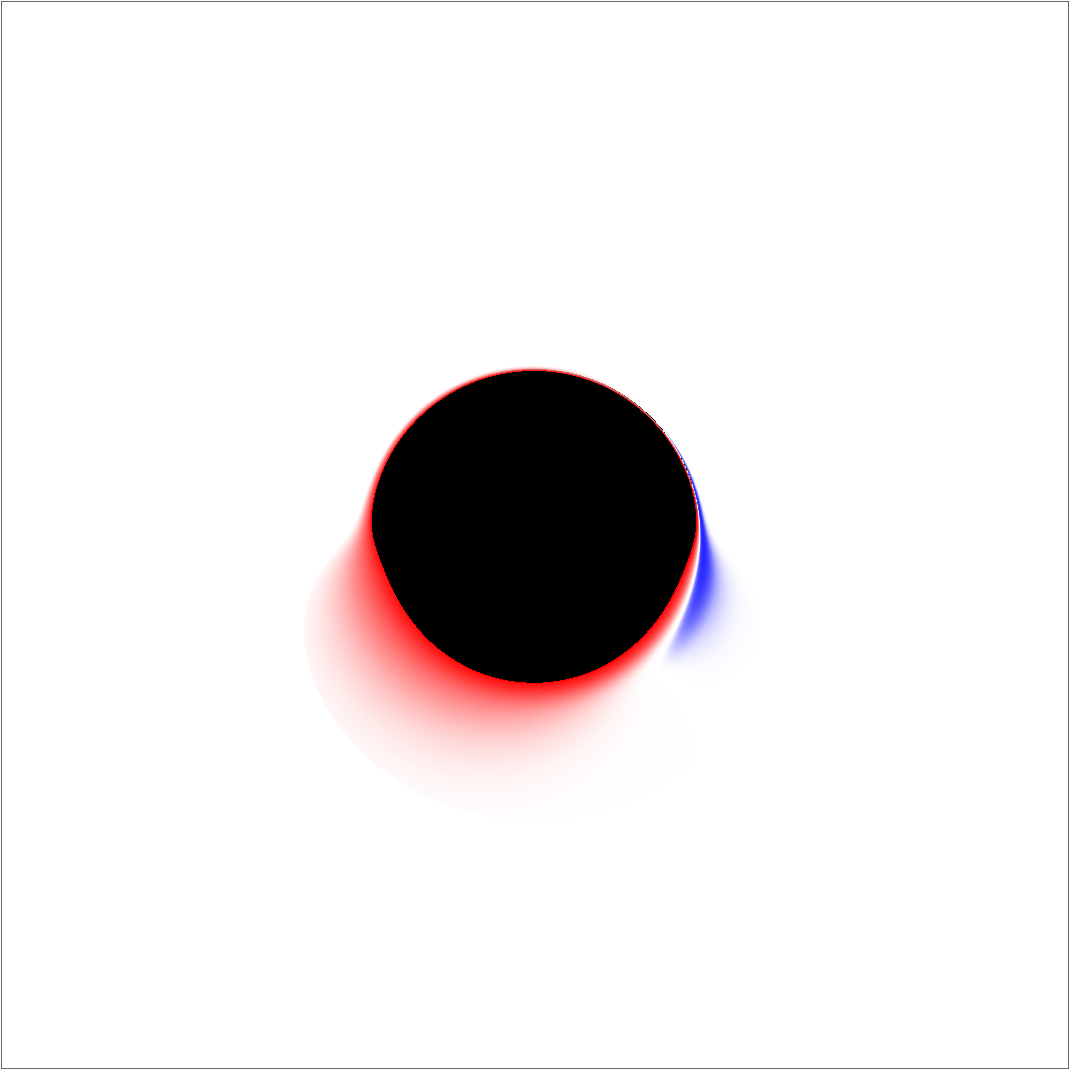}}; 
	\node[below left] at (2.5,1) {\tiny\color{black} $a=0.001\ B=0$};
	\node[below left] at (1.85,0.7)  {\tiny\color{black} Retrograde};
	
	\node at (6,2) {\includegraphics[scale=0.17]{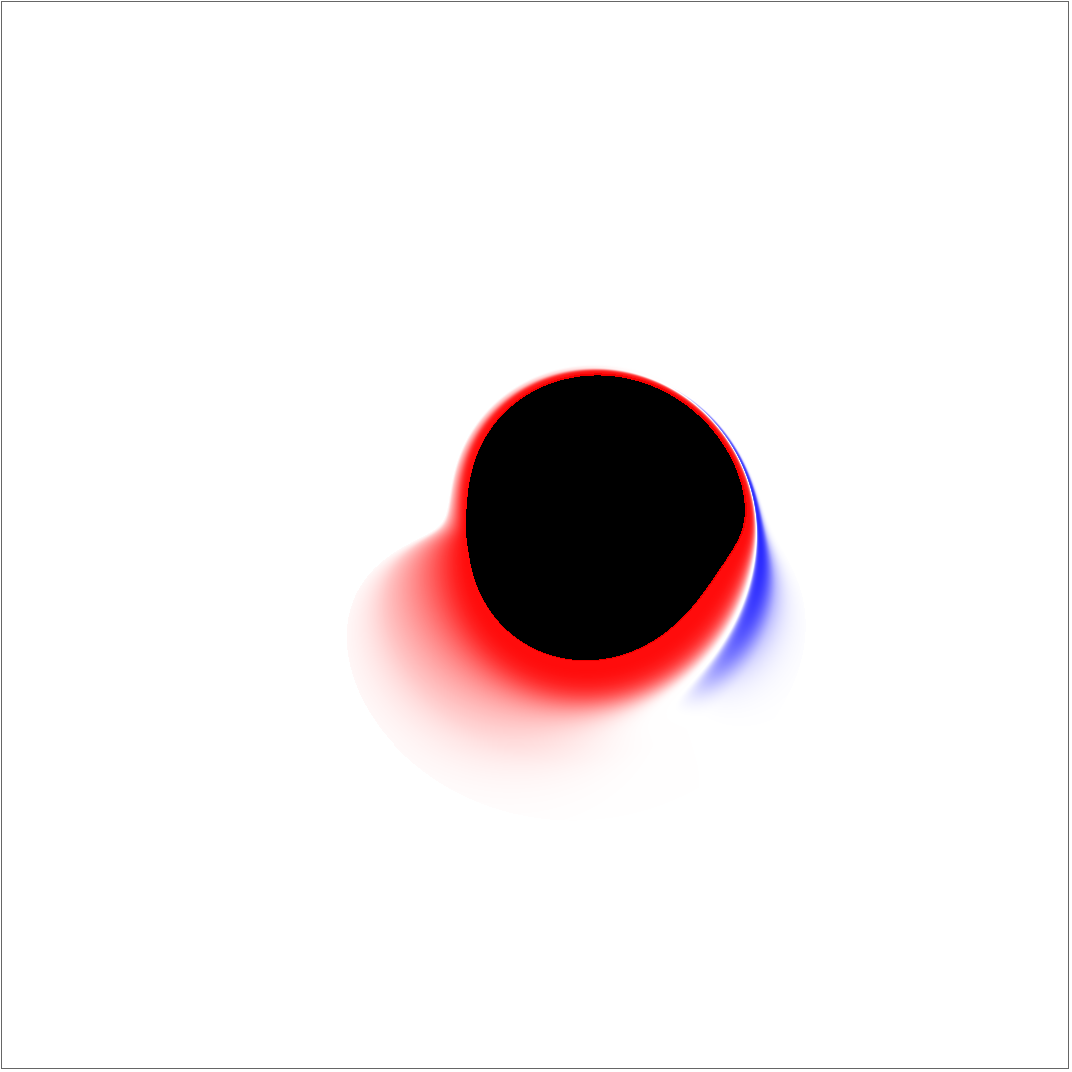}};
	\node[below left] at (6.5,1) {\tiny\color{black} $a=0.998\ B=0$};
	\node[below left] at (5.9,0.7) {\tiny\color{black} Retrograde};

	\node at (-6,-2) {\includegraphics[scale=0.17]{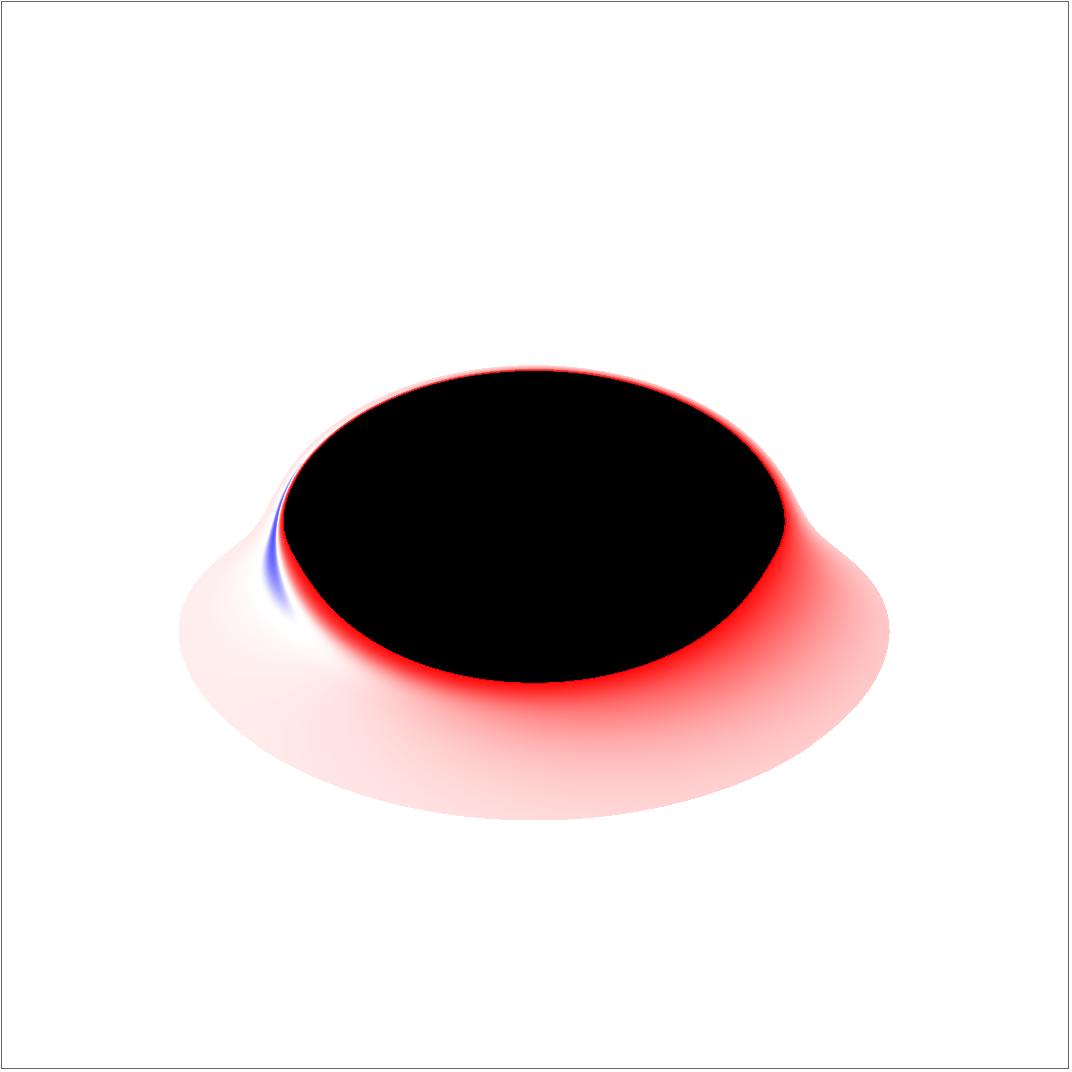}};
	\node[below left] at (-5.17,-3) {\tiny\color{black} $a=0.001\ B=0.01$};
	\node[below left] at (-6.38,-3.3) {\tiny\color{black} Prograde};
	
	\node at (-2,-2) {\includegraphics[scale=0.17]{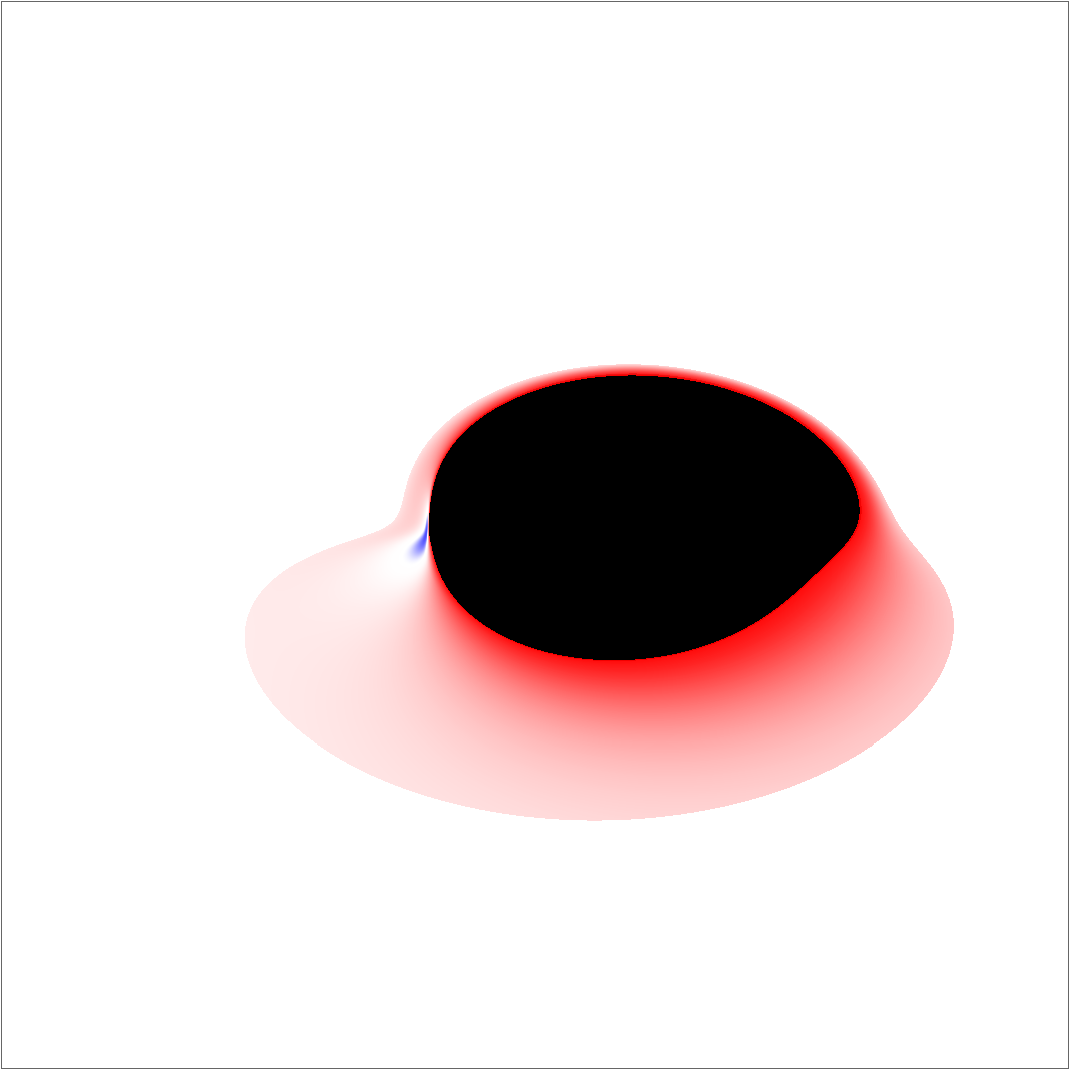}};
	\node[below left] at (-1.18,-3) {\tiny\color{black} $a=0.998\ B=0.01$};
	\node[below left] at (-2.34,-3.3) {\tiny\color{black} Prograde};
	
	\node at (2,-2) {\includegraphics[scale=0.17]{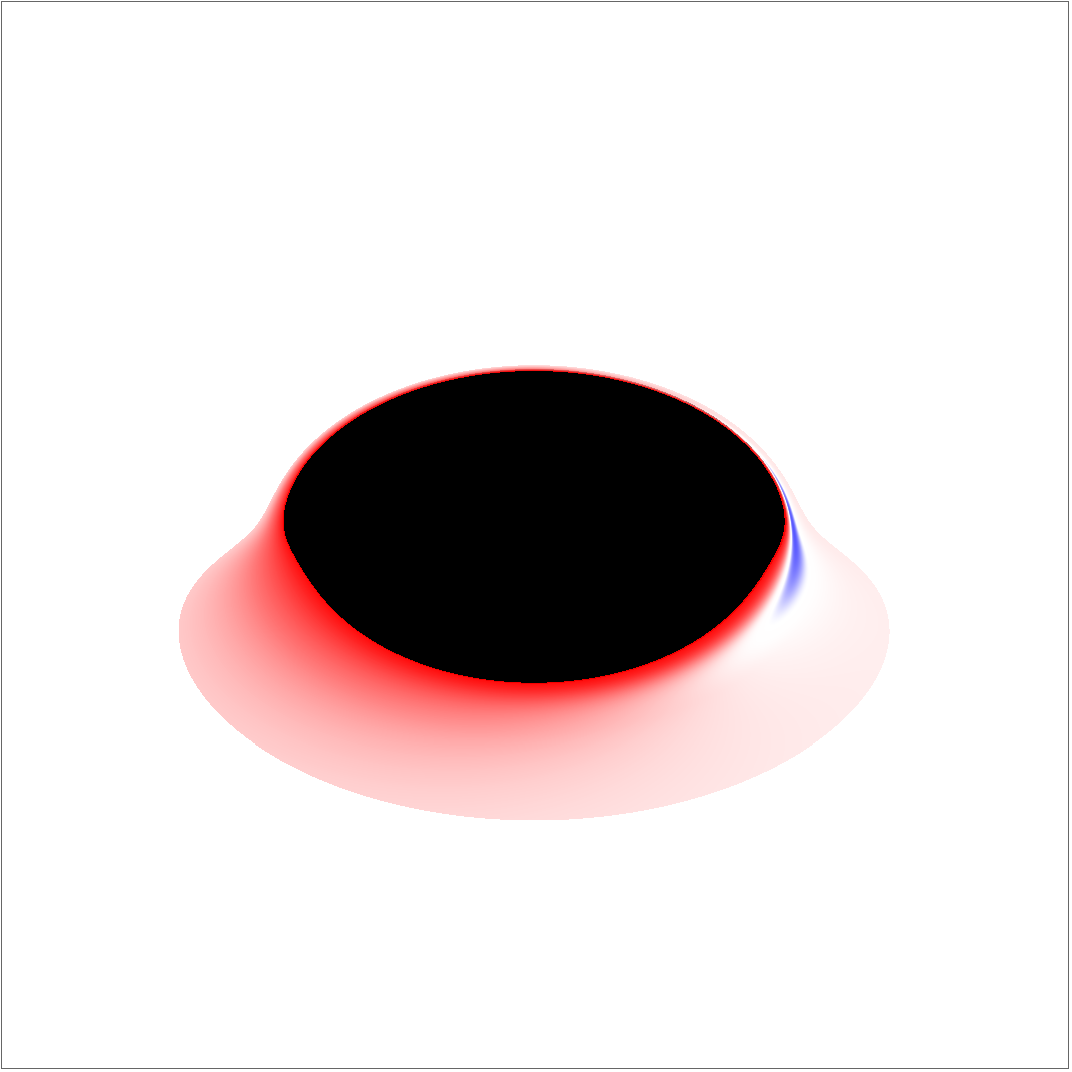}};
	\node[below left] at (2.83,-3) {\tiny\color{black} $a=0.001\ B=0.01$};
	\node[below left] at (1.85,-3.3) {\tiny\color{black} Retrograde};
	
	\node at (6,-2) {\includegraphics[scale=0.17]{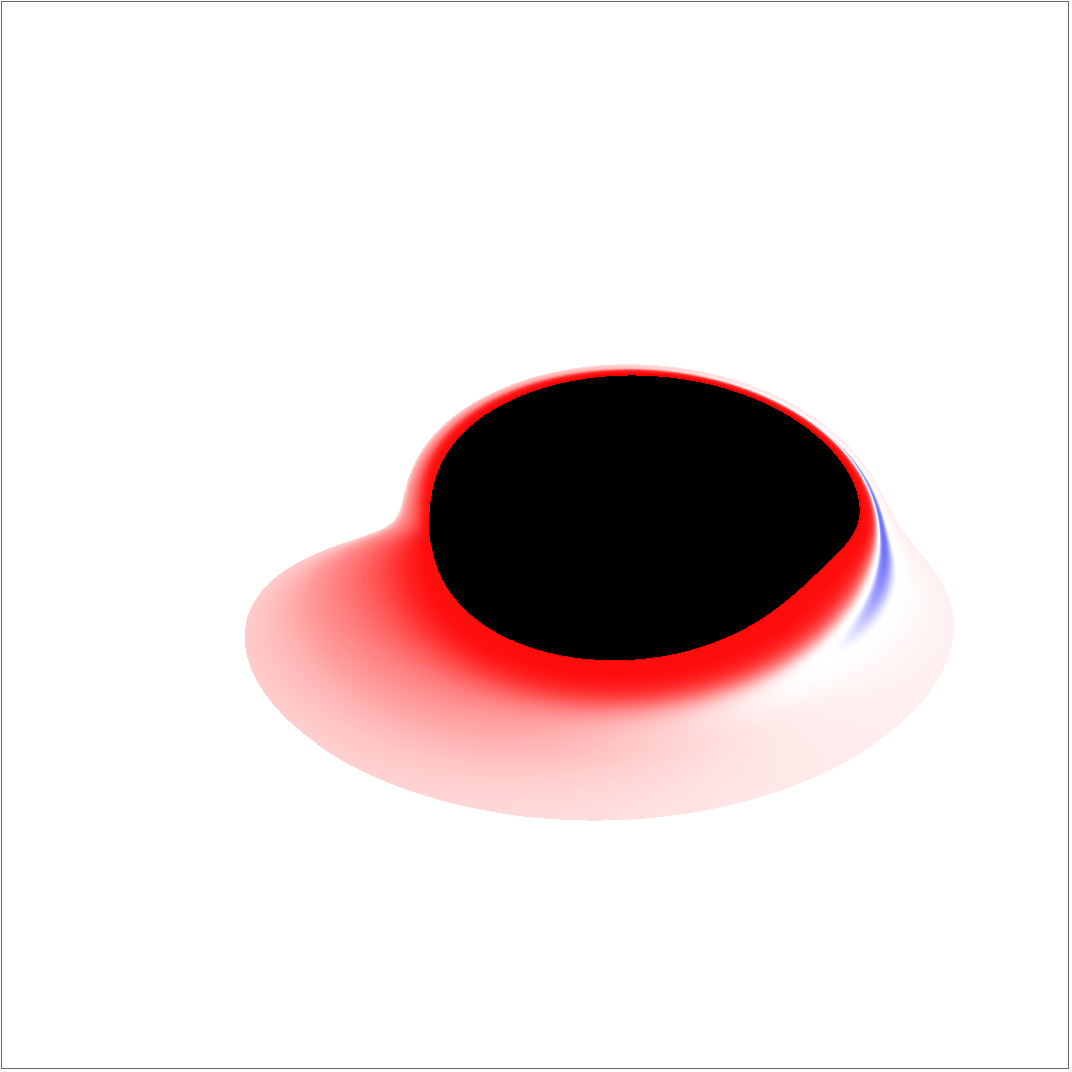}};
	\node[below left] at (6.85,-3) {\tiny\color{black} $a=0.998\ B=0.01$};
	\node[below left] at (5.9,-3.3) {\tiny\color{black} Retrograde};
	\end{tikzpicture}
	\caption{ The redshift factors of lensed images of the accretion disk model. The radii range from $r_H$ to $20$. Red and blue represent redshift and blueshift, respectively. The edges of the black regions are the lensed images of $r_H$.}
	\label{fig:rf2}
\end{figure}

\begin{table}
	\begin{center}
		\begin{tabular}{|c|c|c||c|c|c|c|c|c|}
			\hline
			\multicolumn{3}{|c||}{$a$}&\multicolumn{2}{c|}{0.001}&\multicolumn{2}{c|}{0.6}&\multicolumn{2}{c|}{0.998}\\
			\hline
			\multicolumn{3}{|c||}{$B$} & 0 & 0.01 & 0 & 0.01& 0 & 0.01\\
			\hline
			\hline
			\multirow{3}{*}{P}&
			\multirow{2}{*}{$g_{max}$} &direct&1.455&1.175&1.483  &1.195&1.444&1.163\\
			&&lensed&1.565&1.308&1.627&1.362&1.647&1.336\\
			\cline{2-9}
			&\multicolumn{2}{c||}{$F$}&1&0.774&1.065&0.835&0.559&0.445\\
			\hline
			\multirow{3}{*}{R}&
			\multirow{2}{*}{$g_{max}$} &direct&1.454&1.175&1.43&1.161&1.417&1.153\\
			&&lensed&1.565&1.308&1.525&1.275&1.504&1.26\\
			\cline{2-9}
			&\multicolumn{2}{c||}{$F$}&1&0.774&0.581&0.447&0.087&0.068\\
			\hline
		\end{tabular}
		\caption{ The maximal blueshift ($g_{max}$) and total flux ($F$) under different spin parameters and field strengths. The observational angle is $\t_o=80^{\circ}$. ``P" and ``R" denote prograde and retrograde flow, respectively.}\label{table}
	\end{center}
\end{table}

Moreover, we show the redshifts of the direct and lensed images of the accretion disk in Fig.~\ref{fig:rf1} and Fig.~\ref{fig:rf2} with $\theta_o=80^\circ$. In the upper row we set $B=0$, and in the lower row we set $B=0.01$. The accretion flows are prograde for the first two columns, and they are retrograde in the other twos. Besides, for the first and third columns $a=0.001$ while for the other columns $a=0.998$. From these images, we can see that in addition to the influence of the magnetic field on the size of the image, there are obvious blueshift near the ISCO, which will be suppressed by the Melvin magnetic field. In Table. \ref{table} we present the maximal blueshift factors under different spin parameters and magnetic field strengths. The maximal blueshift factors of both direct and lensed images are significantly suppressed by the magnetic field.

\begin{figure}[h!]
	\centering
	\includegraphics[scale=0.183]{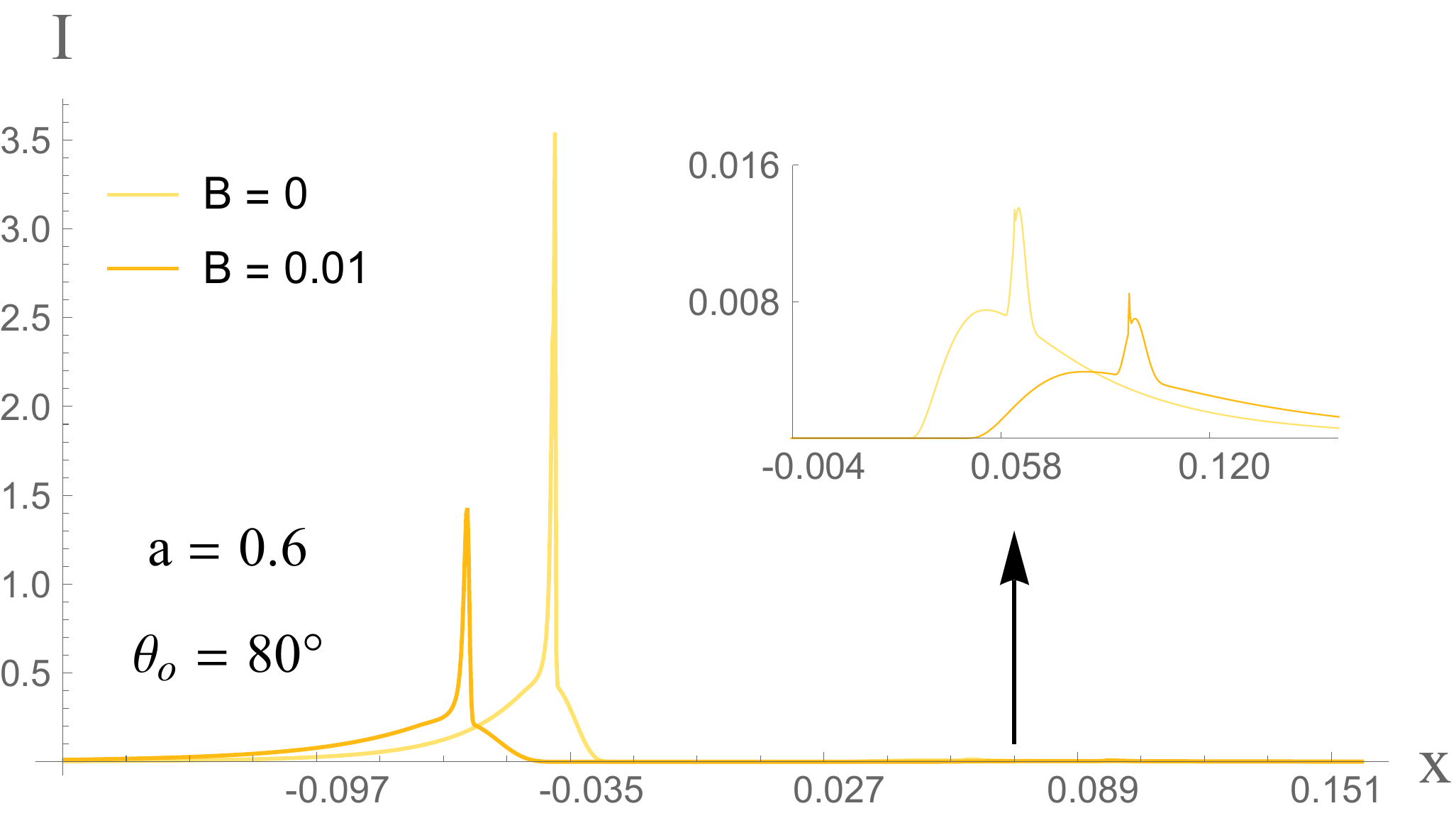}
   \includegraphics[scale=0.183]{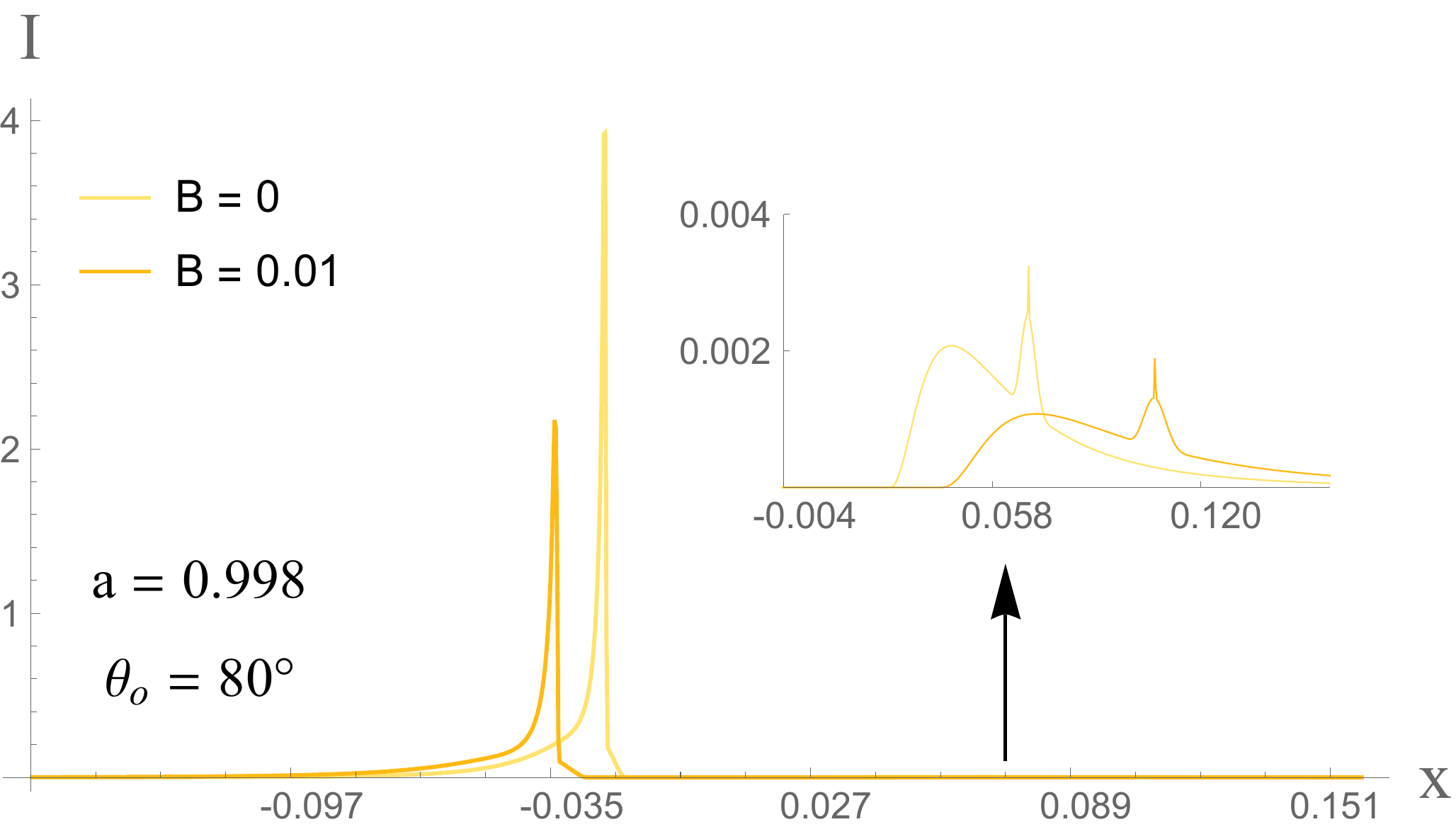} 
   	\includegraphics[scale=0.183]{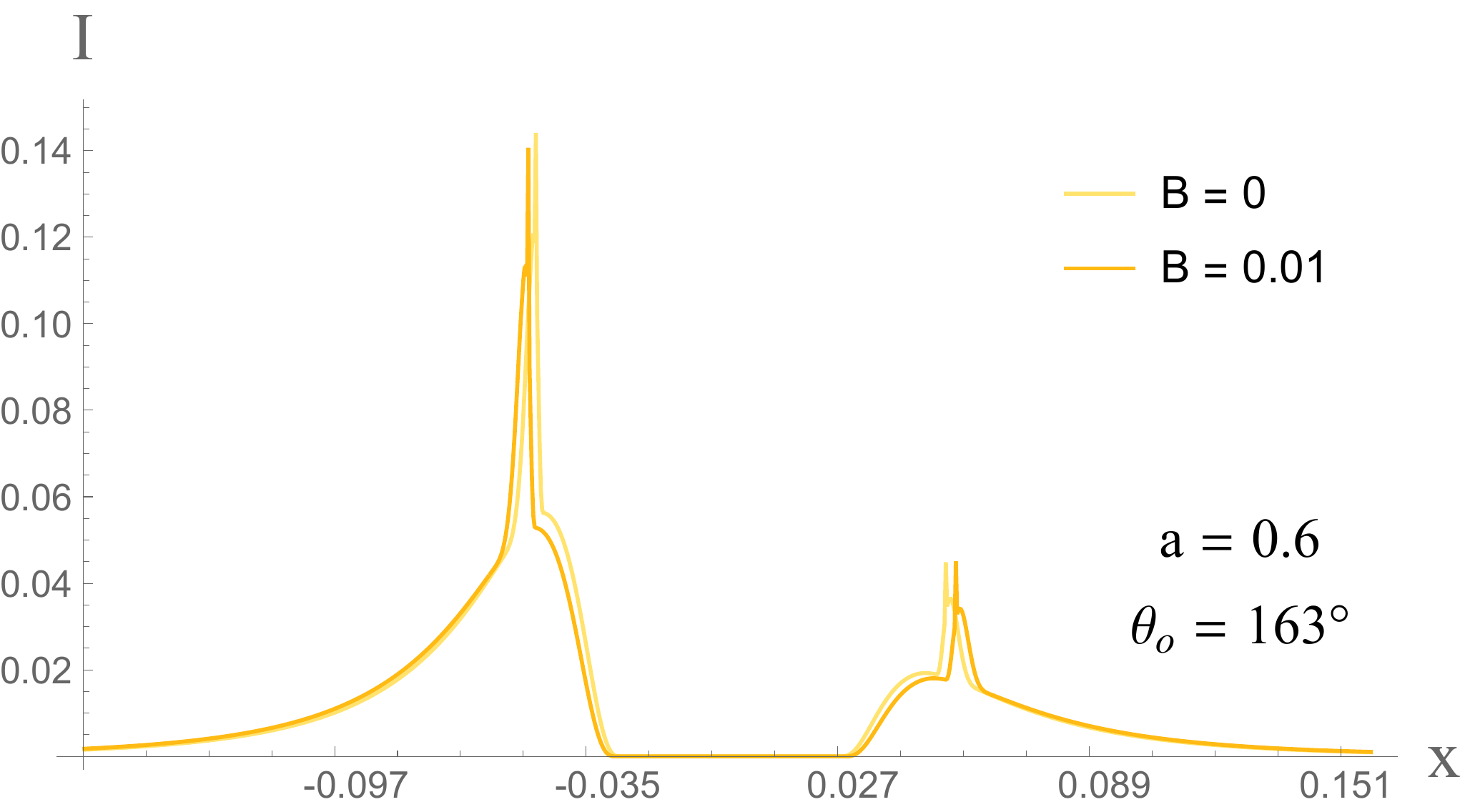}
   \includegraphics[scale=0.183]{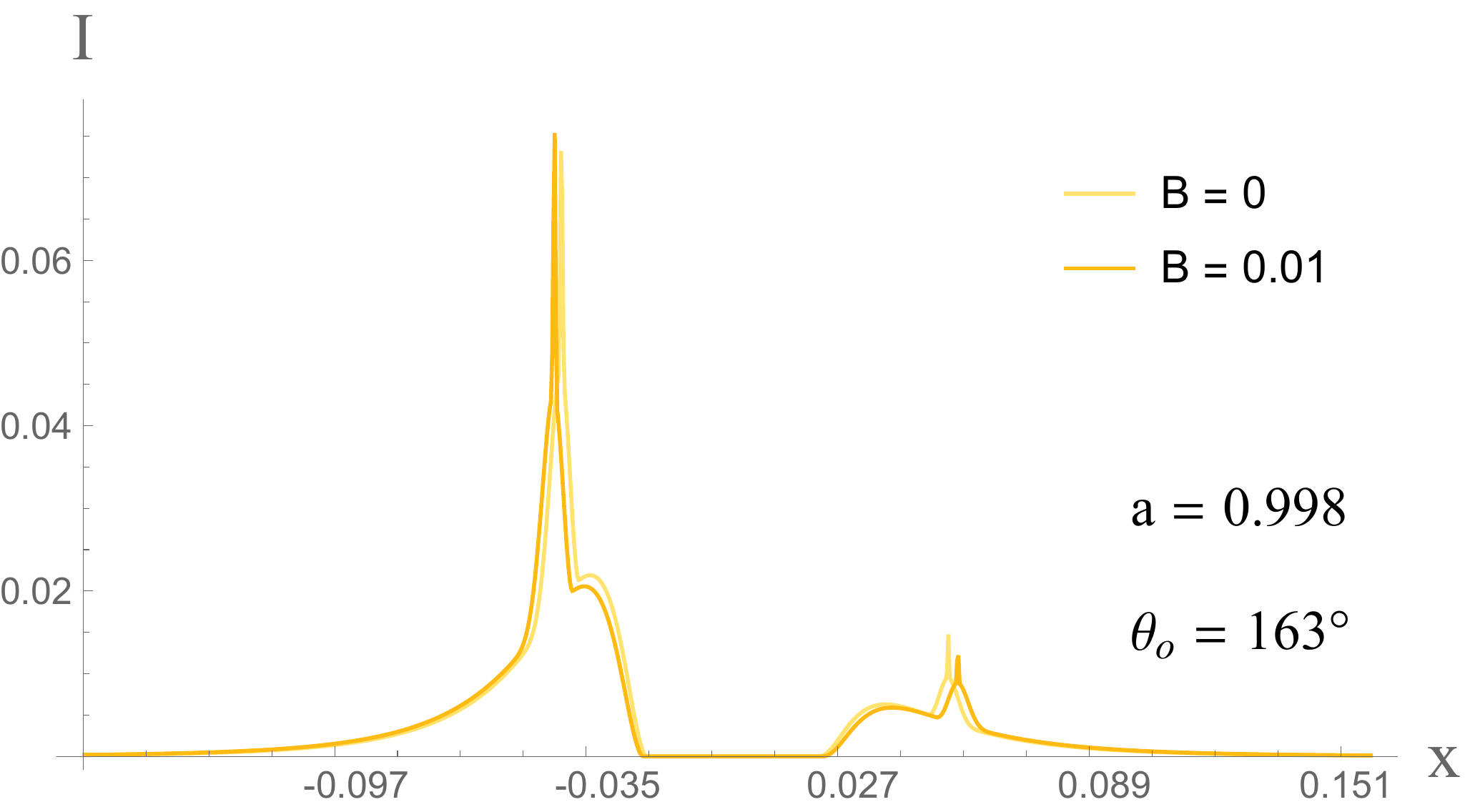} \\
   ~\\
   \includegraphics[scale=0.183]{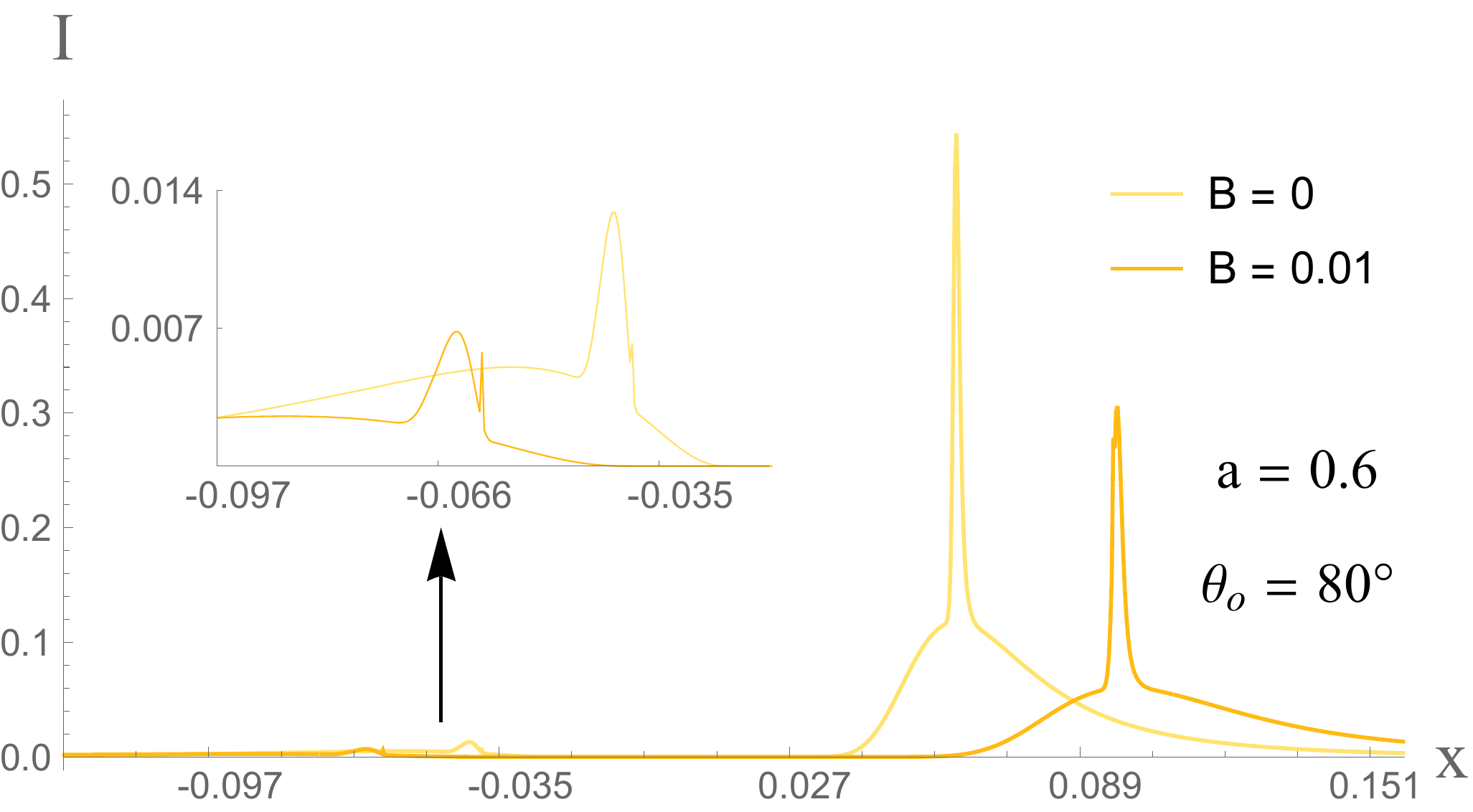}
   \includegraphics[scale=0.183]{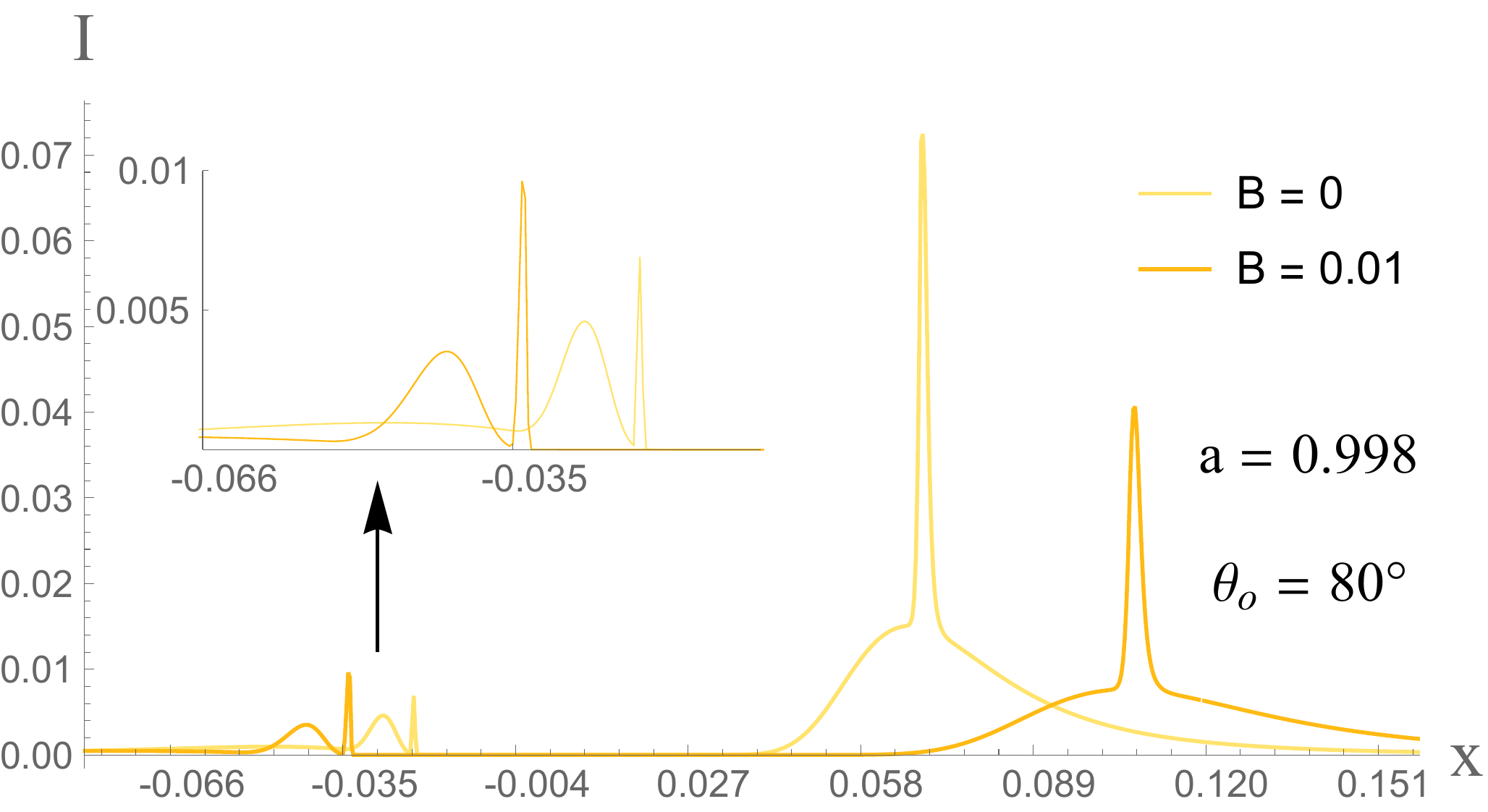} 
   	\includegraphics[scale=0.183]{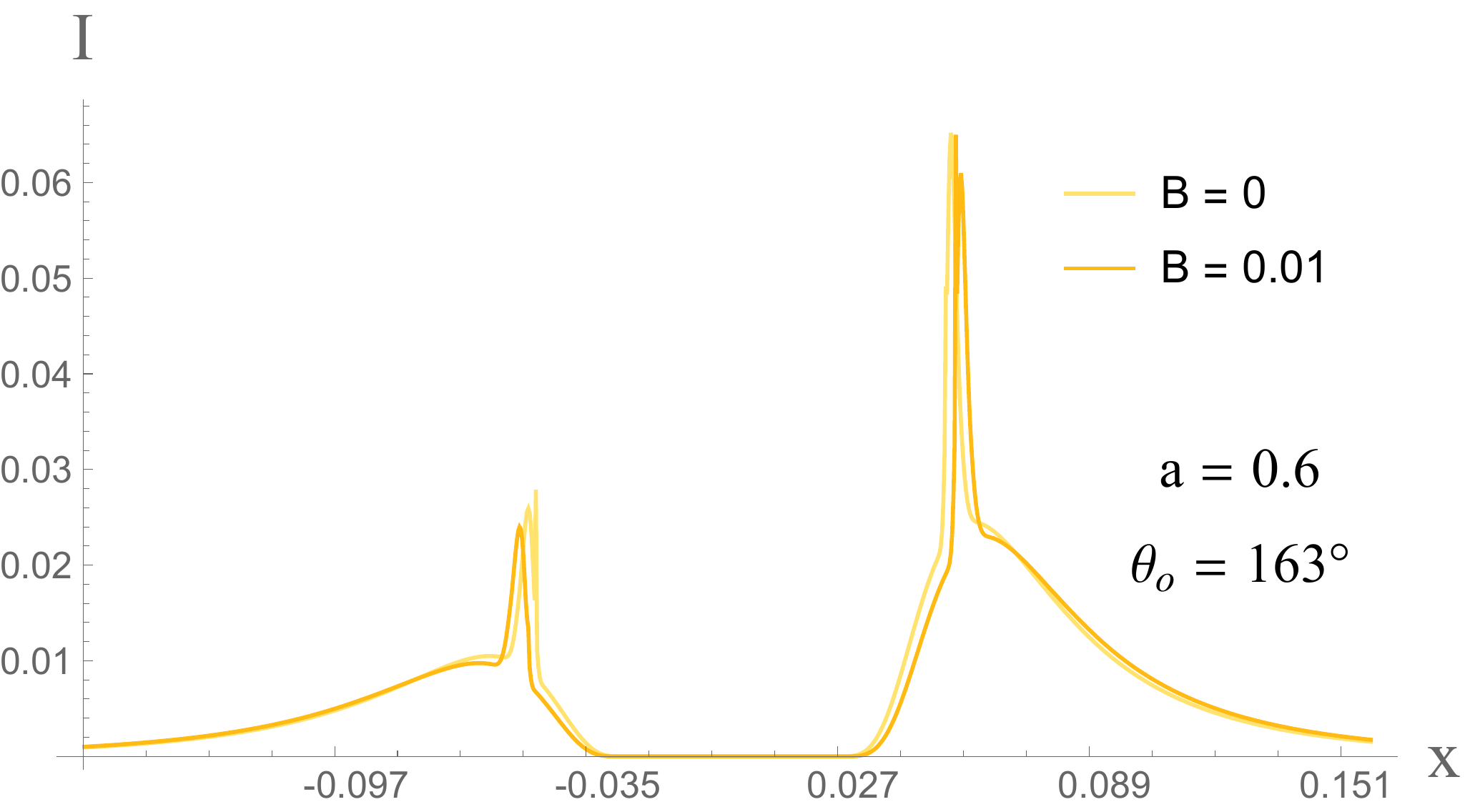}
   \includegraphics[scale=0.183]{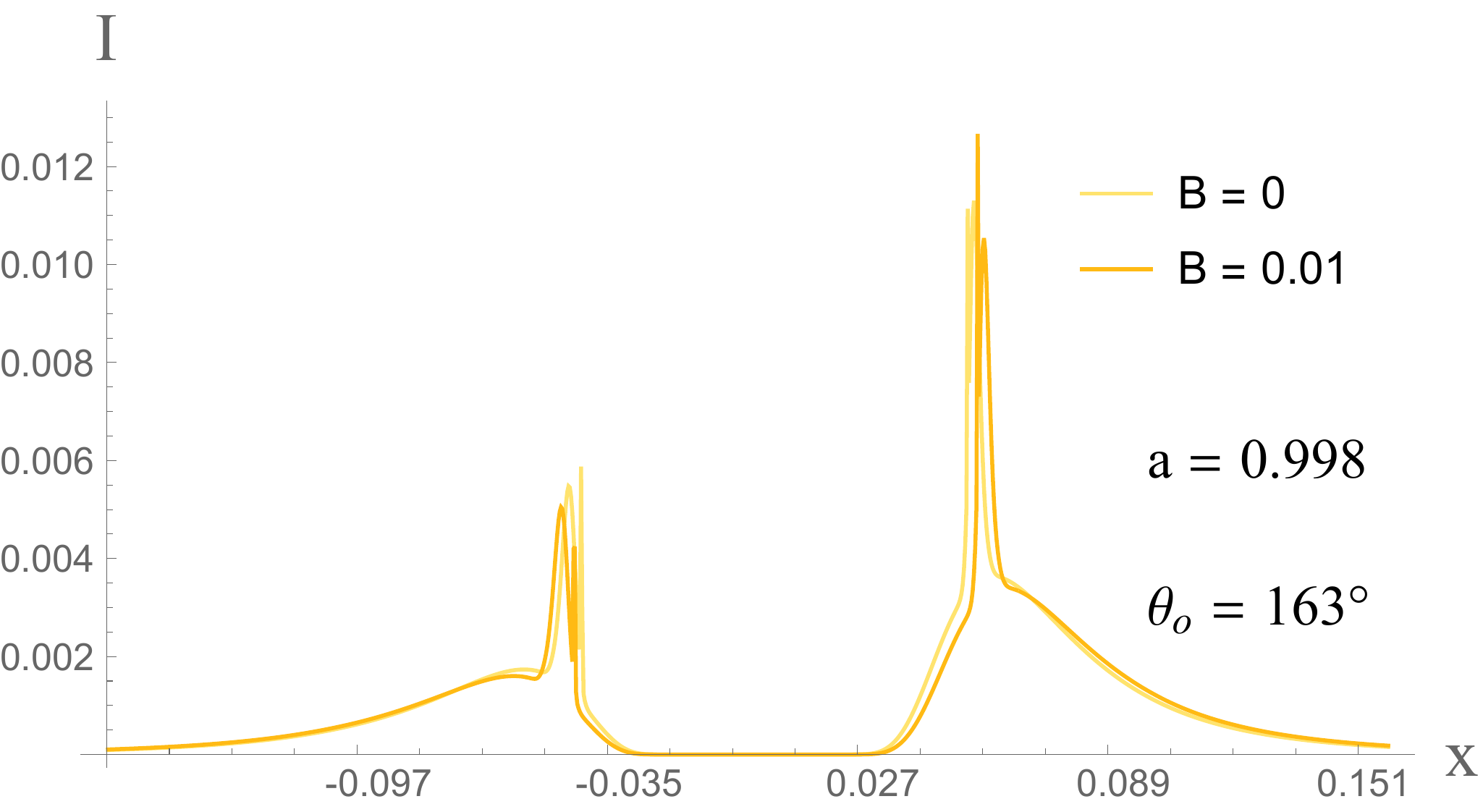} 
	\caption{Intensity distribution along $x$-axis of the screen. The top rows correspond to prograde flow and the bottom rows correspond to retrograde flow.}
	\label{x}
\end{figure}

\begin{figure}[h!]
	\centering
	\includegraphics[scale=0.183]{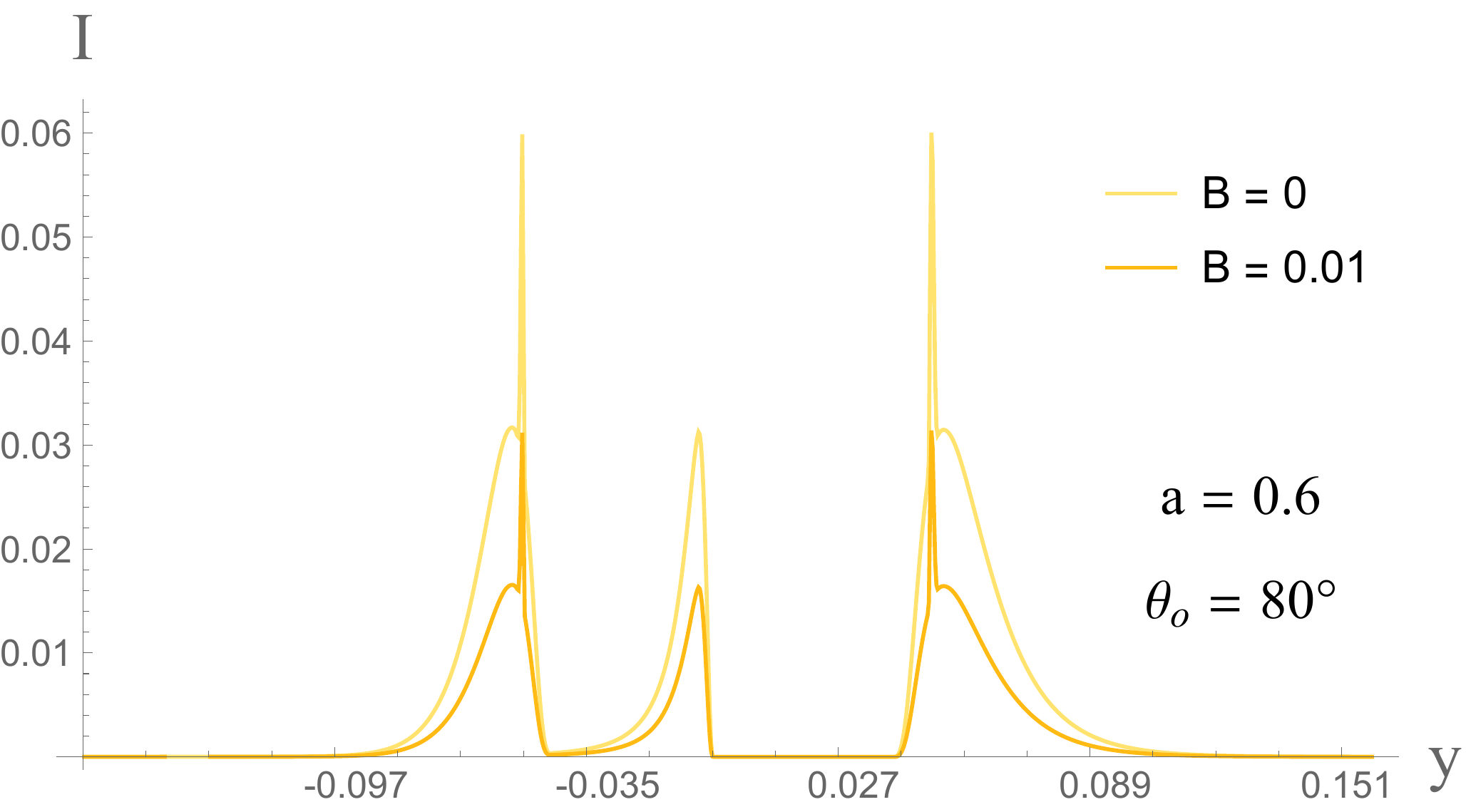}
   \includegraphics[scale=0.183]{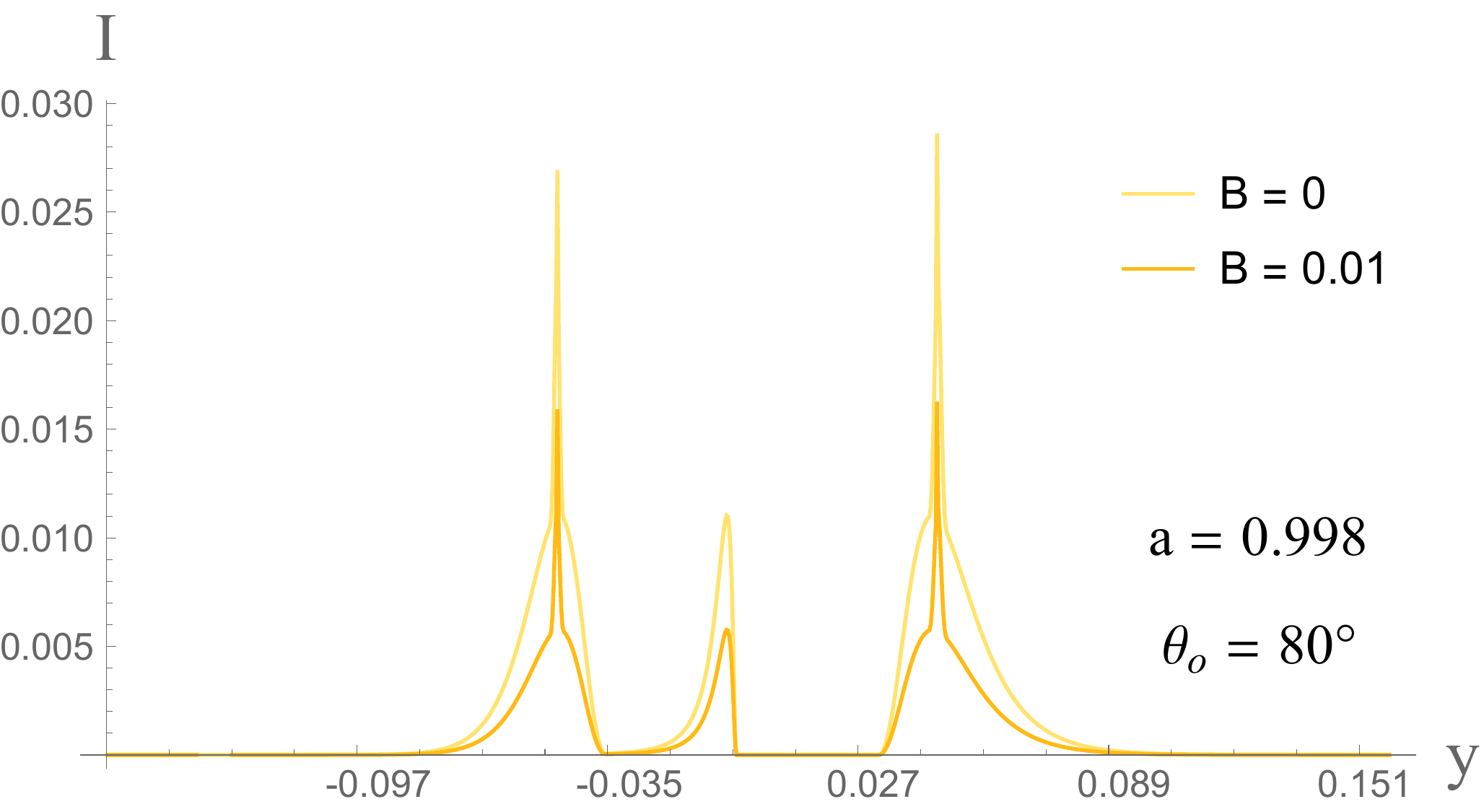} 
   	\includegraphics[scale=0.183]{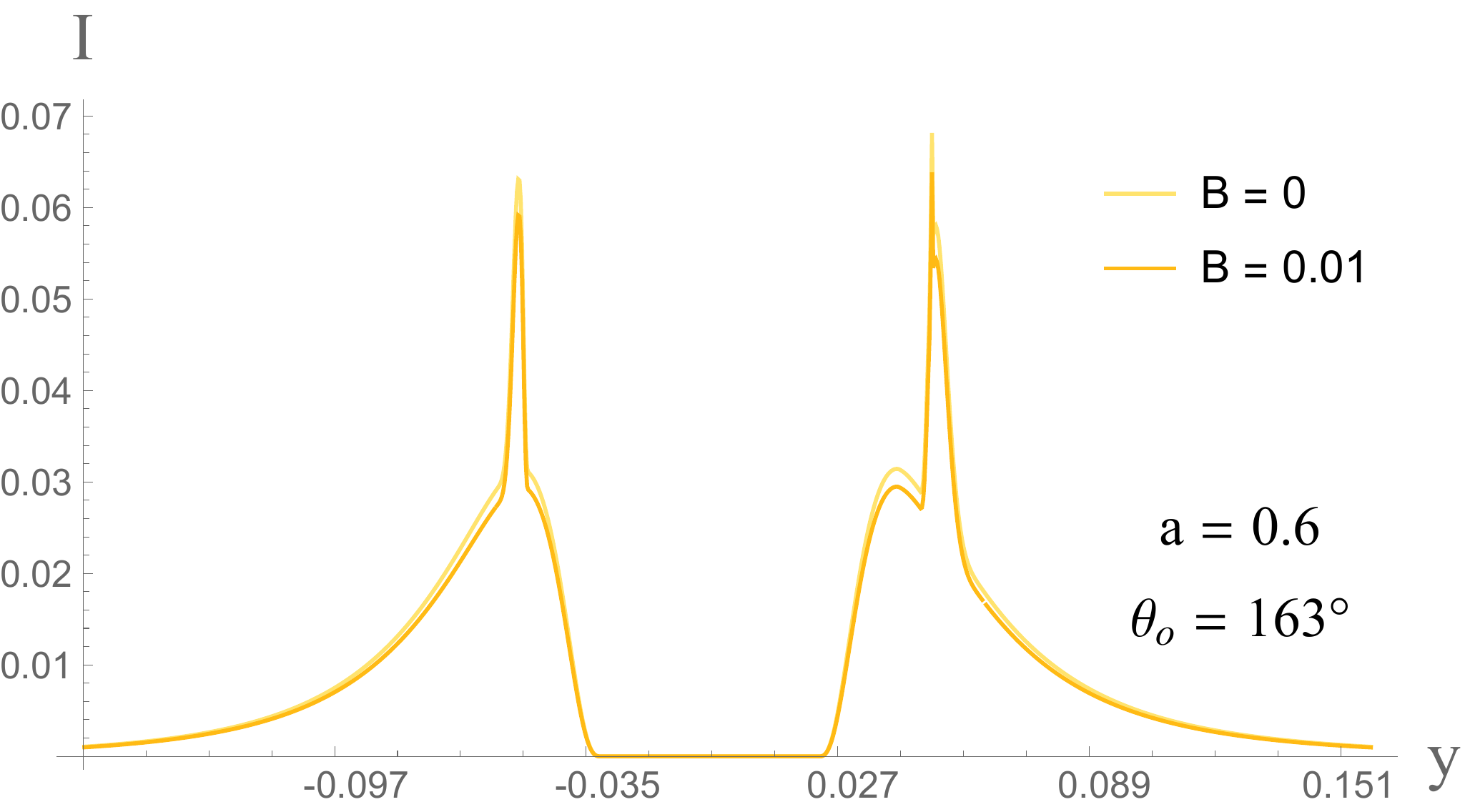}
   \includegraphics[scale=0.183]{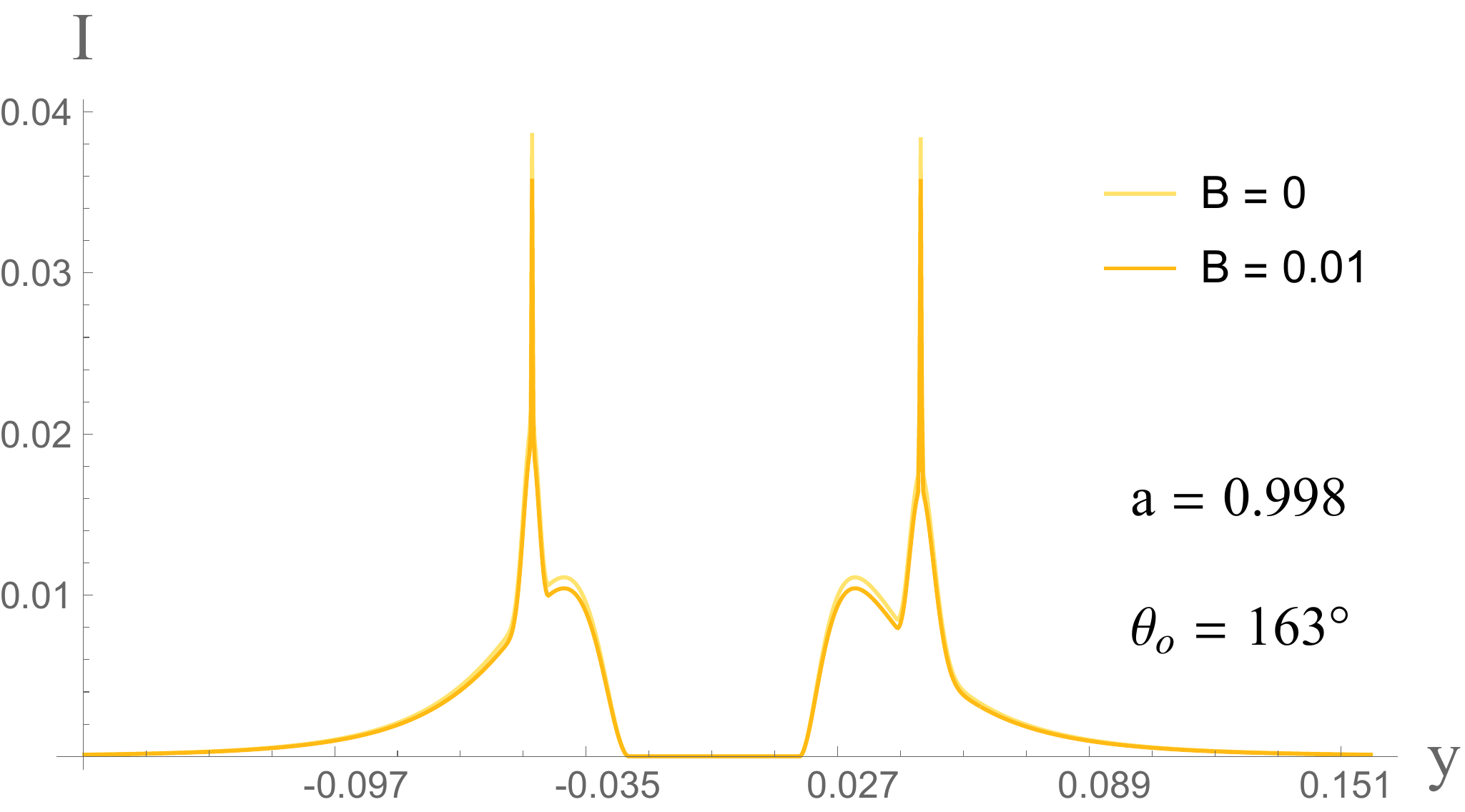} \\
   ~\\
   \includegraphics[scale=0.183]{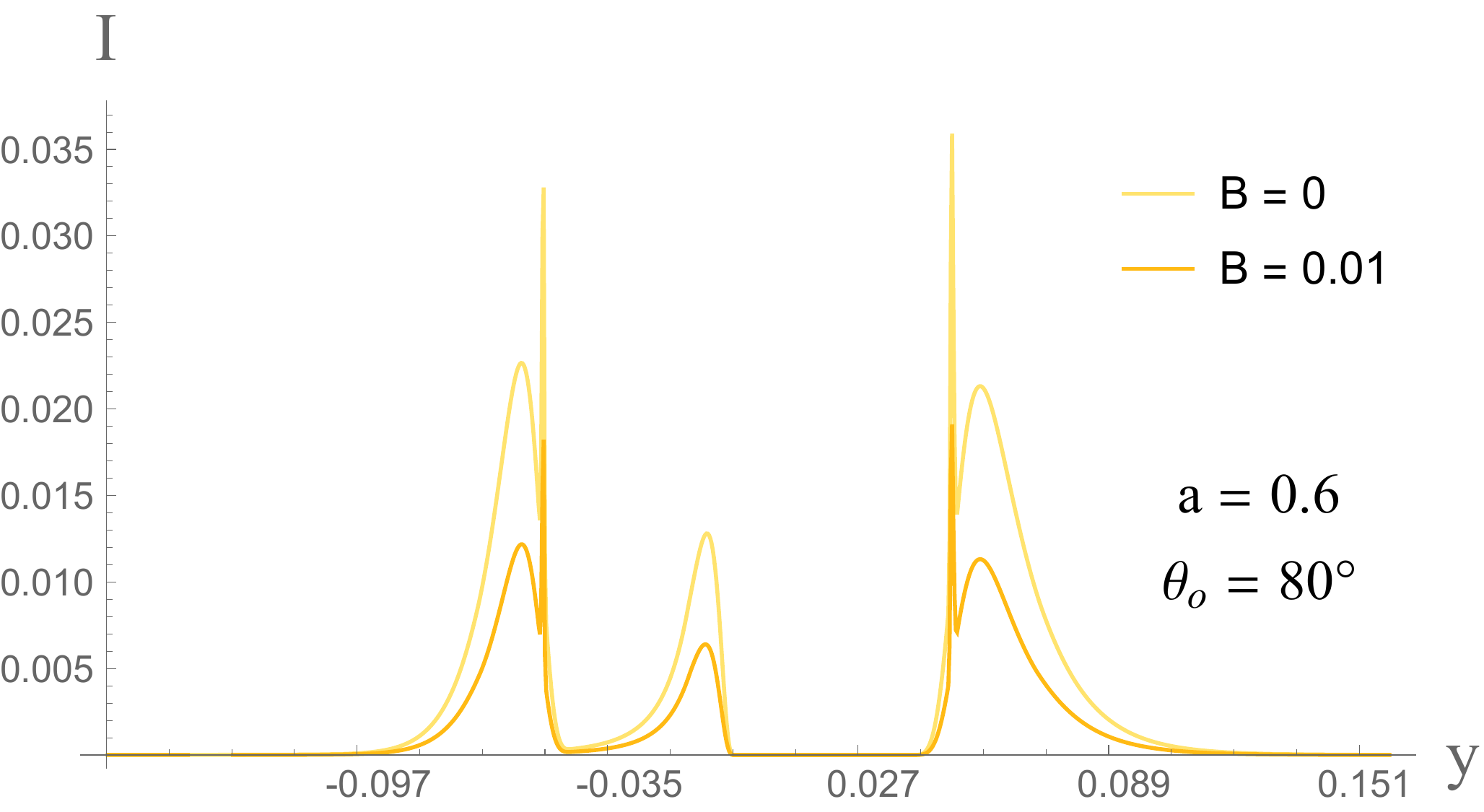}
   \includegraphics[scale=0.183]{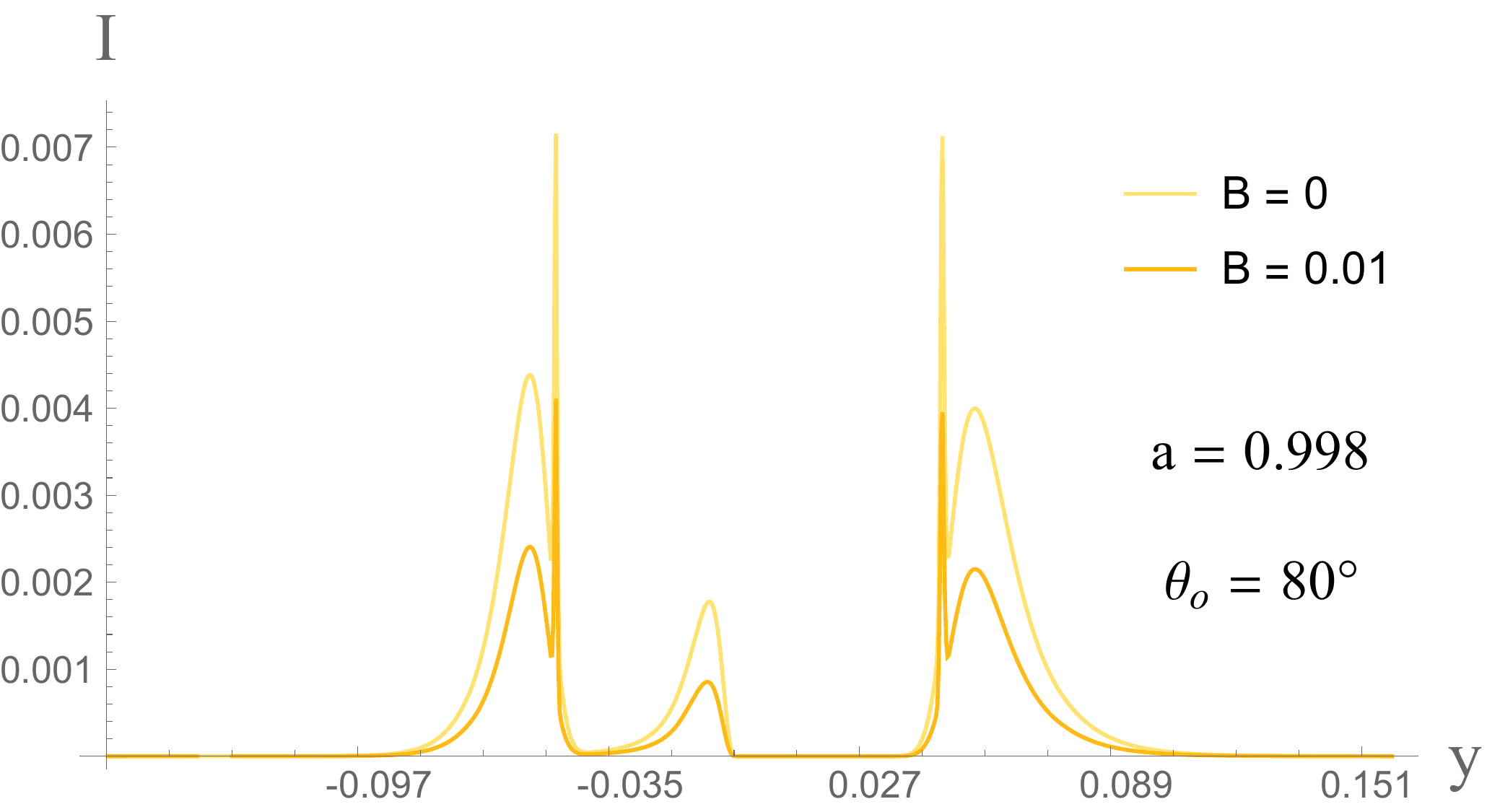} 
   	\includegraphics[scale=0.183]{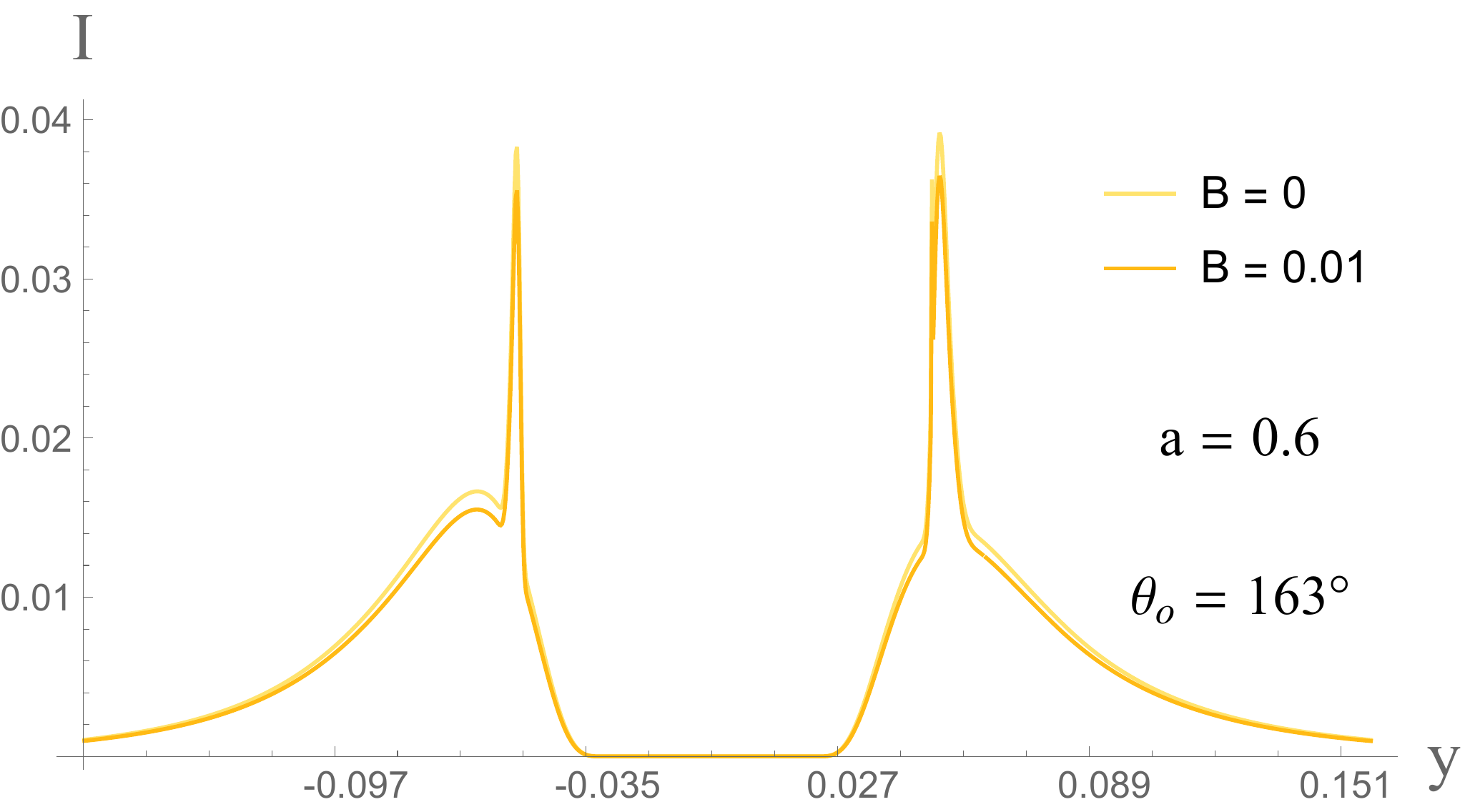}
   \includegraphics[scale=0.183]{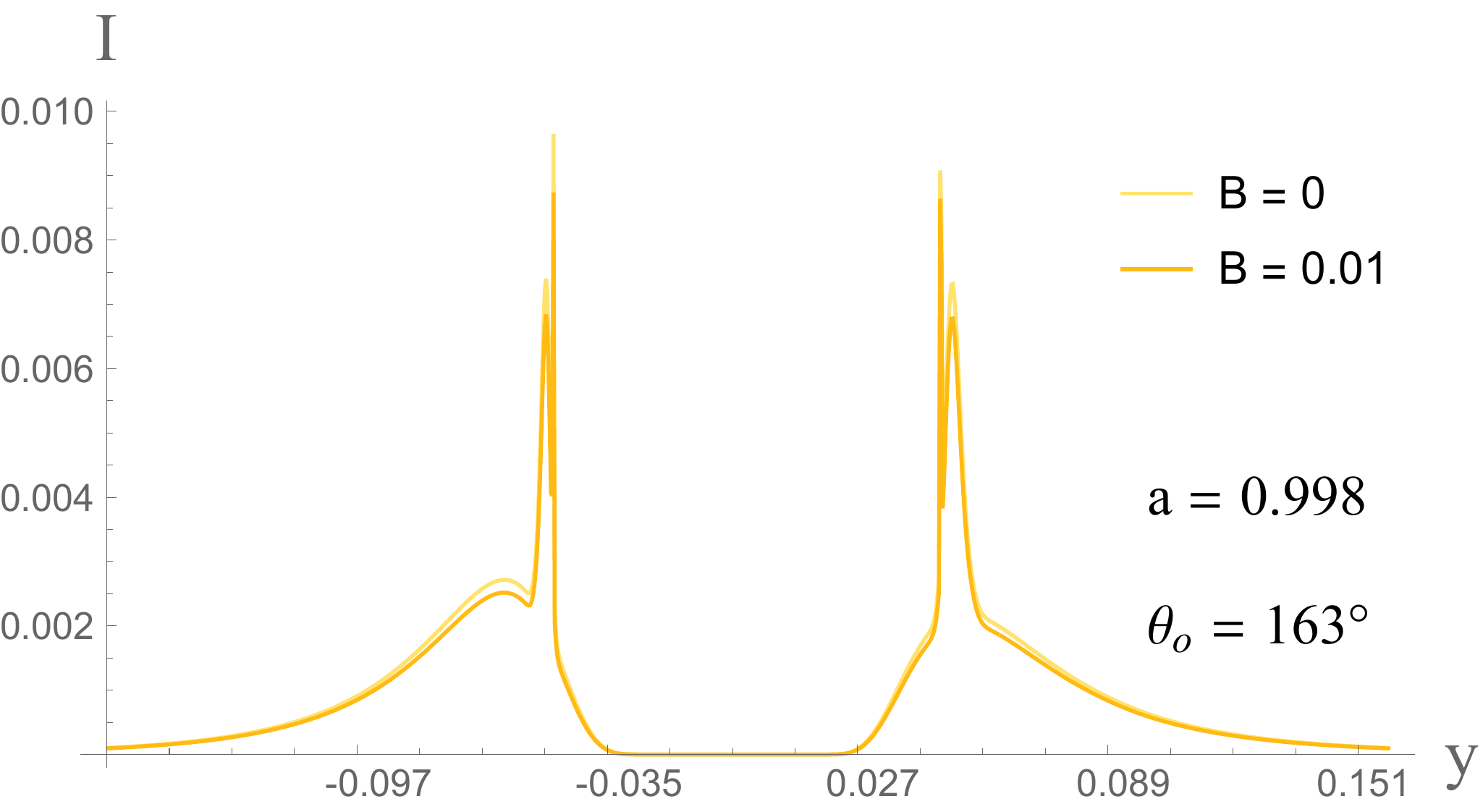} 
	\caption{Intensity distribution along $y$-axis of the screen. The top rows correspond to prograde flow and the bottom rows correspond to retrograde flow.}
	\label{y}
\end{figure}

Next, we turn to discuss the intensity distribution of the image. In Fig. \ref{x}, we show the intensity distribution along the $x$-axis and in Fig. \ref{y} we show the intensity distribution along $y$-axis on the screen. We use the light and deep yellow lines to label $B=0$ and $B=0.01$ separately. From these plots it is not hard to find that the magnetic field has significant influence on the $x$-axis distribution when $\t_o = 80^{\circ}$ while having weaker effect on the $y$-axis distribution. More precisely, the peak of the intensity is significantly reduced and the positions of the peaks move to either side along the $x$-axis in the presence of the magnetic fields. Along the $y$-axis the positions of the peaks keep unchanged and only the intensity at the peak decreases due to the magnetic field. The results become different for $\theta_o=163^\circ$: comparing with the $B=0$ case, the reductions of the peaks of the intensity are very small and there is no variation of the positions of the peaks for $B=0.01$, regardless of whether the intensity distribution is along $x$-axis or along $y$-axis.

In Table. \ref{table} we also present the total flux defined as $F = r_o^{-2}\int{I(x,y)dxdy}$ under different spin parameters and magnetic field strengths, where $I(x,y)=I_{\n_o}$ is the observed intensity in Eq.~\eqref{Io1}, and the results are normalized by the total flux of prograde disk around KMBH with $a = 0$, $B=0$, observed at $\t_o = 80^{\circ}$. The integration region is the entire image within the field of view. When far from the center of the viewpoint $(0,0)$ the intensity is mainly from the emission of direct image, $I(x,y)\sim J(r_1(x,y))$, thus the integral becomes $\int{J(r_1)dxdy} \sim 2\pi\int{J(r)rdr} \sim 2\pi \int{e^{-z^2/2}}dz$, which converges quickly and the contribution outside the field of view can be neglected.
\begin{figure}[h!]
	\centering
   	\includegraphics[scale=0.33]{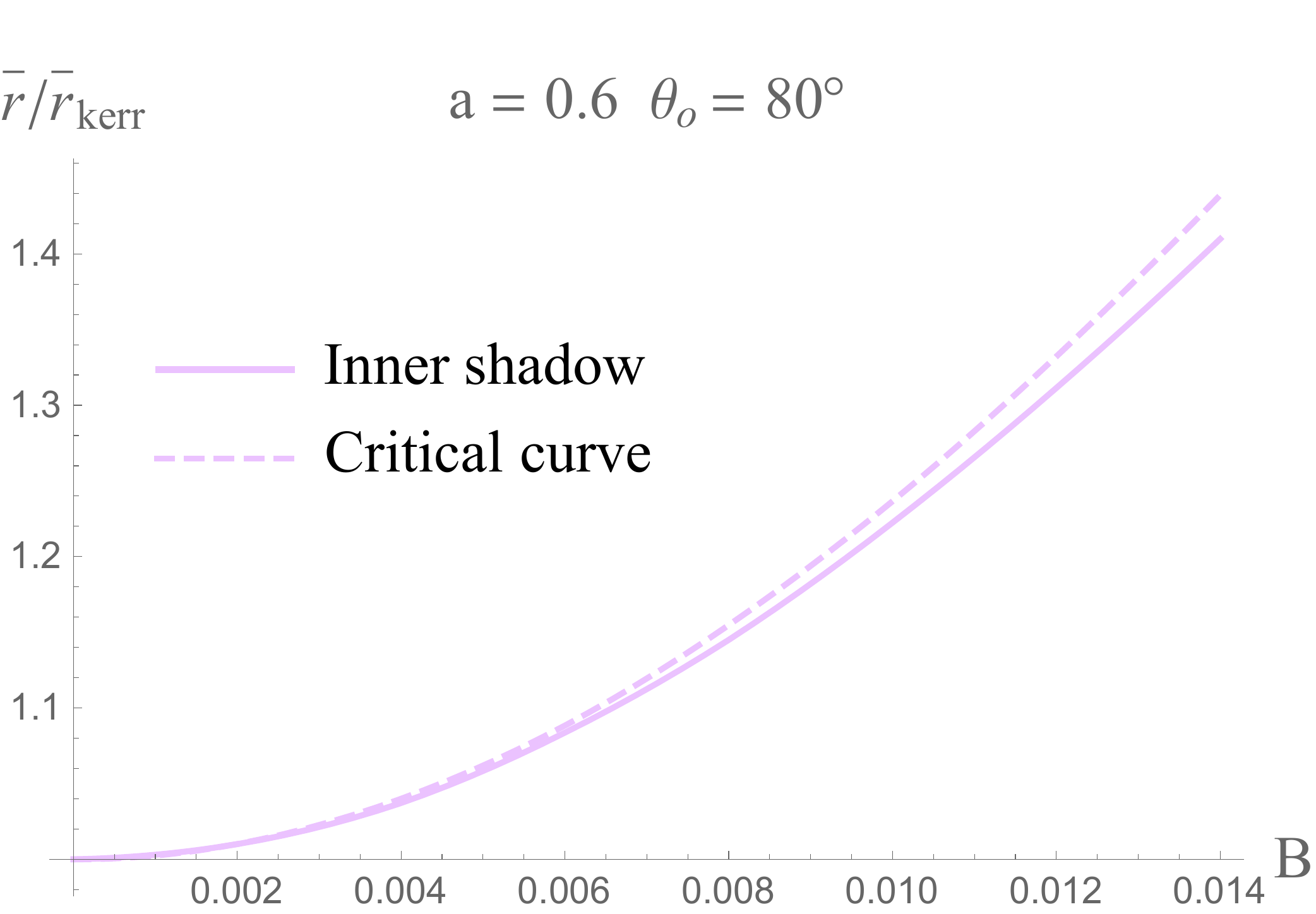}
   \includegraphics[scale=0.33]{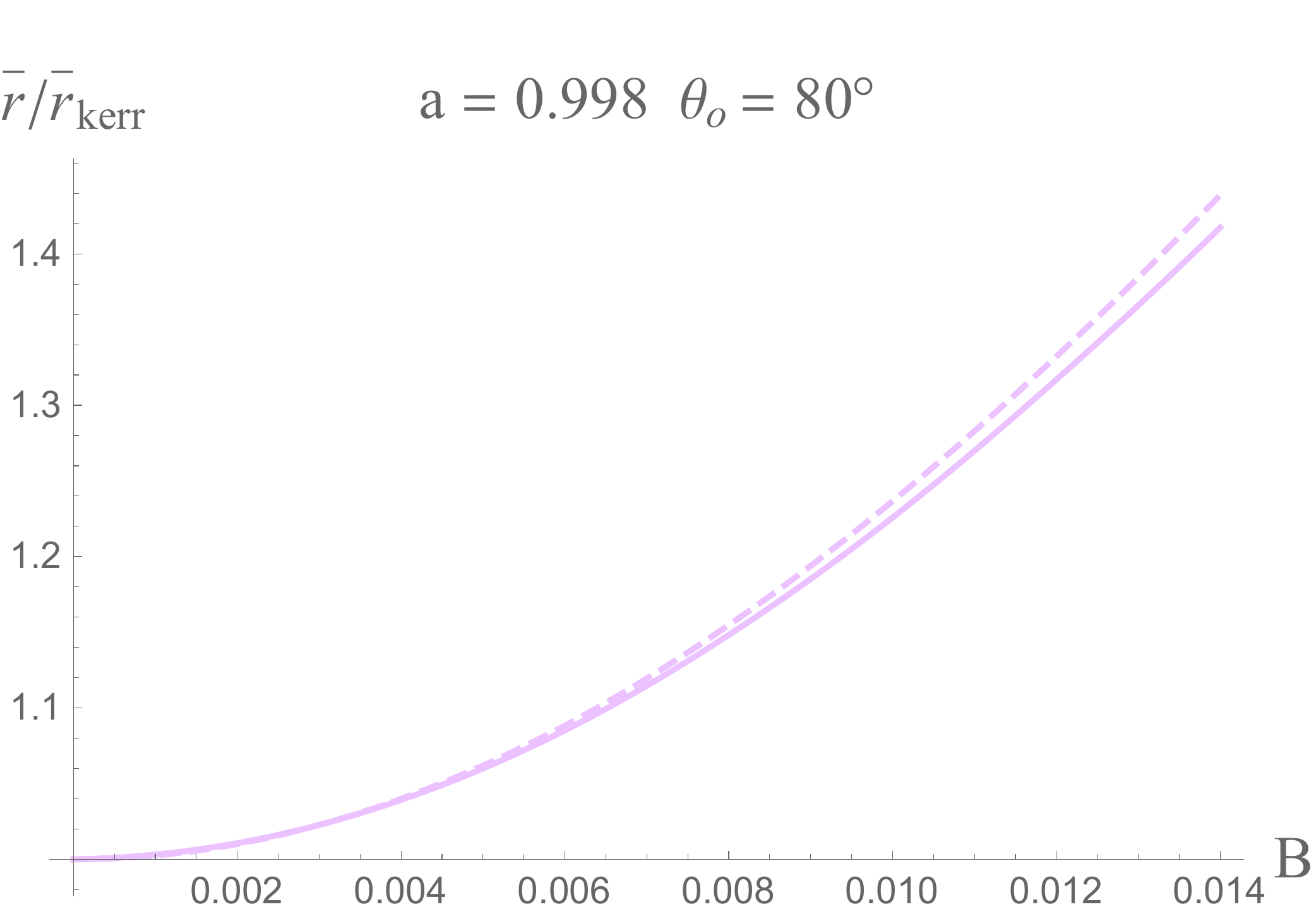} \\
   ~\\
   	\includegraphics[scale=0.35]{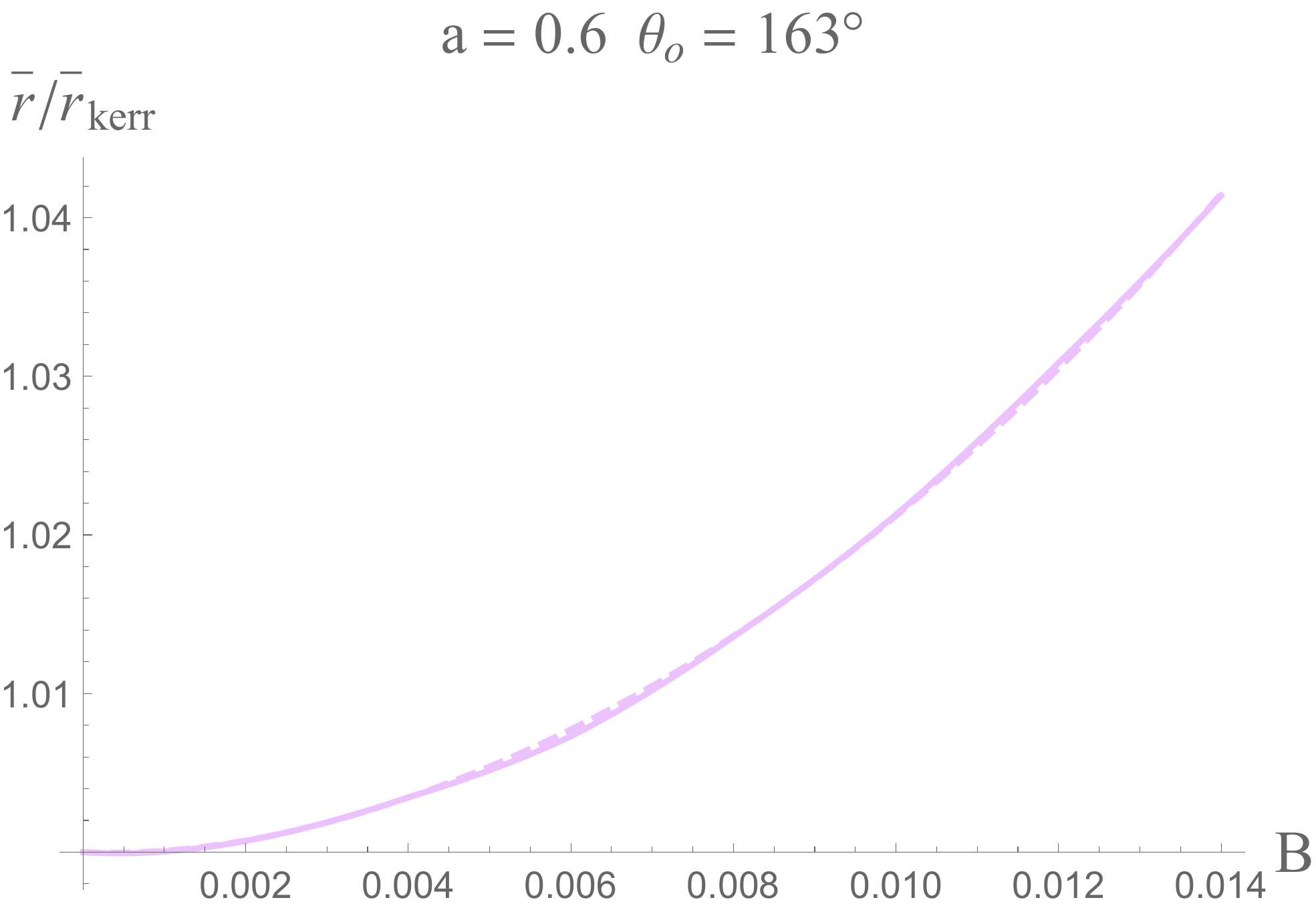}
   \includegraphics[scale=0.35]{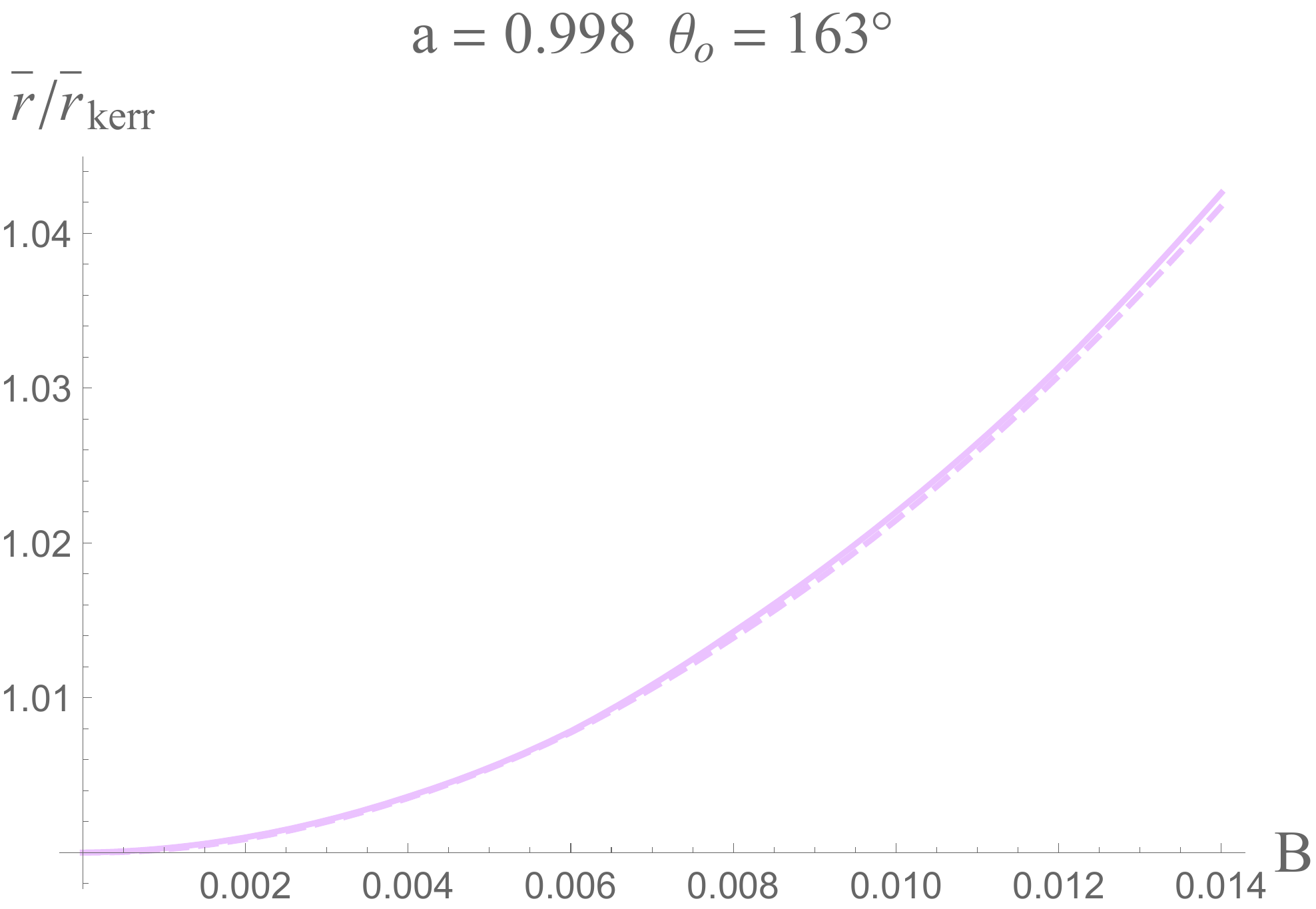} 
	\caption{Rescaled average radius as the functions of $B$.}
	\label{ros}
\end{figure}

At last, we move to study the influence of the magnetic field on the sizes of the critical curve and inner shadow of the KMBH. Recall that the coordinates on the screen of the ZAMO are defined in Eq.~\eqref{xyd}, we can define $(x_c, y_c)$,
\bea
x_c=\frac{x_{\ma}+x_\mi}{2}\,,\quad y_c=\frac{y_\ma+y_\mi}{2}\,,
\eea
as the geometric center of a concerned curve (the critical curve or the inner shadow), where $x_{\ma,\mi}$ is the maximal/minimal horizontal coordinate, and $y_{\ma,\mi}$ is the maximal/minimal vertical coordinate, respectively. Then, we are allowed to introduce the polar coordinates $(\rho, \alpha)$ with the origin at $(x_c, y_c)$ and the average radius is defined as 
\bea
\bar{r}=\frac{1}{2\pi}\int_0^{2\pi} \rho(\alpha)d\alpha\,.
\eea
In order to show the difference between the KMBH and Kerr black hole, it is convenient to calculate a rescaled quantity $\sigma=\bar{r}/\bar{r}_{\text{Kerr}}$, where $\bar{r}_{\text{Kerr}}$ is the average radius of the closed curve for the Kerr black hole with the same spin parameter as the KMBH. We show several results of $\bar{r}/\bar{r}_{\text{Kerr}}$ with respect to $B$ in Fig. \ref{ros}, under different spin and observational angle. One can see that the profiles of the critical curve and the inner shadow overlap, but certainly do not equate. This means that both of them can be used to estimate the magnetic field around the black hole. In addition, we also see that the difference between the critical curve and the inner shadow becomes large with the increasing of the strength of the magnetic field $B$ from Fig. \ref{ros}. Moreover, when the observational angle $\theta_o$ is closer to $\pi/2$ other than the pole, the difference becomes more significant. 

\section{Conclusion } \label{sec5}
In this paper, we considered a geometrically and optically thin accretion disk as a light source and studied the appearance of the KMBH. Employing the numerical backward ray-tracing method, we obtained the direct and lensed images of KMBHs with different parameters including the spin, the observational angle and the strength of the magnetic field. We also took into account of both the prograde and retrograde flows of accretion disks. In an image of the KMBH, we mainly focused on four characteristic curves, that is, the direct and lensed images of the outer contour of the accretion disk, the critical curve and the inner shadow. Quantitatively, we studied the redshift of the direct and lensed images of the accretion disk, the intensity distribution along the $x$-axis and $y$-axis, and the variations of the average radii of the critical curve and the inner shadow with respect to the strength of the magnetic field. From our results, we found that the magnetic field would give a significant influence on the shapes of the four curves and the intensity distribution when the observational angle is closer to $\pi/2$. This suggests most importantly that the critical curve and the inner shadow could be used to estimate the magnetic field around a black hole in addition to the spin and mass of the black hole.

\section*{Acknowledgments}
The work is in part supported by NSFC Grant  No. 11735001, 11775022 and 11873044. MG is also supported by ``the Fundamental Research Funds for the Central Universities'' with Grant No. 2021NTST13.

\appendix

\section{Circular orbits in Melvin universe}\label{app:Melvin}
The metric of the Melvin universe in spherical coordinates is of the form \cite{Melvin:1965zza}
\be
ds^2 = \L_0^2(-dt^2 +dr^2+r^2d\t^2) + \f{1}{\L_0^2}r^2\sin^2{\t}d\phi^2,
\ee
where
\be
\L_0 = \L\big|_{a = 0}= 1 + \f{B^2r^2}{4}\sin^2{\t}.
\ee
This spacetime is stationary and axisymmetric due to the existence of two Killing vectors along the $t$-direction and $\phi$-direction, respectively. In addition, it has a translational  symmetry along the direction of the magnetic field, which can be seen directly from the cylindrical coordinates, i.e., $(\rho, z ) = r(\sin{\t}, \cos{\t})$.
For timelike equatorial geodesic, the equations of motion are
\bea
u^r = - \sqrt{-\f{V_0(r, E, L )}{g_{rr}}}\bigg|_{\t=\f{\pi}{2}} , \ u_t = -E, \ u_{\phi} = L, \ \t = \f{\pi}{2}, 
\eea
where
\be
V_0(r, E, L ) = -1 + \f{16E^2}{(4+B^2r^2)^2} - \f{(4+B^2r^2)^2L^2}{16r^2}
\ee
is the effective potential. The circular orbits should satisfy $V_0 = \partial_r V_0 = 0$, which determine their conserved quantities as follows
\bea\label{ELM}
E_{0}(r) = \f{\sqrt{4-B^2r^2}(4+B^2r^2)}{4\sqrt{4-3B^2r^2}} \ , \quad
L_{0}(r) = \pm \f{4\sqrt{2}|B|r^2}{\sqrt{(4+B^2r^2)^2 (4-3B^2r^2)}} \
\eea
with $\pm$ denoting the prograde/retrograde orbits relative to the orientation of magnetic field. Obviously, this circular orbits can only exist in
\be\label{range}
0 \leq r \leq \f{2}{\sqrt{3}|B|}.
\ee
Actually, the center of the circular orbit can be located anywhere along the $z$-direction due to the translation symmetry. The stability of the circular orbits means the quantity
\bea\label{d2v0}
d_2V_0(r) =  \partial^2_r V_0 \big|_{E=E_{0}(r) , L = L_{0}(r)}= - \f{8B^2(32-12B^2r^2+3B^4r^4)}{(4+B^2r^2)^2(4-3B^2r^2)}
\eea
should be non-negative. But within the range Eq.~\eqref{range}, $d_2V_0$ is always negative and thus there is no stable timelike circular orbit in the Melvin universe.

\section{Ray-tracing process and intensity formula}\label{app:ray-tracing}

To derive the observed intensity from the transfer equation Eq.~\eqref{transfer}, we define
\be
\mI = \f{I_{\nu}}{\nu^3} , \hspace{3ex}  \mJ = \f{J_{\nu}}{\nu^2} , \hspace{3ex} \mK = \kappa_{\nu}\nu,
\ee
 all of which are scalars \cite{Lindquist:1966igj}. In the ray-tracing perspective, we set $\lambda \to -\lambda$ and thus $J \to -J$, $\kappa \to -\kappa$ in Eq.~\eqref{transfer}. We use $F_n$ in Sec.~\ref{sec3} to label the rest frame of the disk on $r_n$. When tracing back and passing through the disk medium at $F_n$, the light ray undergoes an emission and absorption process
\be
\f{d\mI}{d\lambda} = -\mJ_n + \mK_n \mI,
\ee
and changes from $\mI_{n-1}$ to $\mI_{n}$. The initial scalar, $\mI_0$, is represented by the quantities measured on the screen, i.e., $\mI_0=I_o/\n_o^3$. Because the disk is very thin, the emission and absorption coefficients can be taken as constants such that
\be
\mI_n = e_n \mI_{n-1} + \d_n
\ee
with the coefficients
\be
e_n = e^{\mK_n\D\lambda_n}, \ \d_n = (1-e_n)\f{\mJ_n}{\mK_n}.
\ee
Thus, the final intensity can be expressed as
\bea
\mI_{N} &=& e_N\mI_{N-1} + \d_N = e_N(e_{N-1}\mI_{N-2}+\d_{N-1}) + \d_N\nn \\
&=&...... = e_N...e_1\mI_0 + e_N...e_2\d_1 + ... + e_N\d_{N-1} + \d_N
\eea
with $N = N_{max}$ the maximal crossing number. Since the end of the ray in ray-tracing perspective is actually the point of its origin, which is at the horizon or infinity with no initial intensity, $\mI_{N}$ must be zero and then we get a covariant expression of the observed intensity
\be
\mI_0 = -\sum_{n=1}^{N} \f{\d_n}{e_1...e_{n}}.
\ee
Replacing the three scalars with physical quantities $I_{n},J_{n},\kappa_{n}$ and $\nu_n = \nu_o/g_n$ observed in $F_n$, we can get Eq.~\eqref{Io}. Note that it works for geometrically thin disks with finite optical depth. When the optical depth is vary large, i.e., the disk medium interacts strongly with photons, Eq.~\eqref{Io} no longer applies and one should consider the emission profile as black body radiation.

\section{Chaos in the image}\label{appB}
\begin{figure}[h!]
	\centering
	\includegraphics[scale=0.4]{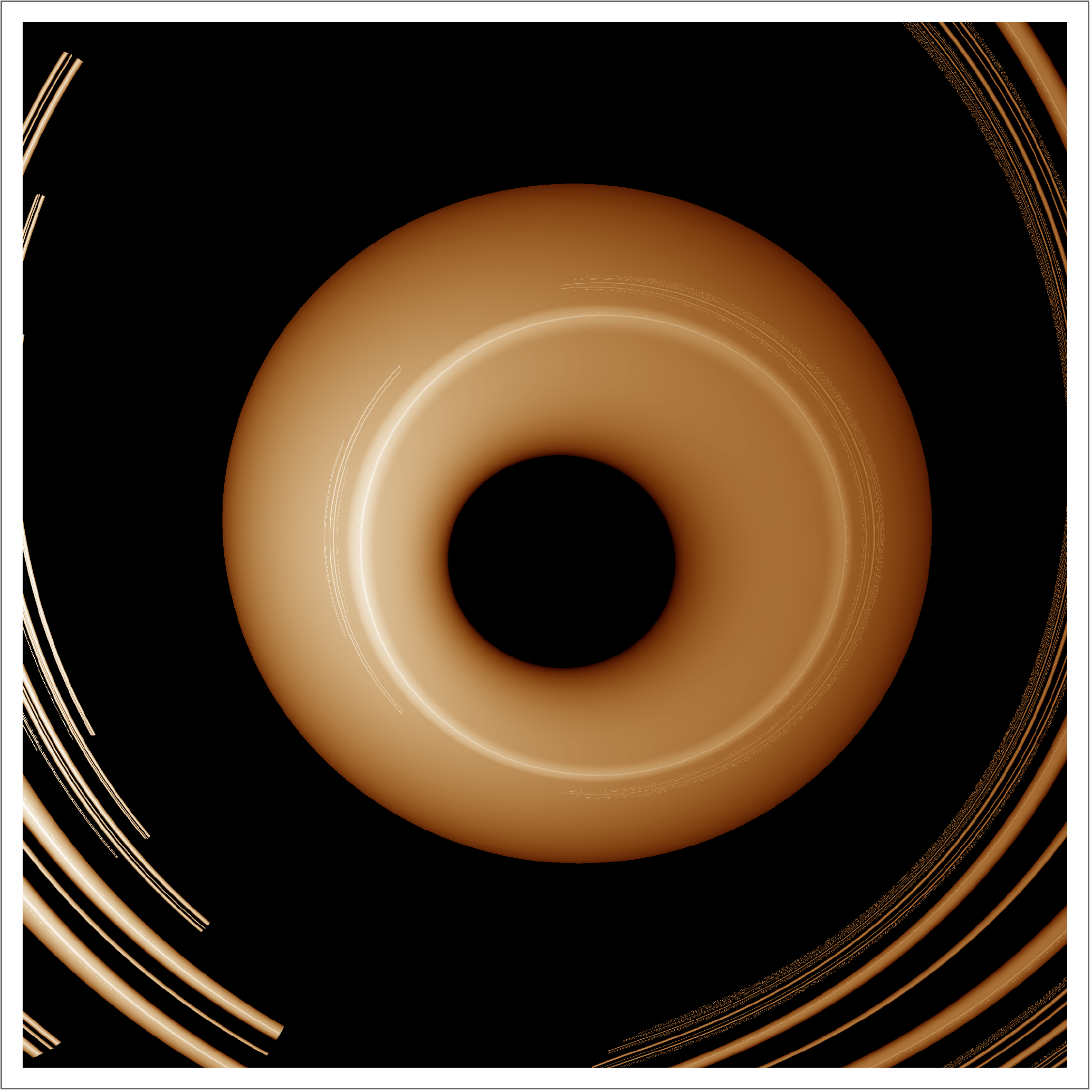} 
	\caption{An Image of KMBH illuminated by prograde flows. We set $r_{\text{ir}}=r_H$, $r_{\text{or}}=r_o=5$, $B=0.2$, $a=0.998$ and $\theta_o=163^\circ$.}
	\label{chaos}
\end{figure}

In this section, we would like to briefly discuss the chaos in the image of KMBH with thin accretion disk for a theoretical comparison with the results in \cite{Junior:2021dyw, Wang:2021ara}. We give an example in Fig. \ref{chaos}.

From the Fig. \ref{chaos}, we can see that there are some irregular bright lines at the edge of the picture and also some irregular circles and arcs appear in the middle of the image of the accretion disk. These results suggest chaotic behaviors of the gravitational lensing in the KMBH spacetime when the magnetic field becomes strong enough. As a result, we can conclude that the chaos in the image not only happen for spherical source \cite{Junior:2021dyw, Wang:2021ara}, but also happen for the case illuminated by an accretion disk in the KMBH spacetime with a strong enough magnetic field. 

\bibliographystyle{utphys}
\bibliography{note}

\end{document}